\documentclass[12pt]{article}

\newlength{\abstractwidth}
\usepackage{amsmath, nccmath, amsthm}
\usepackage{amsfonts}
\usepackage{amssymb}
\usepackage{latexsym}
\usepackage{shuffle}

\usepackage{color}
\usepackage{graphicx}
\usepackage{tikz}
\usepackage{dsfont}

\usepackage{tikz}
\usetikzlibrary{calc} 
\usetikzlibrary{patterns,snakes} 
\usetikzlibrary{decorations.pathreplacing} 
\usetikzlibrary{decorations.markings} 
\usetikzlibrary{decorations.pathmorphing} 
\usetikzlibrary{positioning}
\usetikzlibrary{arrows.meta}

\usetikzlibrary{arrows,shapes,positioning}
\usetikzlibrary{decorations.markings}
\usepackage[rightcaption]{sidecap}
\tikzstyle arrowstyle=[scale=1]
\tikzstyle directed=[postaction={decorate,decoration={markings,
    mark=at position .65 with {\arrow[arrowstyle]{stealth}}}}]
\tikzstyle reverse directed=[postaction={decorate,decoration={markings,
    mark=at position .65 with {\arrowreversed[arrowstyle]{stealth};}}}]
\usetikzlibrary{positioning}

\usepackage{hyperref}
\definecolor{darkred}{rgb}{0.8,0.1,0.1}
\hypersetup{colorlinks=true, linkcolor=darkred, citecolor=blue, linktoc=page}

\numberwithin{equation}{section}

\renewcommand{\thefootnote}{\fnsymbol{footnote}}
\renewcommand{\thanks}[1]{\footnote{#1}}
\newcommand{\starttext}{
\renewcommand{\thefootnote}{\arabic{footnote}}}

\newcommand{\bea}{\begin{eqnarray}}
\newcommand{\eea}{\end{eqnarray}}
\newcommand{\be}{\begin{eqnarray}}
\newcommand{\ee}{\end{eqnarray}}
\newcommand{\bma}{\begin{matrix}}
\newcommand{\ema}{\cr\end{matrix}}

\newtheorem{thm}{Theorem}[section]
\newtheorem{lem}[thm]{Lemma}
\newtheorem{prop}[thm]{Proposition}
\newtheorem{cor}[thm]{Corollary}

\newtheorem{deff}[thm]{Definition}
\newtheorem{rmk}[thm]{Remark}
\newtheorem{DP}[thm]{Definition-Proposition}

\def\beq{\begin{equation}}
\def\eeq{\end{equation}}

\def\cC{{\cal C}}
\def\cD{{\cal D}}
\def\cE{{\cal E}}
\def\cF{{\cal F}}
\def\cG{{\cal G}}

\def\cJ{{\cal J}}
\def\cK{{\cal K}}
\def\cL{{\cal L}}

\def\cR{{\cal R}}
\def\cS{{ \theta}}
\def\cT{{\cal T}}
\def\cU{{\cal U}}

\def\cW{{\cal W}}

\def\bG{{\bf G}}

\def\bJ{{\bf J}}
\def\bK{{\bf K}}

\def\mA{\mathfrak{A}}
\def\mB{\mathfrak{B}}

\def\mF{\mathfrak{F}}
\def\mG{\mathfrak{G}}
\def\mH{\mathfrak{H}}
\def\mI{\mathfrak{I}}
\def\mJ{\mathfrak{J}}

\def\mL{\mathfrak{L}}
\def\mM{\mathfrak{M}}
\def\mN{\mathfrak{N}}

\def\mP{\mathfrak{P}}
\def\mQ{\mathfrak{Q}}
\def\mR{\mathfrak{R}}
\def\mS{\mathfrak{S}}

\def\mf{\mathfrak{f}}
\def\mg{\mathfrak{g}}

\def\mt{\mathfrak{t}}

\def\ZZ{{\mathbb Z}}

\def\CC{{\mathbb C}}

\def\Im{{\rm Im \,}}

\def\half{{1\over 2}}

\def\p{\partial}

\def\a{\alpha}
\def\b{\beta}

\def\f{\varphi}

\def\om{\omega}

\def\XX{{\bf X}}

\def\hf{{\gamma}}

\def\vI{{\vec{I}}}
\def\vJ{{\vec{J}}}
\def\vK{{\vec{K}}}

\def\vP{{\vec{P}}}
\def\vQ{{\vec{Q}}}
\def\vR{{\vec{R}}}
\def\vS{{\vec{S}}}
\def\vT{{\vec{T}}}
\def\vU{{\vec{U}}}
\def\vV{{\vec{V}}}
\def\vW{{\vec{W}}}
\def\vX{{\vec{X}}}
\def\vY{{\vec{Y}}}
\def\vZ{{\vec{Z}}}

\def\no{\nonumber}
\def\sm{\smallskip}

\def\pbx{\p_{\bar x}}

\def\xx{{\boldsymbol x}}
\def\yy{{\boldsymbol y}}

\begin{document}
\starttext
\setcounter{footnote}{0}

\begin{flushright}
2026 February 09 \\
UUITP--03/26
\end{flushright}

\vskip 0.1in

\begin{center}

{\Large \bf Equivalence of flat connections and  Fay identities}

\vskip 0.1in

{\large \bf on arbitrary Riemann surfaces}

\vskip 0.2in

{\large Eric D'Hoker${}^{a}$, Oliver Schlotterer${}^{b}$} 

\vskip 0.15in

{ \sl ${}^{a}$Mani L. Bhaumik Institute for Theoretical Physics}\\
{\sl  Department of Physics and Astronomy}\\
{\sl University of California, Los Angeles, CA 90095, USA}

\vskip 0.1in

{\sl ${}^b$Department of Physics and Astronomy,} \\
  { \sl Department of Mathematics,} \\
  { \sl Centre for Geometry and Physics,} \\ 
  {\sl Uppsala University, 75120 Uppsala, Sweden}

\vskip 0.15in 

{\tt \small dhoker@physics.ucla.edu, oliver.schlotterer@physics.uu.se}

\vskip 0.2in

\begin{abstract}
\vskip 0.1in
A flat connection on a Riemann surface with values in an infinite dimensional Lie algebra provides a systematic and effective tool for generating an infinite family of polylogarithms via iterated integrals. The recent literature offers different types of connections, in one or several variables, on compact Riemann surfaces with or without punctures, and in the meromorphic or single-valued categories. In this work, we show that the flatness conditions for  the single-valued and modular DHS connection in multiple variables, which was introduced in the companion paper arXiv:2602.01461, are equivalent to the union of all the \textit{interchange  and Fay identities} among DHS integration kernels that were proven in arXiv:2407.11476. Based on the same combinatorial techniques, the flatness conditions on the multivariable Enriquez connection is shown to imply the union of all the \textit{interchange and Fay identities} for Enriquez kernels.
\end{abstract}
\end{center}

\newpage

\setcounter{tocdepth}{2} 

\baselineskip=15pt
\setcounter{equation}{0}
\setcounter{footnote}{0}

\newpage

\section{Introduction}
\setcounter{equation}{0}
\label{sec:1}

A flat connection on an arbitrary compact Riemann surface $\Sigma$, valued in an infinite dimensional Lie algebra $\mg$, provides a systematic and effective tool for generating an infinite family of homotopy invariant iterated integrals that are collectively referred to as polylogarithms.

\sm

For a given choice of $\Sigma$ and $\mg$, various different categories of polylogarithms may be introduced depending on the number of variables entering the connection, the number of punctures allowed on $\Sigma$, and the analytic properties of the associated flat connection. Not all selections of these properties are consistent with one another. For example, meromorphicity can be in conflict with single-valuedness on $\Sigma$ and with good transformation properties under the modular group. For each consistent selection, the corresponding space of polylogarithms is constructed with the goal of closing under the operations of addition, multiplication, differentiation and taking primitives.

\sm

For example, a meromorphic multiple-valued connection in one variable on  a Riemann surface with one puncture that takes values in an infinite-dimensional,  freely generated, Lie algebra $\mg= \hat t_{h,1,1}$  was introduced by Enriquez in~\cite{Enriquez:2011}. The same paper provided a meromorphic connection $\cK_\text{E}$ in $n \geq 2$ variables that takes values  in an infinite-dimensional Lie algebra $\mg = \hat \mt_{h,n}$ that is, however, not freely generated. Both of these constructions hold for Riemann surfaces of arbitrary genus $h \geq 1$ and generalize their  $h=1$ counterparts  derived in \cite{Levin:2007, Calaque:2009}.

\sm

A single-valued and modular invariant, but non-meromorphic, connection in one variable on a Riemann surface with one puncture, that takes values in the Lie algebra $\mg= \hat \mt _{h,1,1}$, was introduced by D'Hoker, Hidding and Schlotterer (DHS)  in \cite{DHoker:2023vax} and generalizes the Brown-Levin connection \cite{BrownLevin} to arbitrary genus $h\geq 1$. The DHS connection involving one variable and puncture was related to the Enriquez connection by the composition of a gauge transformation and an automorphism of the Lie algebra in  \cite{DHoker:2024desz}. A generalization of this connection to the case of $n\geq 2$ variables, with or without punctures, was presented in the companion paper \cite{paper.1}. The same reference related the DHS connection without punctures to the multivariable Enriquez connection by the composition of a gauge transformation and an automorphism of the Lie algebra $\hat \mt_{h,n}$.

\sm

The DHS and Enriquez connections for several variables are both built out of the same integration kernels as their single-variable counterparts.
 Nonetheless,  while the case of a single variable and one puncture is governed by the freely generated Lie algebra $\hat \mt_{h,1,1}$,  the multivariable case is based on the Lie algebra $\hat \mt_{h,n}$ which is not freely generated for $n\geq 2$. As a result of the flatness conditions of the multivariable connections, the corresponding integration kernels can no longer be independent as functions on the Riemann surface. The explicit form and the counting of the number of resulting identities among the integration kernels is determined by the non-trivial relations between Lie algebra generators in $\hat \mt_{h,n}$. 
  
\sm

In this paper, we shall show that the resulting relations induced on the integration kernels are precisely the union of all the \textit{interchange identities} and \textit{Fay identities} (collectively referred to as \textit{Fay identities} for brevity in the title) that were derived from a different point of view and with different methods in \cite{DHoker:2024ozn}. More concretely, the single-valued DHS connection $\cJ_\text{DHS}$ on the configuration space of $n$ variables $x_1, \cdots, x_n \in \Sigma$ and the multiple-valued Enriquez connection $\cK_\text{E}$ admit the following decompositions, 
\bea
\label{1.a}
\cJ_\text{DHS} = \sum _{i=1}^n \Big ( J^{(1,0)}_i \, dx_i + J^{(0,1)}_i \, d\bar x_i \Big )
\hskip 1in 
\cK_\text{E} = \sum_{i=1}^n K_i \, dx_i 
\eea
The components take values in the Lie algebra $\hat \mt_{h,n}$ and admit a Lie series expansion whose coefficients are the corresponding DHS and Enriquez kernels, respectively, including Abelian differentials. The flatness conditions of these connections on configuration space are the Maurer-Cartan equations, 
\bea
\label{1.b}
d \cJ_\text{DHS}  - \cJ_\text{DHS}  \wedge \cJ_\text{DHS} =0 
\hskip 1in 
d \cK_\text{E} - \cK_\text{E}  \wedge \cK_\text{E} =0
\eea
and may be similarly decomposed according to the different types of differential forms. The components of the flatness condition for $\cJ_\text{DHS}$ that involve  $J^{(0,1)}_i $ readily follow from the Massey structure of the system of differential equations for DHS kernels \cite{DHoker:2023vax, DHoker:2024ozn}. Similarly, the contributions $\p_i J^{(1,0)}_j - \p_j J^{(1,0)}_i$ and $\p_i K_j - \p_j K_i$ with $i,j =1, \cdots, n$ vanish  by the basic properties of the DHS and Enriquez kernels, respectively \cite{Enriquez:2011, DHoker:2024ozn, paper.1}. 
 From this point of view, the only ``non-trivial" components of the flatness conditions consist of the commutator relations, for $1\leq i < j \leq n$,
 \bea
 \label{1.c}
 [J_i^{(1,0)}, J_j ^{(1,0)}]=0 \hskip 1in [K_i, K_j]=0
 \eea 
For the case of $n=1$ variable, the commutator conditions are clearly empty and impose no relations between the kernels. For $n \geq 2$, however, the flatness conditions of (\ref{1.c}) together with the structure relations of the Lie algebra $\hat \mt_{h,n}$ imply non-trivial relations between bilinears in the respective kernels.

\sm

We will show that, for $n \geq 3$, the flatness conditions for $\cJ_\text{DHS}$ are equivalent to the union of all the \textit{interchange identities} $\mP=0$ and \textit{Fay identities} $\mF=0$ proven for DHS kernels in~\cite{DHoker:2024ozn}. While the \textit{interchange identities} are bilinears in integration kernels that depend on two variables $x_i,x_j \in \Sigma$, the \textit{Fay identities} are bilinears in integration kernels that depend on three variables $x_i,x_j,x_k \in \Sigma$. We will further prove that the flatness conditions for $\cK_\text{E}$ imply the analogous interchange identities $\mQ=0$ and Fay identities $\mG=0$ for Enriquez kernels proposed in \cite{DHoker:2024ozn}, but equivalence, if it holds, remains to be proven. Thus, the flatness conditions $[K_i, K_j] = 0$ will lead to a unified proof of all interchange identities (proven in \cite{DHoker:2024ozn}) and all Fay identities (conjectured in \cite{DHoker:2024ozn} and later proven in \cite{Baune:2024ber}) among Enriquez~kernels. 

\sm

The correspondences between the flat DHS and Enriquez connections for $n \geq 3$ on the one hand,  and the interchange and Fay identities among the respective kernels on the other hand are schematically rendered in the Figure below. The advertised alternative proof of Enriquez-kernel identities represented by the arrow in the second line follows from the three other arrows in the Figure.

\vskip 0.3in

\begin{align}
& \hskip 0.43in \fbox{\begin{minipage}{3.4 cm} 
\begin{center}
{ Flatness of $\cJ_\text{DHS}$} \\ \vskip 0.07in $[J_i^{(1,0)}, J^{(1,0)}_j]=0$ 
\end{center}
\end{minipage}}
& \Longleftrightarrow & &
\fbox{\begin{minipage}{3.1 cm} 
\begin{center}
{Flatness of $\cK_\text{E}$} \\ \vskip 0.07in $[K_i, K_j]=0$ 
\end{center}
\end{minipage}} \hskip 0.5in 
\no \\ 
& \hskip 1.05in \Updownarrow &&& \Downarrow \hskip 1.1in
\no \\
& \fbox{\begin{minipage}{5.7 cm} 
\begin{center}
{Interchange and Fay identities for DHS kernels} \\ \vskip 0.07in $\mF = \mP= \mL  =0$ 
\end{center}
\end{minipage}}
&  \Longrightarrow & &
\fbox{\begin{minipage}{5.7 cm} 
\begin{center}
{Interchange and Fay identities for Enriquez kernels} \\ \vskip 0.07in $\mG=\mQ= \mM=0$ 
\end{center}
\end{minipage}}
\no 
\end{align}

\begin{center}
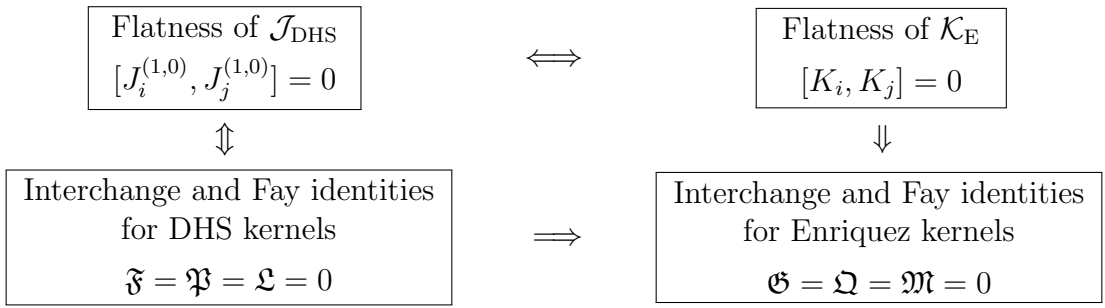
Figure 1: Schematic representation of the relations between the flatness conditions and the interchange and Fay identities for both DHS and Enriquez kernels.
\end{center}

\vskip 0.1in

The equations $\mL=0$ for DHS kernels and $\mM=0$ for Enriquez kernels refer to relations that are derived from  the coincident limits of the Fay identities combined  with certain discrete symmetries. These symmetries are proven in the DHS case but still remain conjectural for the ingredients of the coincident limit of Enriquez kernels, see section 9.4 of \cite{DHoker:2024ozn}. Finally, we note that the  case $n=2$ is special in that it has insufficient room to accommodate the full Fay identities, which critically depend on three distinct points.

\subsection*{Organization}

The remainder of this paper is organized as follows. In section \ref{sec:2}, we review the structure relations of the  Lie algebra $\hat \mt_{h,n}$, the definition and properties of the DHS kernels, the multivariable DHS connection $\cJ_\text{DHS}$, the interchange and Fay identities for DHS kernels, and their coincident limits. Various  combinatorial  definitions and identities used throughout this paper are collected in appendices \ref{sec:A} and \ref{sec:B}, while a technical proof related to section \ref{sec:2.5} is relegated to appendix \ref{sec:F}. In section \ref{sec:4}, we present the proof of equivalence between the flatness conditions on $\cJ_\text{DHS}$ for $n\geq 3$ variables and the interchange and Fay identities for DHS kernels, relegating the proof of various auxiliary lemmas to appendices \ref{sec:C}, \ref{sec:D}, and \ref{sec:E}. Finally, in section \ref{sec:5}, we review the definition and properties of the  Enriquez kernels, the multivariable Enriquez connection $\cK_\text{E}$ and the interchange and Fay identities for Enriquez kernels.  We then  use structural parallels between the multivariable connections $\cJ_\text{DHS}$ and $\cK_\text{E}$ to show that flatness conditions on $\cK_\text{E}$ imply all the interchange and Fay identities on Enriquez kernels for $n\geq 3$ variables. We briefly discuss the open conjectures which, at present, prevent the correspondence for Enriquez kernels from being promoted to an equivalence.

\subsection*{Acknowledgments}

We are grateful to Benjamin Enriquez, Yoann Sohnle and Federico Zerbini for valuable discussions and various collaborations on related topics. The research of ED was supported in part by NSF grant PHY-22-09700. The research of OS is funded by the European Union under ERC Synergy Grant MaScAmp 101167287. Views and opinions expressed are however those of the author(s) only and do not necessarily reflect those of the European Union or the European Research Council. Neither the European Union nor the granting authority can be held responsible for them.

\newpage

\section{Review of the DHS connection and Fay identities}
\label{sec:2}

In this section, we provide a brief review of the multi-variable DHS connection on a compact Riemann surface $\Sigma$ of arbitrary genus without punctures \cite{paper.1}. The multi-variable DHS connection is built out of the same DHS kernels used to formulate the connection in a single variable with one fixed puncture in \cite{DHoker:2023vax}. These kernels satisfy an infinite family of \textit{interchange identities} and \textit{Fay identities}, which were derived in \cite{DHoker:2024ozn}, and will be reviewed here as well. 
 
\subsection{The Lie algebra}
\label{sec:2.1}

The Lie algebra $\hat \mt_{h,n}$ in which both the DHS connection and the Enriquez connection for $n$ variables $x_i$ with $i=1,\cdots, n$ on a compact Riemann surface $\Sigma$ of arbitrary genus $h \geq 1$ without punctures and its universal cover $\tilde \Sigma$ take values was constructed in \cite{Enriquez:2011}. Contrarily to the single-variable case with one puncture, $\hat \mt_{h,n}$ is no longer freely generated when $n \geq 2$. 

{\deff
\label{2.def:1}
The Lie algebra $\mt_{h,n}$ may be presented in terms of the generators $a_j^J, b_{iI}$ and $t_{ij}= t_{ji}$, for $I,J=1,\cdots, h$ and $i,j = 1, \cdots, n$ with $j \not= i$ as follows \cite{Enriquez:2011},\footnote{For $h \geq 2$, the fourth line follows from the first two lines. Indeed,  for any value of $I$  we can choose a $K \not= I$ such that $[ a^I_i, t_{jk} ] \delta ^J_K =  [ b_{kK}, [ a^I_i, a^J_j    ]  ] +  [ a^J_j, [b_{kK}, a^I_i ]  ]$ with the help of the Jacobi identity.  Both terms on the right vanish by the first two lines of (\ref{11.1}) in view of $j \not = i$ and  $K\not= I$. The vanishing of $[b_{iI}, t_{jk}]$ follows from analogous~arguments.} 
\begin{align}
& \big [ a^I_i, a^J_j \big ]  =  \big [ b_{iI} , b_{jJ} \big ] =  0 && \hbox{$i,j$ distinct} 
\no \\
&  \big [b_{iI } , a^J _j \big ]  =  \delta ^J _I \, t_{ij} && \hbox{$i,j$ distinct} 
\no \\
& \big [ b_{iI} , a^I _i \big ] + \sum_{j \not = i} t_{ij}  =0
\no  \\
& \big [ a^I_i, t_{jk} \big ]  = \big [b_{iI}, t_{jk} \big] = 0 && i,j,k \hbox{ distinct}
\label{11.1}
\end{align}
The Lie algebra $\mt_{h,n}$ admits a positive bi-grading that assigns the bi-degrees $|a_i^I|=(1,0)$, $|b_{iI}|=(0,1)$ and $| t_{ij}|=(1,1)$. The Lie algebra $\hat \mt_{h,n}$ is the degree completion of $\mt_{h,n}$.\footnote{The connections take values in the degree completion $\hat \mt_{h,n}$ rather than in $\mt_{h,n}$ since they are expressed as an infinite Lie series whose convergence is subject to the completion condition. }}

\sm

Immediate consequences of Definition \ref{2.def:1} are as follows.
\begin{align}
\label{2.prop.1}
& \big [ a^I_i + a^I_j, t_{ij} \big ] =   \big [ b_{iI} + b_{jI} , t_{ij} \big ] = 0  && \hbox{$i,j$ {\it distinct}} 
\no \\
&  \big [ t_{ij} + t_{ik} , t_{jk} \big ] = 0 && \hbox{$i,j,k$ {\it distinct}} 
\no \\
& \big [ t_{ij}, t_{k\ell} \big ] =0  && \hbox{$i,j,k, \ell$ {\it distinct}}
\end{align}
These relations may be proven using the relations of Definition \ref{2.def:1} and the Jacobi identity~\cite{Enriquez:2011}.

\subsection{The DHS kernels}
\label{sec:2.2}

The homology group $H_1(\Sigma, \ZZ)$ of a compact Riemann surface $\Sigma$ of arbitrary genus $h\geq 1$ is equipped with an  intersection pairing $\mJ$. A canonical basis of homology cycles $\mA^I$ and $\mB_I$ has intersection pairings  $\mJ(\mA^I, \mA^J) = \mJ(\mB_I, \mB_J) =0$ and $\mJ(\mA^I, \mB_J) = \delta ^I_J$ for $I,J=1,\cdots, h$. Its dual basis of $h$ holomorphic Abelian differentials $\om_I$ is chosen by normalizing the $\mA$-periods while their $\mB$-periods give the components $\Omega_{IJ}$ of the period matrix  $\Omega$, 
\bea
\label{2.a.1}
\oint _{\mA^I} \om_J = \delta ^I_J \hskip 1in \oint _{\mB_I} \om_J = \Omega _{IJ}
\eea 
The Riemann relations ensure symmetry $\Omega ^t = \Omega$ and positivity of $\Im(\Omega)$. Choosing the cycles $\mA^I$ and $\mB_I$ such that they share a base point $p \in \Sigma$, we  may also use them to generate the fundamental group $\pi_1 (\Sigma, p)$ with base point $p$. 

\sm

The Arakelov Green function $\cG(x,y)$ is a single-valued symmetric function of $x,y \in \Sigma$ which may be defined by the following differential equation and vanishing surface integral, 
\bea
\pbx \p_x \cG(x,y) = - \pi \delta (x,y) + \pi \kappa(x) 
\hskip 1in 
\int _\Sigma d^2 x \, \kappa (x) \cG(x,y)=0
\eea
where $\kappa (x)$ is the pull-back of the canonical normalized translation invariant K\"ahler form on the Jacobian $\CC^h / ( \ZZ^h {+} \Omega \ZZ^h)$ given in terms $\om_I$ as follows,\footnote{Throughout, a pair of a repeated upper and lower index is understood to be summed over by the Einstein convention, namely without writing the sum symbol,  unless stated otherwise. Indices $I,J=1,\cdots, h$ may be lowered and raised using the metric $Y = \Im (\Omega)$ and its inverse $Y^{-1}$ with components $Y_{IJ}$ and $Y^{IJ}$, respectively. For example, with these notations we have $Y_{IJ} Y^{JK} = \delta ^K_I$, $\bar \om^I =  Y^{IJ} \, \bar \om_J$ and $\om_I = Y_{IJ} \, \om^J$. Unless stated otherwise, the dependence on the moduli $\Omega _{IJ}$ of the compact Riemann surface $\Sigma$ will not be exhibited. Finally, in a local system of complex coordinates $x, \bar x$ on $\Sigma$, we use the notations $\om_I = \om_I(x) dx$,  $\bar \om^I= \bar \om^I(x) d \bar x$, $d^2 x = {i \over 2} dx \wedge d\bar x$ and normalize the Dirac $\delta(x,y)$ distribution by $\int _\Sigma d^2 x \, \delta(x,y) = 1$.}
\bea
\kappa(x) = { 1 \over h} \om_I(x) \bar \om^I(x) \hskip 1in
\int _\Sigma d^2 x \, \kappa(x) =1
\eea
An explicit construction of the Arakelov Green function may be given in terms of the prime form (see for example \cite{DHoker:2017pvk,DHoker:2023vax}) but will not be needed for our purposes. The DHS kernels may be defined recursively in terms of the Arakelov Green function, the holomorphic Abelian differentials $\om_I$ and their complex conjugates $\bar \om^I$ as follows, 
\bea
f^I{}_J(x,y) & = & \int _\Sigma d^2 z \, \p_x \cG(x,z) \Big ( \bar \om^I(z) \om_J(z) - \delta (z,y) \delta^I_J \Big )
\no \\
f^{I_1 \cdots I_r} {}_J(x,y) & = & \int _\Sigma d^2 z \, \p_x \cG(x,z) \, \bar \om^{I_1} (z) \, f^{I_2 \cdots I_r}{}_J(z,y) 
\hskip 1in r \geq 2
\label{deffkers}
\eea
and setting $f^\emptyset {}_J(x,y) = \om_J(x)$ for the case corresponding to $r=0$. It will be useful to decompose the DHS kernels into their trace and traceless parts with respect to the indices $I_r$ and $J$, in terms of the single-valued functions $\Phi ^{I_1 \cdots I_r} {}_J (x)$ and $\cG^{I_1 \cdots I_{r-1}} (x,y)$ in $x,y \in \Sigma$, 
\bea
\label{2.b.4}
f^{I_1 \cdots I_r} {}_J(x,y)  =  \p_x \Phi ^{I_1 \cdots I_r} {}_J (x) - \p_x \cG^{I_1 \cdots I_{r-1}} (x,y) \, \delta ^{I_r}_J
\eea
Further properties of the DHS kernels may be found in \cite{DHoker:2023vax,DHoker:2024ozn}.

\subsection{The multivariable DHS connection}

The multivariable DHS connection $\cJ_\text{DHS}(\xx; a,b,t)$  takes values in the Lie algebra $\hat \mt_{h,n}$ of Definition \ref{2.def:1} and is defined as a single-valued one-form on the configuration space,
\bea
\label{2.b.0}
\text{Cf}_n(\Sigma) =  \Sigma ^n \setminus \{ \hbox{diagonals} \}
\eea 
of $n$ points $\xx= (x_1, \cdots, x_n)$. Here, we shall consider only the case without punctures and refer to section 4.4 of \cite{paper.1} for the case with punctures. The connection $\cJ_\text{DHS}(\xx; a,b,t)$ may be decomposed into a sum of $(1,0)$ forms $dx_i$ and $(0,1)$ forms $d \bar x_i$ in the variables $x_i \in \Sigma$, 
\bea
\label{2.b.1}
\cJ_\text{DHS} (\xx;a,b,t) &= \sum_{i=1}^n \Big ( J_i^{(1,0)} (\xx;a,b,t) \, dx_i  + J_i^{(0,1)} (\xx;a,b,t) \, d \bar x_i  \Big )
 \eea
where the components $J_i^{(0,1)} (\xx;a,b,t)$ are given by,
\bea
\label{2.b.1a}
J_i^{(0,1)} (\xx;a,b,t) = - \pi \bar \om^I(x_i) b_{iI} 
\eea
and the triplet  $(a,b,t)$ obeys the structure relations of $\hat \mt_{h,n}$ in Definition \ref{2.def:1}. To express the components $J_i^{(1,0)} (\xx;a,b,t)$ in terms of DHS kernels, it will be convenient to introduce the following generating functions for the DHS kernels and their trace part,\footnote{\label{ftwhsum} Throughout, we shall denote by $\cW_h$ the set of all words in the alphabet of $h$ letters $I = 1,\cdots,h$, including the empty word $\emptyset$,  and use  the notation $\vI= I_1 \cdots I_r$ for a word of length $r \geq 1$. By contrast to the Einstein convention for the summation over single indices, we shall exhibit explicitly the summation over words $\vI \in \cW_h$ to represent the double summation over the length $r$ of the words $\vI$ and, for given $r$, the sum over all its individual letters $I_1, \cdots ,I_r$, as illustrated in the right equalities of (\ref{2.b.3}). The symbols $f^\vI{}_J(x,y)$, $\cG^\vI(x,y)$ and $B_\vI$ stand for  $f^{I_1 \cdots I_r}{}_J(x,y)$, $\cG^{I_1 \cdots I_r}(x,y)$ and for the concatenation product $B_\vI = B_{I_1} \cdots B_{I_r}$, respectively. A variety of combinatorial definitions and identities, including for instance the concatenation and shuffle products and the antipode, are collected in appendix \ref{sec:A}. } 
\bea
\label{2.b.3}
\bJ_K(x,y;B) & = & \sum _{r=0}^\infty f^{I_1 \cdots I_r}{}_K(x,y) B_{I_1} \cdots B_{I_r}
= \sum _{\vI \in \cW_h} f^\vI{}_K(x,y) B_{\vI} 
\no \\
\bG (x,y;B) & =  & \sum _{r=0}^\infty \p_x \cG ^{I_1 \cdots I_r}  (x,y) B_{I_1} \cdots B_{I_r}
\, = \sum _{\vI \in \cW_h} \p_x \cG ^\vI (x,y) B_{\vI}
\eea
where we set $B_I X = [b_I,X]$ for arbitrary $X \in \hat \mt_{h,n}$.  In terms of these generating functions, the components of the $(1,0)$ part of the multivariable DHS connection take the form,   
\bea
\label{2.b.5}
J_i^{(1,0)} (\xx;a,b,t)  &= 
\bJ_K(x_i,z; B_i) a_i^K + \sum_{j \neq i} \big \{  \bG (x_i,x_j;B_i) - \bG (x_i,z;B_i)  \big \}  t_{ij}  
\eea
where the dependence on $z$ drops out by the decomposition (\ref{2.b.4}) of DHS kernels
and the $\hat \mt_{h,n}$ relation in the third line of (\ref{11.1}).
On the configuration space $\text{Cf}_n(\Sigma)$, defined in (\ref{2.b.0}), the flatness condition, 
\bea
d \cJ_\text{DHS} -  \cJ_\text{DHS} \wedge \cJ_\text{DHS} =0
\eea 
for the total differential $d = d_{x_1}+\cdots + d_{x_n}$ reduces to the following set of relations among the  components $J_i^{(1,0)}$ and $J_j^{(0,1)}$ for $i,j=1,\cdots, n$,
\bea
\label{2.b.2a}
d\, J^{(0,1)}_j =0 
\hskip 1in
\bar \p_j J^{(1,0)} _i - [J^{(0,1)}_j , J^{(1,0)} _i]=0
\eea
and the following set of relations that involves only the components $J^{(1,0)} _i$,
\bea
\label{2.b.2}
\p_i J^{(1,0)} _i - \p_j J^{(1,0)} _i = 0
\hskip 1in
{}  [ J^{(1,0)} _i , J^{(1,0)} _j  ] = 0
\eea
The set (\ref{2.b.2a}) contains all the ``easy relations" that follow from the Massey structure of the system of differential equations satisfied by the DHS kernels \cite{DHoker:2023vax,DHoker:2024ozn}. The first equation of (\ref{2.b.2})  is a consequence of the following relations, 
\bea
\label{2.b.7}
\cG^{I_1 \cdots I_r} (x,y) & = & (-)^r \cG ^{I_r \cdots I_1} (y,x)
\no \\
B_{i I_1} \cdots B_{i I_r} t_{ij} & = & (-)^r B_{jI_r} \cdots B_{jI_1} t_{ij}
\eea
the first of which was proven in \cite{DHoker:2023vax} while the second follows from repeated use of (\ref{2.prop.1}) and $B_{iI} B_{jJ} = B_{jJ}  B_{iI}$ for $i\neq j$.
The second relation  of (\ref{2.b.2}) is at the heart of this paper. Its proof requires various intricate arguments that were presented in \cite{paper.1}. In section \ref{sec:4}, we will prove that the commutators $ [ J^{(1,0)} _i , J^{(1,0)} _j  ] $ serve as generating functions for all the interchange and Fay identities among DHS kernels of \cite{DHoker:2024ozn}.

\subsection{The interchange identities for DHS kernels}
\label{sec:2.4}

DHS kernels obey an  infinite family of \textit{interchange identities} that are bilinear in the DHS kernels,  depend on two points $x_i,x_ j \in \Sigma$. The functions $\mP^{ \vI}{}_K (i,j) = \mP^{ \vI}{}_K (x_i,x_j)$  are $(1,0)$ forms in $x_i$ and $x_j$ defined by,
\bea
\label{2.c.2}
 \mP^{ \vI}{}_K (i,j) = 
 \sum_{\vec{I} = \vP \vQ} \Big \{ 
 f^{\cS(\vQ) L }{}_K (i,j) \, f^{ \vP }{}_L(j,i) 
 +  f^{ \cS(\vQ) }{}_L(i,j) \, f^{ \vP L }{}_K (j,i)
\Big \}
\eea
where  $\vI = \vP \vQ$ stands for summing over all possible deconcatenations of the given word $\vI$ into words $\vP$ and $\vQ$ including the cases $(\vP, \vQ) = (\vI,\emptyset)$, $(\emptyset,\vI)$ that involve the empty word~$\emptyset$. The functions $\mP^{ \vI}{}_K (i,j) $  satisfy the following symmetry property,
\bea
\label{2.c.3}
\mP^{ \vI}{}_K (j,i) =  \mP^{ \cS(\vI)}{}_K (i,j) 
\eea
where $\theta$ is the antipode, whose definition is given in Definition-Proposition \ref{A.def:4} of appendix~\ref{sec:A}.

{\thm
\label{2.intlem}
The interchange identities among DHS kernels correspond to the relations,
\bea
\label{2.c.1}
 \mP^{ \vI}{}_K (i,j) =0
 \eea
for arbitrary word $\vI \in \cW_h$, arbitrary index $K=1,\cdots,h$ and distinct points $x_i, x_j \in \Sigma$.}

\sm

The interchange identities for DHS kernels were established  for various special cases in~\cite{DHoker:2020tcq,DHoker:2020uid} and were proven  for the general case  in \cite{DHoker:2024ozn}.

\subsection{The Fay identities for DHS kernels}

The Fay identities for DHS kernels are bilinear relations among DHS kernels that involve three distinct points $x_i,x_j, x_k \in \Sigma$ and were proven  in Theorems 6.2 and 6.3 of \cite{DHoker:2024ozn}. They are defined modulo linear combinations with the interchange identities of section \ref{sec:2.4}. 

\sm

The Fay identities for DHS kernels may be expressed in terms of two functions $\mF_0$ and~$\mF_1$ both of which are $(1,0)$ forms in $x_i$ and $x_j$. The function $\mF_0$ is a scalar in $x_k$   defined by,\footnote{The expressions (\ref{2.d.2a}) and (\ref{2.d.2b}) were obtained from formula (6.11) of Theorem 6.2 in  \cite{DHoker:2024ozn} by  renaming the word $\vP \to \vJ$ and the index $J \to L$; relabeling the points $(x,y,z) \to (x_i, x_j, x_k)$; absorbing the first term on the right of (6.11) in  \cite{DHoker:2024ozn} as the $\vR= \emptyset$ in the first term in (\ref{2.d.2b}), and the $f^{\vI M}{}_L(i,j) f^{\vJ L}{}_K(j,i) $ term in the $\vS= \emptyset$ in the second term of (\ref{2.d.2b}).} 
\bea
 \label{2.d.2a}
\mF_0 ^{\vI |  \vJ | M}  {}_{\! K}   (i,j;k) 
& =  &
f^{\vI L}{}_{\! K}(i,k) \Big ( f^{\vJ M}{}_{\! L}(j,k) - f^{\vJ M}{}_{\! L}(j,i)  \Big )
- \! \! \sum_{\vI = \vP \vQ }  \!\!  f^{\vP}{}_L(i,j) f^{(\vJ \shuffle L \vQ )M}{}_{\! K}(j,k) 
 \\ && 
-f^{\vI M}{}_{\! L}(i,j) \Big ( f^{\vJ L}{}_{\! K}(j,k)  - f^{\vJ L}{}_{\! K}(j,i)  \Big )
- \! \! \sum_{\vJ = \vR \vS } \!\! f^{\vR}{}_L(j,i) f^{(\vI \shuffle L  \vS)M}{}_{\! K}(i,k)
 \qquad \quad
\no
\eea
while the function $\mF_1$ is independent of $x_k$,
\bea
 \label{2.d.2b}
\mF_1 ^{\vI |  \vJ | M}  {}_{\! K}   (i,j)
= - \sum_{\vJ = \vR \vS } 
\Big (  f^{\vR}{}_L(j,i) f^{\vI M \cS(\vS)  L}{}_{\! K}(i,j) 
+f^{\vI M \cS(\vS)}{}_L(i,j)   f^{\vR L}{}_{\! K}(j,i)  \Big )
\eea
In terms of the decomposition of (\ref{2.b.4}) for $f$ into its traceless and trace parts, and making use of the following  convenient shorthand below, 
\bea
 \label{2.d.1}
\cG^\vI (i ; j,k)  = \cG^\vI(i,j) -  \cG^\vI(i,k) 
\eea
the first line in the combination $\mF_0 ^{\vI |  \vJ | M}  {}_{\! K}   (i,j;k) $ of (\ref{2.d.2a}) may be simplified and produces the following more economical expression for this function, 
\bea
 \label{2.d.3}
\mF _0 ^{\vI |  \vJ | M}  {}_K   (i,j;k) 
& = &   \p_i \cG^\vI (i ; k, j ) \, \p_j \cG^\vJ (j ; k, i ) \, \delta^M_K
-  \sum_{\vI = \vP \vQ }   f^{\vP}{}_L(i,j) f^{(\vJ \shuffle L \vQ )M}{}_K(j,k) 
\no \\ &&
 -  \sum_{\vJ = \vR \vS } f^{\vR}{}_L(j,i) f^{(\vI \shuffle L  \vS)M}{}_K(i,k)
\eea
In terms of the functions $\mF_0$ and $\mF_1$, we define, 
\bea
\label{2.d.5}
\mF ^{\vI | \vJ | M} {}_K (i,j;k) = \mF_0 ^{\vI | \vJ | M} {}_K (i,j;k) + \mF_1 ^{\vI |  \vJ | M}  {}_{\! K}   (i,j) 
\eea
for arbitrary $M,K\in \{ 1,\cdots,h\}$, $\vI,\vJ \in \cW_h$ and pairwise distinct points $x_i, x_j, x_k \in  \Sigma$. 

{\thm
\label{2.thm:1}
The Fay identities for DHS kernels correspond to the relations \cite{DHoker:2024ozn}, 
\bea
\label{2.thm.1}
\mF ^{\vI | \vJ | M} {}_K (i,j;k) =0
\eea
for arbitrary indices $K,M$, words $\vI, \vJ$ and pairwise distinct points $x_i, x_j, x_k \in  \Sigma$.}

\subsection{Transformation under swapping $ i$ and $j$}

The transformation of $\mF_0, \mF_1$ and $\mF$ under swapping $i$ and $j$ and simultaneously swapping $\vI$ and $\vJ$ is given by the following proposition.

{\prop 
\label{2.prop:1}
Under swapping $(i \, \vI) $ and $(j \, \vJ)$ the functions $\mF_0, \mF_1$ and $ \mF$ behave as,
\bea
\label{2.prop.1a}
\mF_0^{\vI | \vJ | M} {}_K(i,j) - \mF_0^{\vJ | \vI | M} {}_K(j,i) & = & 0 
\no \\
\mF_1^{\vI | \vJ | M} {}_K(i,j) - \mF_1^{\vJ | \vI | M} {}_K(j,i) & = & \mP^{ \vJ M \cS(\vI)} {}_K(i,j)
\no \\
\mF^{\vI | \vJ | M} {}_K(i,j;k) - \mF^{\vJ | \vI | M} {}_K(j,i;k) & = & \mP^{ \vJ M \cS(\vI)} {}_K(i,j)
\eea
where  $\mP^{\vI} {}_J(i,j)$ is the protagonist of  the interchange identities and was defined in (\ref{2.c.2}).}

\begin{proof}
\vskip -0.05in
The first relation in (\ref{2.prop.1a}) follows from (\ref{2.d.2a}) or (\ref{2.d.3}) by inspection, while the third relation follows from the first two with the help of (\ref{2.d.5}).  To prove the second relation, we evaluate its right side using formula (\ref{2.c.2}) to express the function $\mP^{ \vJ M \cS(\vI)} {}_K(i,j)$, 
\bea
\label{2.d.6}
\mP^{ \vJ M \cS(\vI)} {}_K(i,j)
= 
\sum_{\vJ M \cS(\vI)= \vR '\vS' } \bigg\{ 
 f^{\cS(\vS ') L }{}_K(i,j) \, f^{ \vR' }{}_L(j,i) 
 +  f^{ \cS(\vS ') }{}_L(i,j) \, f^{ \vR 'L }{}_K(j,i) \bigg \}
 \eea
 In the deconcatenation of $\vJ M \cS(\vI)$, two sets of contributions arise, distinguished by whether the letter $M$ belongs to the factor $\vR'$ or to the factor $\vS'$. In terms of the deconcatenations $\vI = \vP \vQ$ and $\vJ = \vR \vS$, the factors $\vR'$ and $\vS'$ are given by the union of contributions,
 \bea
 \label{2.d.7}
( \vR' , \vS' ) = \big ( \vJ M \cS(\vQ), \cS(\vP) \big ) 
\hskip 1in
( \vR'  , \vS') = \big ( \vR  , \vS M \cS(\vI) \big ) 
\eea
Rendering the contributions from these two sets of deconcatenations explicit, 
\bea
\label{2.d.8}
\mP^{ \vJ M \cS(\vI)} {}_K(i,j)
& = &
\sum_{\vI = \vP \vQ}
\Big \{ 
 f^{\vP L }{}_K(i,j) \, f^{ \vJ M \cS(\vQ) }{}_L(j,i) +  f^{ \vP }{}_L(i,j) \, f^{ \vJ M \cS(\vQ) L }{}_K(j,i) \Big \}
\no \\ && 
- \sum_{\vJ = \vR \vS}  
\Big \{ 
 f^{\vI M \cS(\vS)  L }{}_K(i,j) \, f^{ \vR }{}_L(j,i) +  f^{\vI M \cS(\vS)  }{}_L(i,j) \, f^{ \vR L }{}_K(j,i) \Big \}
 \qquad
\eea
and comparing with $\mF_1$ in (\ref{2.d.2b}), shows that (\ref{2.d.8}) equals the left side of the second equation  in (\ref{2.prop.1a}), thereby  completing the proof of Proposition \ref{2.prop:1}. 
 \end{proof}

\begin{rmk}
While the function $\mF^{\vI | \vJ | M} {}_K(i,j;k) $ fails to be invariant under swapping $(i \, \vI) \leftrightarrow (j \, \vJ)$, one may construct a linear combination of the function $\mF^{\vI | \vJ | M} {}_K(i,j;k) $ that determines the Fay identities and the function $\mP^\vI{}_K(i,j)$ that determines the interchange identities to obtain an equivalent representation of the Fay identities that is swapping symmetric. In view of the last line in (\ref{2.prop.1a}) of Proposition \ref{2.prop:1} this  combination is given by,
\bea
\mF^{\vI | \vJ | M} {}_K(i,j;k)  - \half \mP^{ \vJ M \cS(\vI)} {}_K(i,j)
\eea
One verifies swapping symmetry by using $\mP^{ \vI M \cS(\vJ)} {}_K(j,i)= \mP^{\cS( \vI M \cS(\vJ))} {}_K(i,j)$ via  (\ref{2.c.3}).
\end{rmk}

\subsection{Trace decomposition}

In order to obtain a convenient decomposition of the Fay identities into trace and traceless parts, we begin by establishing the following proposition.\footnote{At genus $h=1$, traceless parts of arbitrary tensors vanish trivially, such that $\mH ^{\vI | \vJ | M } {}_K (i,j)$ in (\ref{2.d.10}) below or the $\Phi$-tensors in (\ref{2.b.4}) only exist for $h\geq 2$.}

{\prop
\label{2.def:2}
The traceless part of $\mF^{\vI | \vJ | M}{}_K (i,j;k)$ in the indices $K$ and $M$ is independent of $k$ and is given by,
\bea
\label{2.d.10}
\mH ^{\vI | \vJ | M } {}_K (i,j) =  \mF ^{\vI |  \vJ | M}  {}_K   (i,j;k) - { 1 \over h} \mF ^{\vI |  \vJ | L}  {}_L   (i,j;k)  \, \delta ^M_K
\eea
The Fay identities are equivalent to the combined vanishing of the traceless and trace parts, 
\bea
\label{2.d.11}
\mF ^{\vI |  \vJ | M}  {}_K   (i,j;k) =0 \ \ \
\Longleftrightarrow \ \ \ \left\{ \begin{array}{r} 
 \mH^{\vI | \vJ | M } {}_K (i,j) =0 \\
  \mF ^{\vI |  \vJ | L} {}_L(i,j;k) =0  \end{array} \right.
\eea
}

\begin{proof}
\vskip -0.1in
The only non-trivial statement in the proposition is the independence on $k$ of the traceless part of $\mF^{\vI | \vJ | M}{}_K (i,j,k)$ which follows by inspection of (\ref{2.d.5}) and (\ref{2.d.3}).\footnote{The decomposition (\ref{2.b.4}) of DHS kernels implies that the $x_k$ dependence of $f^{(\vJ \shuffle L \vQ )M}{}_K(j,k) $ and $f^{(\vI \shuffle L  \vS)M}{}_K(i,k)$ in (\ref{2.d.3}) drops out from the traceless part with respect to $M,K$.} 
\end{proof}

\subsection{Coincident limits of DHS kernels and Fay identities} 
\label{sec:2.5}

The interchange and Fay identities admit convergent limits as some of the points coincide. Clearly this must be the case since the equations (\ref{2.c.1}) and (\ref{2.thm.1}) for the functions $\mP^\vI{}_K(i,j)$ and $\mF^{\vI | \vJ |M}{}_K(i,j;k)$ hold true for all values of the points $x_i, x_j, x_k \in \Sigma$. Here we shall make use only of one particular  coincident limit of the function that enters the Fay identities, 
\bea
\label{2.h.1}
\mF^{\vI | \vJ |M}{}_K(i,j;j) = \lim _{x_k \to x_j} \mF^{\vI | \vJ |M}{}_K(i,j;k)
\eea
While the limits of the individual contributions to $\mF_1^{\vI | \vJ |M}{}_K(i,j)$ given by (\ref{2.d.2b}) trivially exist (since they are independent of $x_k$), the limit does not exist for every individual term that contributes to $\mF_0^{\vI | \vJ |M}{}_K(i,j;k)$. The reason is that the function $\p_x \cG(x,y)$ has a simple  pole $-1/(x{-}y)$ while the function $\partial_x\cG^I(x,y)$ has an angular singularity as $y \to x$. It was shown in section 8 of \cite{DHoker:2024ozn} that the following limits converge  for $\vI =I_1 \cdots I_r$ and $r \geq 2$, 
\bea
\label{2.h.2}
\lim _{y \to x} \Big ( \p_x \cG^J (x,y) - { \pi \over x-y} \int ^x _y \bar \om^J \Big ) = \hf ^J (x)
\hskip 0.6in
\lim _{y \to x} \Big (  \p_x \cG^\vI  (x,y) \Big ) = \hf^\vI(x) 
\eea

\sm
 
 A convenient expression for the coincident limits $\hf^\vI(x) $ in terms of DHS kernels may be obtained by introducing the modular tensors $\hat \mN^{I_1 \cdots I_r} $, which are  defined as follows. For $r=0$ and $r=1$, we set $\hat \mN= \hat \mN^{I_1}=0$; for $r=2$ they are given by,\footnote{For $r \geq 3$, the integrals in (\ref{defhatn}) are absolutely convergent for fixed moduli, while for $r=2$ the second term under the parentheses in (\ref{defhatn2}) is required to achieve absolute convergence of an integral that would be only conditionally convergent without it.}
\bea
\label{defhatn2}
\hat \mN^{I_1I_2} = \int_\Sigma d^2 z_1 \, \bar \omega^{I_1}(z_1) \,  \int_\Sigma d^2 z_2 \, \bar \omega^{I_2}(z_2)\, \Big( \partial_{z_1} \cG(z_1,z_{2}) \partial_{z_2} \cG(z_2,z_{1}) -  \partial_{z_1} \partial_{z_2} \cG(z_1,z_{2})  \Big) 
\eea
while for arbitrary $r \geq 3$ the tensors $\hat \mN^{I_1 \cdots I_r} $ are defined by $n$ convolution integrals over  $\Sigma$, 
\bea
\label{defhatn}
\hat \mN^{I_1 \cdots I_r} =   \prod_{i=1}^r \int _\Sigma d^2 z_i \, \bar \omega^{I_i}(z_i) \, \partial_{z_i} \cG(z_i,z_{i+1})  \bigg|_{z_{r+1}= z_1}
\eea
The resulting $\hat \mN^{I_1 \cdots I_r}$ are readily seen to be modular tensors for all $r\geq 2$ that solely depend on the moduli of the compact Riemann surface $\Sigma$. The following proposition, which was proven in \cite{DHoker:2024ozn},  expresses 
the coincident limits $\hf ^\vI(x)$ in terms of DHS kernels and the modular tensors $\hat \mN^{I_1 \cdots I_r}$:

{\prop
\label{2.prop:66}
The functions $\gamma ^\vI(x)$ defined by (\ref{2.h.2}) admit the following decomposition,
\bea
\label{2.h.3}
\gamma ^\vI (x) & = &  \sum_{\vI = \vX \vY \vZ} \Big ( \p_x \Phi ^{\vX \shuffle \cS(\vZ)}{}_M(x) \, \hat \mN^{M \vY} 
+ \p_x \Phi ^{(\vX \shuffle \cS(\vZ)) M \vY }{}_M(x) \Big ) 
\eea 
for $\vI \neq \emptyset$.
For $\vX \shuffle \cS(\vZ) = \emptyset$ we set $\p_x \Phi ^\emptyset {}_M(x)=\om_M(x) $; for $\vJ \not= \emptyset$ the functions $\p_x \Phi ^\vJ{}_M(x)$ are the traceless parts of $f^\vJ{}_M(x,y)$ defined by (\ref{2.b.4}); and the coefficients $\hat \mN^{M \vY} $ are invariant under the dihedral group, whose action is generated by the~relations,
\bea
\label{2.h.4}
\hat  \mN^{M \vY} = \hat \mN^{\vY M} \hskip 1in \hat \mN^{M \vY} = \hat \mN^{ \cS(M \vY)}
\hskip 0.6in \hbox{ for all } \vY \in \cW_h
\eea}
As detailed in section 8 of \cite{DHoker:2024ozn}, the decomposition (\ref{2.h.3}) is proven
by taking $\partial_{\bar x}$ derivatives and integrating both sides against $d^2 x \, \bar \omega^J(x)$.
The dihedral symmetry (\ref{2.h.4}) is an immediate consequence of (\ref{defhatn2}) and (\ref{defhatn}) along with integration by parts.

\sm

The coincident limit of the function $\mF_0^{\vI | \vJ |M}{}_K(i,j;k)$  will play a key role in the sequel, and was evaluated in \cite{DHoker:2024ozn} as well. In the notations of the present paper, it is given by, 
\bea
\label{2.h.5}
\mF_0^{\vI | \vJ |M}{}_K(i,j;j)
& = &
 \p_i \p_j \cG^\vI (i,j) \, \delta _{\vJ, \emptyset} \,  \delta^M_K 
-  \sum_{\vI = \vP \vQ }   f^{\vP}{}_L(i,j) \, \phi^{(\vJ \shuffle L \vQ )M}{}_K(j) 
\no \\ &&
 -  \sum_{\vJ = \vR \vS } f^{\vR}{}_L(j,i) \, f^{(\vI \shuffle L  \vS)M}{}_K(i,j)
\eea
where we have defined $\phi^{\vI M} {}_K(x) = \p_x \Phi ^{\vI M} {}_K(x) - \gamma ^\vI (x) \, \delta ^M_K$
which equals the coincident limit $f^{\vI M} {}_K(x,x)$ for words $\vI = I_1\cdots I_r$ of length $\geq 2$. 

\sm
 
 The manner in which the result of Proposition \ref{2.prop:66} will enter the proof of equivalence between the flatness of the multivariable DHS connection and the set of all  interchange and Fay identities is given by the following proposition. 
{\prop
\label{prop:dih}
The combination $\mL^\vI(j,i) $, defined for arbitrary $\vI \not= \emptyset$ by,\footnote{Note that the restriction of the sum in (\ref{dih1}) to $\vY \neq \emptyset$ is actually redundant since the contributions $\sum_{\vI = \vX \vZ} f^{ (\vX \shuffle \cS(\vZ)) M}{}_M(j,i)$ from $\vY= \emptyset$ cancel by (\ref{std.sa}) for the words $\vI \neq \emptyset$ under consideration.}
\bea
\label{dih1}
\mL^\vI(j,i)  = \hf^{\vI + \cS(\vI)}(j) 
- \p_j \cG^{\vI + \cS(\vI)}(j,i)   - \sum_{\vI = \vX \vY \vZ; \vY \neq \emptyset }
f^{ (\vX \shuffle \cS(\vZ)) M (\vY + \cS( \vY)) }{}_M(j,i)
\qquad
\eea
with $\hf^\vI(x)$ given by the decomposition of (\ref{2.h.3}), vanishes identically if and only if $\hat \mN$ obeys the 
reflection identities of (\ref{2.h.4}), 
\bea
\label{dih2}
\Big \{\hat  \mN^{N \vY} =  \hat \mN^{\cS(N \vY)} \Big \} _{\vY \in \cW_h} 
\qquad  \Longleftrightarrow \qquad
\Big \{ \mL^\vI(j,i) =0 \Big \} _{\vI \in \cW_h} 
\eea
By inspection of (\ref{dih1}), one verifies the relation $\mL^\vI(j,i) = \mL^{\cS(\vI)}(j,i)$ for all $\vI \not= \emptyset$. }

\sm

The proof of Proposition \ref{prop:dih} is presented in appendix \ref{sec:F}.

\newpage

\section{Flatness of DHS connection  $\Longleftrightarrow $ Fay identities}
\label{sec:4}

In this section, we shall show that flatness of the multivariable DHS connection $\cJ_\text{DHS}$ of (\ref{2.b.1}) is equivalent to the combination of the interchange and Fay identities of Theorem~\ref{2.intlem} and Theorem~\ref{2.thm:1}, respectively. This equivalence (between the left boxes in Figure 1), is understood as follows. As was noted earlier already, the flatness conditions on the components of $\cJ_\text{DHS}$ given in (\ref{2.b.2a}) merely follow from the form of $J^{(0,1)}_i$ and the Massey structure of the system of differential equations satisfied by DHS kernels \cite{DHoker:2023vax,DHoker:2024ozn}. The first equation of (\ref{2.b.2}) in turn readily follows from combining reflection properties of the trace part of the DHS kernels with the Lie algebra relations among $B_i, B_j$ and $t_{ij}$.  

\sm 

The proof of the second equation $[J_i^{(1,0)} ,J_j^{(1,0)} ]=0$ of (\ref{2.b.2}), however, required a considerably more sophisticated set of arguments
\cite{DHoker:2024ozn}. In this section, we shall show that the decomposition of these commutator conditions for all $i \not= j \in \{ 1, \cdots, n\}$ onto independent Lie algebra generators is equivalent to the combination of all the interchange and Fay identities obtained in \cite{DHoker:2024ozn}. This will identify the vanishing commutator $[J_i^{(1,0)} ,J_j^{(1,0)} ]$ as the generating function of the interchange and Fay identities. Throughout this section and the associated appendices, we shall assume that $n\geq 3$ in order to capture the full content of the Fay identities. Modifications, required for the case $n=2$, will be provided in the form of comments and footnotes.
    
\subsection{Decomposition of the commutator }
\label{sec:3.1}

To proceed, we begin by substituting the expressions for $J_i^{(1,0)}$ and $J_j^{(1,0)}$ given in (\ref{2.b.5}) and repeated here for convenience, 
\bea
\label{3.a.1}
J_i^{(1,0)} (\xx;a,b,t)  &=& 
\bJ_M(x_i,z; B_i) a_i^M + \sum_{k \neq i} \big \{  \bG (x_i,x_k;B_i) - \bG (x_i,z;B_i)  \big \}  t_{ik} 
\no \\
J_j^{(1,0)} (\xx;a,b,t)  &=& 
\bJ_N(x_j,z'; B_j) a_j^N + \sum_{\ell \neq j} \big \{  \bG (x_j,x_\ell;B_j) - \bG (x_j,z';B_j)  \big \}  t_{j\ell}  
\eea
into the commutator $[J_i^{(1,0)} ,J_j^{(1,0)} ]$. Here, $z$ and $z'$ are arbitrary points in $ \Sigma$ upon which $J_i^{(1,0)}$ and $J_j^{(1,0)} $ do not depend. For given values of $i \not= j \in \{ 1, \cdots, n\}$, we shall choose $z= x_j$ and $z'=x_i$ so that the sums over $k$ and $\ell$ may both be restricted to $k, \ell \not= i,j$. Furthermore, we shall decompose the result according to the number of exposed $a$-generators that appear in the Lie algebra elements of each term. The resulting decomposition takes the form,
\bea
\label{3.a.2}
{} [J_i^{(1,0)} ,J_j^{(1,0)} ]= \cC_{ij} ^{(0)} + \cC_{ij} ^{(1)} + \cC_{ij} ^{(2)} 
\eea
where components are indexed by the total number of exposed $a$-generators and are given~by, 
\bea
\label{3.a.3}
\cC_{ij} ^{(0)}  & = &
\sum _{k,\ell \not = i,j} \Big [ \big (  \bG (x_i,x_k;B_i) - \bG (x_i,x_j;B_i) \big ) \, t_{ik},
\big (  \bG (x_j,x_\ell;B_j) - \bG (x_j,x_i;B_j)  \big ) \, t_{j\ell} \Big ]
\no \\
\cC_{ij} ^{(1)}  & = & 
\sum _{k \not = i,j} \Big [ \bJ_M(x_i,x_j ;B_i)  \, a^M_i,  \big (  \bG (x_j,x_\ell;B_j) - \bG (x_j,x_i;B_j)  \big ) \, t_{jk} \Big ]
- ( i \leftrightarrow j) 
\no \\
\cC_{ij} ^{(2)}  & = & 
\Big [ \bJ_M(x_i,x_j;B_i)  \, a^M_i, \bJ_N(x_j,x_i ;B_j)  \, a^N_j \Big ]
\eea
Since the Lie algebra $\hat \mt_{h,n}$ is not freely generated for $n \geq 2$, the Lie algebra elements in the above decomposition are not all linearly independent. Their linear interrelations will be at the source of the equivalence with the interchange and Fay identities. The remainder of this section, along with appendices \ref{sec:C}, \ref{sec:D} and \ref{sec:E}, will be devoted to the precise formulation and proof of this statement.

 \subsection{Reducing the component $\cC^{(0)}_{ij}$}
 \label{sec:3.2}

Substituting the expansion given in (\ref{2.b.3}) for the generating series $\bG$ into the expression (\ref{3.a.3}) for $\cC_{ij} ^{(0)}$ and using the notation of (\ref{2.d.1}) for $\cG^\vI (i;k,j) = \cG^\vI (i,k) - \cG^\vI (i,j)$, we obtain, 
\bea
\label{3.b.0}
\cC_{ij} ^{(0)}  & = &
\sum _{k,\ell \not = i,j} \sum_{\vI, \vJ} \p_i \cG^\vI (i;k,j) \, \p_j \cG^\vJ (j;\ell,i) 
\big [ B_{i\vI} \, t_{ik}, B_{j \vJ} \, t_{j \ell} \big ]
\eea
Here and throughout the remainder of this work, $\sum_{\vI, \vJ} $ refers to an infinite sum over words $\vI, \vJ \in \cW_h$ as defined in appendix \ref{sec:A.1}, also see footnote \ref{ftwhsum}.
The relations $B_{i I } \, t_{j\ell} = B_{j J} \, t_{ik} =0$ on the last line of Definition \ref{2.def:1}, for arbitrary $I,J \in \{1,\cdots,h\}$, $k, \ell \not = i,j$ and $i\neq j$, imply  the following relation by iteratively peeling off factors of $B_{iI}$ and $B_{jJ}$,
\bea
\label{3.b.1}
[B_{i \vec{I}} \, t_{ik} ,B_{j \vec{J}} \, t_{j\ell}] = B_{i \vec{I}} \, B_{j \vec{J}} \, [ t_{ik} , t_{j\ell}]
\eea
Terms with $\ell \not= k$ in the sum of (\ref{3.b.0}) cancel since the indices $i,j,k,\ell$ are then all mutually distinct and the 
commutator $[ t_{ik} , t_{j\ell}]$ vanishes by the last line of (\ref{2.prop.1}), which proves the following proposition.

{\prop
\label{3.lem:1}
The double sum over $k, \ell \not= i,j$  in (\ref{3.a.3}) reduces to the sum over the terms with $k=\ell$ and yields the following expression,
\bea
\label{3.b.2}
\cC_{ij} ^{(0)}  = \sum _{k \not = i,j} \sum_{\vI,\vJ}   
\p_i \cG ^\vI (i ; k ,j) \, \p_j \cG^\vJ(j ; k, i) \, \big [B_{i \vec{I}} \, t_{ik} ,B_{j \vec{J}} \, t_{jk} \big ]
\eea}
We note that the functional dependence $\p_i \cG ^\vI (i ; k ,j) \, \p_j \cG^\vJ(j ; k, i)$ in (\ref{3.b.2}) coincides with that of the first term on the right of the expression for the trace part of $\mF_0^{\vI | \vJ |M}{}_K(i,j;k)$ in (\ref{2.d.3}).

 \subsection{Reducing the component $\cC^{(1)}_{ij}$}
 \label{sec:3.3}
 
Substituting the expansions for $\bJ_M$ and $\bG$ given in (\ref{2.b.3}) into the expression (\ref{3.a.3}) 
for  $\cC^{(1)}_{ij}$,
\bea
\label{3.c.1}
\cC_{ij} ^{(1)}  & = & 
\sum _{k \not = i,j} \sum_{\vI, \vJ} f^\vI{}_M (i,j) \, \p_j \cG ^{\vJ} (j;k,i) \big [ B_{i\vI} \, a^M_i, B_{j\vJ} \, t_{jk} \big ]
- ( i \leftrightarrow j) 
\eea
 produces commutators of the form $[ B_{i\vI} \, a_i^M, B_{j \vJ} \, t_{jk}]$ with $k \not= i,j$. Although these commutators are presented with the help of the generator $a_i^J$, they may be re-expressed, with the help of the structure relations (\ref{11.1}),  as linear combinations of the commutators $[B_{i \vec{P}} \, t_{ik} ,B_{j \vec{Q}} \, t_{jk}]$ which no longer involve any exposed generators $a$, as shown in the following proposition.

{\prop
\label{3.lem:2}
The component $\cC^{(1)}_{ij}$ reduces to the following sum,
\bea
\label{3.c.2}
\cC_{ij} ^{(1)} &= &    \sum_{k \neq i,j} \sum_{\vI,\vJ}  
  \bigg\{
\sum_{\vI = \vP \vQ} f^\vP{}_M(i,j) \, \p_j \cG ^{\vJ \shuffle M  \vQ }(j ; k, i) 
\no \\ &&  \hskip 0.7in
+ \sum_{\vJ = \vR \vS} f^\vR {}_M(j,i) \, \p_i \cG ^{\vI \shuffle M  \vS }(i ; k, j)  
\bigg\}
\big[ B_{i \vI} \, t_{ik} , B_{j \vJ} \, t_{jk} \big]
\eea}

The proof of Proposition \ref{3.lem:2} is given in appendix \ref{sec:C}.

 \subsection{Reducing the component $\cC^{(2)}_{ij}$}
 \label{sec:3.4}

Substituting the expansion of $\bJ_M$ given in (\ref{2.b.3}) into the expression (\ref{3.a.3}) for  $\cC^{(2)}_{ij}$,
\bea
\label{3.d.1}
\cC_{ij} ^{(2)}  & = & 
\sum_{\vI, \vJ} f^\vI{}_M (i,j) f^\vJ {}_N (j,i)  \big [ B_{i\vI} \, a^M_i, B_{j\vJ} \, a_j^N \big ]
\eea
produces commutators of the form $\big [ B_{i\vI} \, a^M_i, B_{j\vJ} \, a_j^N \big ]$ with $i\neq j$. To lowest order in the expansion in powers of $B_i$ and $B_j$, the commutator reduces to $[a_i^M, a_j^N]$ and vanishes in view of the structure relations (\ref{11.1}). In fact, all terms bilinear in $a$ either vanish or may be reduced to commutators that involve one or zero exposed $a$, as stated in the following proposition.

{\prop
\label{3.lem:3}
The component $\cC^{(2)}_{ij}$ reduces to the following sum,
\bea
\label{3.d.2}
\cC_{ij} ^{(2)} &= & \cC_{ij}^{(2\mR)} + \cC_{ij}^{(2\mS)}
\eea
separating different type of brackets among $\hat \mt_{h,n}$ generators. The first term $\cC_{ij}^{(2\mR)}$ is given by, 
\bea
\label{3.d.3}
 \cC^{(2\mR)} _{ij} & = & 
 - \sum _{\vJ} \mP^\vJ {}_K(i,j) \big [ a_i ^K, B_{i \vJ} t_{ij} \big ]
 \no \\ &&
+ \sum_{\vI, \vJ} \sum_{\vJ = \vR \vS}  \mH^{\vI \shuffle \cS(\vS) | \vR | M}{}_K(i,j) \,
 \Big [ B_{i \vI} \, \Big ( B_{iM} a^K_i 
- {1 \over h}  \delta ^K_M  \, B_{i N} a^N_i \Big ) , B_{j \vJ} \, t_{ij} \Big ] 
\no \\ &&
- {1 \over h} \sum_{\vI, \vJ} \, \sum_{\vJ = \vR \vS} \Xi ^{\vI \shuffle \cS(\vS)| \vR } (i,j)   \,  \Big [ B_{i \vI} \,  B_{i N} a^N_i , B_{j \vJ} \, t_{ij} \Big ] 
\eea
in terms of the constituents $ \mP^\vJ {}_K(i,j) $ and $\mH^{\vI \shuffle \cS(\vS) | \vR | M}{}_K(i,j)$
of interchange and Fay identities in (\ref{2.c.2}) and (\ref{2.d.10}) as well as
the function $\Xi$ defined by (see (\ref{2.d.2b}) for $\mF_1^{\vI | \vJ | L}{}_L(i,j) $),
\bea
\label{3.f.2}
\Xi ^{\vI |\vJ}(i,j) & =  & 
- h \sum_{\vI = \vP \vQ }  f^{\vP}{}_L(i,j) \, \p_j \cG ^{\vJ \shuffle L \vQ} (j,i) - \mF_1^{\vI | \vJ | L}{}_L(i,j) 
\no \\ && 
- h \sum_{\vJ = \vR \vS } f^{\vR}{}_L(j,i) \, \p_i \cG ^{\vI \shuffle L  \vS}(i,j)
\eea
Note that the traceless projection $\sim (B_{iM} a^K_i - {1 \over h}  \delta ^K_M   B_{i N} a^N_i)$ in the second line of (\ref{3.d.3}) is absent for genus $h=1$. The second term $\cC_{ij}^{(2\mS)}$ of (\ref{3.d.2}) is given by,
\bea
\label{D.99}
\cC^{(2 \mS)}_{ij} & = & - \cC^{(01)}_{ij} + 
\sum_{\vI, \vJ}  \mf^{\vI|\vJ} (i,j) \, \big [ B_{i \vI} \, t_{ij} , B_{i \vJ} \,  t_{ij} \big ] 
\eea
in terms of the function $\mf^{\vI|\vJ} (i,j) $ which is defined by (see (\ref{dih1}) for the function $\mL^{ \vI }(j,i) $),
\bea
\label{3.f.f}
\mf^{\vI|\vJ} (i,j) =  \sum_{\vJ = \vR \vS}  \Bigg \{ \frac{1}{2} \sum_{\vI =  \vP \vQ}  
f^{\vP \shuffle \vR}{}_M(i,j) \, \mL^{ \cS(\vS)L \vQ }(j,i) 
 - \frac{1}{h}  
 \mF^{ \vI \shuffle \vR | \cS(\vS) | M }{}_L(i,j;j) \Bigg \} 
 \qquad
\eea
and the combination,
\bea
\label{3.f.4}
\cC^{(01)}_{ij} & = &  
  - {1 \over h} \sum_{\vI, \vJ}   \cR^{\vI M |\vJ} _{ij}{}_M  \,   \big [ B_{i \vI} \, t_{ij} , B_{j \vJ} \,  t_{ij} \big ] 
\eea
where $ \cR^{\vI M |\vJ} _{ij}{}_M$ is determined by,
\bea
 \label{D.c.3}
\cR ^{ \vI | \vJ}_{ij} {}_M = 
- \sum_{\vI= \vP \vQ} \sum _{\vJ = \vR \vS} 
\Big ( f^{ (\vP \shuffle \cS(\vS) )  L \vQ} {}_M (i,j) \, f^\vR {}_L (j,i) 
 + f^{\vP \shuffle \cS(\vS)} {}_L (i,j) \, f^{\vR L \vQ} {}_M (j,i)  \Big )
 \quad
\eea
}

The proof of Proposition \ref{3.lem:3} is quite involved and will be given in appendix \ref{sec:D}. The combination
$\cR ^{ \vI | \vJ}_{ij} {}_M$ will already appear in early stages of the proof, see Proposition \ref{D.prop:4}, whereas
the emergence of $\Xi ^{\vI |\vJ}(i,j)$ defined by (\ref{3.f.2}) requires more work, see Proposition~\ref{D.prop:3}.

\subsection{Combining the contributions from $\cC^{(0)}_{ij}$, $\cC^{(1)}_{ij}$ and $\cC^{(2)}_{ij}$}

The combination of $\cC^{(0)}_{ij}$, $\cC^{(1)}_{ij}$, which were given in (\ref{3.b.2}) and (\ref{3.c.2}), respectively, may be expressed in terms of the trace of the function $\mF$ defined in (\ref{2.d.5}),  and we have, 
\bea
\label{3.f.1}
\cC^{(0)}_{ij} + \cC^{(1)}_{ij}
= { 1 \over h} \sum _{k \not = i,j} \sum_{\vI,\vJ}   
\Big ( \mF ^{\vI | \vJ | L} {}_L (i,j;k) +  \Xi ^{\vI|\vJ}(i,j) \Big ) 
\big [B_{i \vec{I}} \, t_{ik} ,B_{j \vec{J}} \, t_{jk} \big ]
\eea
where $\Xi ^{\vI|\vJ}(i,j)$ was defined in (\ref{3.f.2}).  In order to rearrange this result in a form that will lend itself to various simplifications, we establish the following lemma next. 
{\lem
\label{2.lem:4}
The structure relations of $\mt_{h,n}$ imply the following linear relation for $i \not= j$,
\bea
\label{2.lem.4}
\sum_{k \not= i,j} \big [ B_{i \vI} \, t_{ik}, B_{j \vJ} \, t_{jk} \big ]
=
- B_{i \vI } \,   \big [ B_{j \vJ} \,  t_{ij}, t_{ij} \big ] - B_{i \vI} \,   \big [ B_{j \vJ} \,  t_{ij}, B_{iM} \, a^M_i \big ]
\eea}

\begin{proof}
\vskip -0.1in
To prove the lemma, we use the relations $[B_{i I} , t_{jk} ] = [ B_{j J} , t_{ik}]=0$ repeatedly, as well as the relation $[t_{ij} + t_{ik}, t_{jk}]=0$,   to recast the left side of (\ref{2.lem.4}) as follows,
\bea
\sum_{k \not= i,j} B_{i \vI} \, B_{j \vJ} \,   \big [ t_{ik}, t_{jk} \big ]
& = &
 \sum_{k \not= i,j} B_{i \vI} \, B_{j \vJ} \,   \big [ t_{ij}, t_{ik} \big ]
= \sum_{k \not= i,j}  B_{i \vI}  \big [ B_{j \vJ} \,  t_{ij}, t_{ik} \big ]
\no \\ &=&
- B_{i \vI } \,   \big [ B_{j \vJ} \,  t_{ij}, t_{ij} \big ] + \sum_{k \not= i} B_{i \vI} \,   \big [ B_{j \vJ} \,  t_{ij}, t_{ik} \big ]
\label{more2lem4}
\eea
Using the relation on the third line of (\ref{11.1}) we replace the sum over $k$ of $t_{ik}$ by $- B_{iM} a^M_i$ to obtain (\ref{2.lem.4}),   thereby completing the proof of the lemma. 
\end{proof}

Applying Leibniz's rule in Proposition \ref{B.prop:1} to $B_{i \vI}$ on the right side of (\ref{2.lem.4}) of Lemma \ref{2.lem:4}, and relabeling the summation variables, the contribution from $\Xi$ to (\ref{3.f.1}) takes the~form, 
\bea
\label{3.f.3}
\cC^{(0)}_{ij} + \cC^{(1)}_{ij}
& = &   \cC^{(01)}_{ij}  + 
 { 1 \over h}  \sum _{k \not = i,j} \sum_{\vI,\vJ} \mF ^{\vI | \vJ | L} {}_L (i,j;k) 
\big [B_{i \vec{I}} \, t_{ik} ,B_{j \vec{J}} \, t_{jk} \big ] 
\no \\ && 
+ {1 \over h} \sum_{\vI, \vJ}   \sum_{\vJ=\vX \vY} 
\Xi ^{\vI \shuffle \cS(\vY) |\vX}(i,j) \, 
 \big [ B_{i \vI} B_{iM} \, a^M_i , B_{j \vJ} \,  t_{ij} \big ] 
\eea
where the first term on the right of (\ref{2.lem.4}) has been identified with the contribution $\cC^{(01)}_{ij}$ defined in (\ref{3.f.4}) and rewritten using the results of Proposition \ref{D.prop:3} as follows,
\bea
\label{3.f.4new}
\cC^{(01)}_{ij} & = &  
   {1 \over h} \sum_{\vI, \vJ}   \sum_{\vJ = \vR \vS} \Xi ^{\vI \shuffle \cS(\vS) | \vR} (i,j)  \,   \big [ B_{i \vI} \, t_{ij} , B_{j \vJ} \,  t_{ij} \big ] 
\eea
The contribution $\cC^{(2)}_{ij}$ is obtained by combining (\ref{D.b.1}) and (\ref{D.b.2}) 
and using the results of Propositions \ref{D.prop:2} and \ref{D.prop:3}. It may be expressed as follows,  
\bea
\label{3.f.5}
\cC^{(2)} _{ij} & = &  \cC^{(2\mS)}_{ij} 
- \sum _{\vJ}  \mP^\vJ {}_K(i,j)  \, \big [ a^K_i , B_{j \vJ} \, t_{ij} \big ] 
 \no \\ &&
+  \sum_{\vI, \vJ} \sum_{\vJ = \vR \vS}  \mH^{\vI \shuffle \cS(\vS) | \vR | M}{}_K(i,j) \,
 \Big [ B_{i \vI} \, \Big ( B_{iM} a^K_i 
- {1 \over h}  \delta ^K_M \, B_{i N} a^N_i \Big ) , B_{j \vJ} \, t_{ij} \Big ] 
\no \\ &&
- {1 \over h} \sum_{\vI, \vJ} \, \sum_{\vJ = \vR \vS} \Xi ^{\vI \shuffle \cS(\vS)| \vR } (i,j)   \,  \Big [ B_{i \vI} \,  B_{i N} a^N_i , B_{j \vJ} \, t_{ij} \Big ] 
\eea
where $\cC^{(01)}_{ij} + \cC^{(2\mS)}_{ij}$ was obtained in (\ref{D.99}). Considering now the full sum 
$\cC^{(0)}_{ij} + \cC^{(1)}_{ij} + \cC^{(2)}_{ij}$ giving the commutator $[J_i^{(1,0)} ,J_j^{(1,0)} ]$, see (\ref{3.a.2}), we see that all the terms $\Xi ^{\vI \shuffle \cS(\vS)| \vR } (i,j) $ cancel between the last lines of (\ref{3.f.3}) and (\ref{3.f.5}). Combining all contributions establishes the following theorem, which constitutes one of the fundamental results of this paper. 

{\thm 
\label{2.thm:4}
The commutator $[J_i^{(1,0)} ,J_j^{(1,0)} ]$ of the $(1,0)_i$, $(1,0)_j$ components of the multivariable
DHS connection $\cJ_{\rm DHS}$ in (\ref{2.b.1}) admits the following decomposition,
\bea
\left  [J_i^{(1,0)} ,J_j^{(1,0)} \right ] & = & 
 { 1 \over h} \sum _{\vI, \vJ}  \sum_{k \not = i,j} 
\mF ^{\vI |  \vJ | M}  {}_M   (i,j;k)  \Big [ B_{i \vI} \, t_{ik} \, , B_{j \vJ} \, t_{jk} \Big ]
 - \sum_{\vJ} \mP^{\vec{J}}{}_K  (i,j) \, \big [ a_i^K , B_{i \vec{J}} \, t_{ij} \big ] 
\no \\ &&
+ \sum_{\vec{I} ,\vec{J}}  
\sum_{\vec{J} = \vR \vS}  \mH^{\vec{I} \shuffle \cS( \vS) | \vR | M }{}_K(i,j)
 \bigg [ B_{i \vI} \, \Big \{ B_{iM} \, a^K_i - { 1 \over h} \delta ^K_M \, B_{iN} a^N_i \Big \}   \, , B_{j \vJ} \,  t_{ij} \bigg ]
\no \\ &&
 + \sum _{\vI, \vJ}  \mf^{\vI|\vJ} (i,j) \, \Big [ B_{i \vI} \, t_{ij}  \, , B_{i \vJ} \, t_{ij} \Big ] 
  \label{2.thm.4} 
\eea
where the combinations $\mF ^{\vI |  \vJ | M}  {}_K   (i,j;k) $, $\mP^{\vec{J}}{}_K  (i,j)$, $\mH^{\vI | \vJ | M }{}_K(i,j)$ and  $\mf^{\vI|\vJ} (i,j) $ are defined in (\ref{2.d.5}), (\ref{2.c.2}), (\ref{2.d.10}) and (\ref{3.f.f}), respectively, and the second line is absent for genus $h=1$}.

\subsection{Linear independence of Lie algebra elements}

The structure relations of the algebra $\mt_{h,n}$ imply many relations between multiple commutators involving the generators $a_j ^J, b_{i I}$ and $ t_{k \ell}$ for various arrangements of $i,j,k,\ell$ and $ I,J$. For example, Lemma \ref{2.lem:4} already provides non-trivial relations between commutators involving one $a$ and one $t$ generator  and commutators which  involve two $t$ generators. The linear independence of the Lie algebra elements that enter into the fundamental Theorem~\ref{2.thm:4} is the subject of the following lemma. 

{\lem
\label{2.lem:5}
For arbitrary fixed values of $i \not= j \in \{ 1, \cdots, n\}$ and arbitrary words $\vI, \vJ \in \cW_h$ the following four sequences of Lie algebra elements in $\mt_{h,n}$ are linearly independent,\footnote{Removing the trace in the indices $K,M$ is imposed  in 4.\ as the trace part is linearly dependent on the sequences of lines 1.\ and 2.\ in view of the relation on the third line of  (\ref{11.1}) and Lemma \ref{2.lem:4}. Note that the second sequence is empty when $n= 2$ and that the fourth sequence is empty at genus $h=1$.} 

\begin{enumerate}
\itemsep=0in
\item  $\big [ B_{i \vI} \, t_{ij}  \, , B_{i \vJ} \, t_{ij} \big ] $
for any antisymmetrized pair of words $\vI,\vJ$;
\item $\big [ B_{i \vI} \, t_{ik} \, , B_{j \vJ} \, t_{jk} \big ]$ further indexed by $k \not = i,j \in \{ 1, \cdots, n\}$;
\item $\big [ a_i^K , B_{i \vec{J}} \, t_{ij} \big ] $ further index by $K \in \{ 1,\cdots,h \}$;
\item $ \big [ B_{i \vI} \,  \big\{ B_{iK} \, a^M_i - h^{-1} \delta ^M_K \, B_{iN} a^N_i  \big\}   \, , B_{i \vJ} \, t_{ij} \big ]$ 
further indexed by  $M,K \in \{ 1,\cdots, h\} $.
\end{enumerate} 
}

The proof of the lemma is relegated to appendix \ref{sec:E}.

\subsection{The equivalence Theorem}

In this subsection, we state and complete the proof of Theorem \ref{2.thm:4} -- the key result of this work -- namely that the vanishing of the commutator $[J_i^{(1,0)} ,J_j^{(1,0)} ]$ is equivalent to the collection of all interchange and Fay identities for DHS kernels.

{\thm 
\label{3.thm:3}
The vanishing of the  commutator $[J_i^{(1,0)} ,J_j^{(1,0)} ]$ of the  components of the DHS connection in $n\geq 3$ variables is equivalent to the set of all interchange and Fay identities,
\bea
\label{3.thm.3}
\left  [J_i^{(1,0)} ,J_j^{(1,0)} \right ] =0 
\qquad \Longleftrightarrow \qquad 
\left\{  \begin{array}{r} \mP^{\vec{J}}{}_K  (i,j) =0 \\ \mF ^{\vI |  \vJ | M}  {}_K   (i,j;k) =0 
\end{array}\right.
\eea
where $\mF ^{\vI |  \vJ | M}  {}_K   (i,j;k) $ and $\mP^{\vec{J}}{}_K  (i,j)$ were defined in (\ref{2.d.5}) and  (\ref{2.c.2}), respectively, for arbitrary $i \not= j \in \{ 1, \cdots, n \}$; $K,M \in \{ 1, \cdots, h \}$ and $\vI, \vJ \in \cW_h$. }

\begin{proof}
The proof will be based on combining the results of Theorem \ref{2.thm:4} and Lemma \ref{2.lem:5}. 

\sm

We begin by proving the implication from right to left in (\ref{3.thm.3}). The vanishing of $\mP^{\vec{J}}{}_K  (i,j) $ and $\mF ^{\vI |  \vJ | M}  {}_K   (i,j;k) $ implies the vanishing of the first, second and third terms in (\ref{2.thm.4}) of Theorem \ref{2.thm:4}, as well as of the second term in the expression (\ref{3.f.f}) for $\mf ^{\vI | \vJ}(i,j)$. Its first term $\mL^{ \cS(\vS)L \vQ }(j,i) $ vanishes by Proposition \ref{prop:dih}. Therefore, we have $[J_i^{(1,0)} ,J_j^{(1,0)} ]=~0$. 

\sm

To prove the implication from left to right in (\ref{3.thm.3}), we use the result of  Theorem \ref{2.thm:4} together with the results of Lemma \ref{2.lem:5}. Therefore, all Lie algebra elements in (\ref{2.thm.4}) are linearly independent so that each one of their coefficients has to vanish independently, 
\bea
\label{3.thm.4}
\left  [J_i^{(1,0)} ,J_j^{(1,0)} \right ]=0 
\qquad \Longrightarrow \qquad 
\left\{ \begin{array}{r}  \mP^{\vec{J}}{}_K  (i,j) =0 \cr 
\mF ^{\vI |  \vJ | M}  {}_M   (i,j;k) =0 \cr 
\sum_{\vec{J} = \vR \vS}  \mH^{\vec{I} \shuffle \cS( \vS) | \vR | M }{}_K(i,j) =0 \cr
\mf^{\vI | \vJ} (i,j) =0 
\end{array} \right.
\eea
The third relation on the right of (\ref{3.thm.4}) implies that $\mH^{\vI | \vJ | M }{}_K(i,j) =0$ for arbitrary $\vI, \vJ \in \cW_h$ and $K, M \in \{ 1, \cdots, h \}$, as proven in the following  lemma. 

{\lem 
\label{3.lem:6}
The following set of relations for arbitrary  $i\not= j \in \{ 1, \cdots, n \}$,  letters $K,M \in \{ 1, \cdots , h \}$ and words $\vI, \vJ \in \cW_h$,
\bea
\label{3.lem:22}
\sum_{\vec{J} = \vR \vS}  \mH^{\vec{I} \shuffle \cS( \vS) | \vR | M }{}_K(i,j) =0
\eea
implies $\mH^{\vI | \vJ | M }{}_K(i,j) =0$ for all $i\not= j \in \{ 1, \cdots, n \}$, $K,M \in \{ 1, \cdots , h \}$ and $\vI, \vJ \in \cW_h$.}

\begin{proof}
\vskip -0in
The lemma is proven by induction on the length $s$  of the word $\vJ$ for arbitrary $\vI$. For $s=0$ we have $\vJ =\emptyset$ so that $\vR =\vS = \emptyset$ and thus $\mH^{\vI | \emptyset |M}{}_K=0$.  
For $s=1$, the sum in (\ref{3.lem:22})  receives contributions from $(|\vR|, |\vS|) =(1,0)$ and $(|\vR|, |\vS|) =(0,1)$. The latter vanishes in view of our result for $s=0$, so that only the contribution 
$(|\vR|, |\vS|) =(1,0)$ survives, which implies that $\mH^{\vI  | \vR | M }{}_K=0$ for all $|\vR|=1$. 
The induction process is now clear. Assuming that the relation (\ref{3.lem:22}) for all $\vJ$ of length $s$ implies $\mH^{\vI  | \vQ | M }{}_K=0$ for all $|\vQ| \leq s$, we consider the relation (\ref{3.lem:22}) for $\vJ$ of length $s+1$. 
The sum over $\vR$ and $\vS$ with $\vR \vS=\vJ$ decomposes into a sum over $\vR$ with $0 \leq |\vR| \leq s+1$. But the contributions from all terms with $0 \leq |\vR| \leq s$ vanish by the induction hypothesis, leaving only those with $\vR=\vJ=\vQ$ for $|\vQ|=s+1$. This concludes the proof of Lemma \ref{3.lem:6} by induction.
\end{proof}

The union of the sets of conditions $\mF ^{\vI |  \vJ | L}  {}_L(i,j;k) =0$ and $\mH^{\vI | \vJ | M } {}_K (i,j) =0$ deduced in (\ref{3.thm.4}) and Lemma \ref{3.lem:6}, respectively, implies $\mF^{\vI | \vJ | M}{}_K(i,j;k)=0$ in view of Proposition \ref{2.def:2}. Thus, the conditions of (\ref{3.thm.4}) imply the full set of interchange and Fay identities. Since $\mf^{\vI | \vJ} (i,j)=0$ is a consequence of the interchange and Fay identities,  no extra conditions are being implied. This completes the proof of Theorem \ref{3.thm:3}. 
\end{proof}

\newpage

\section{Flatness of Enriquez connection $\Longrightarrow $ Fay identities}
\label{sec:5}

In this section, we shall show that the flatness conditions for the Enriquez connection in $n \geq 3$ variables imply the collection of all the interchange and Fay identities on Enriquez kernels. The derivation will make heavy use of the results we have already proven in section \ref{sec:4} expressing the flatness conditions in terms of interchange and Fay identities for DHS kernels, since the case for Enriquez kernels may be derived from the DHS case by simple algebraic substitutions.

\sm

We shall present brief reviews of the Enriquez connection and kernels; of the equivalence between the DHS and Enriquez connections shown in Figure 1 and proven in \cite{paper.1}; and of the interchange and Fay identities for Enriquez kernels conjectured in \cite{DHoker:2024ozn} and later proven in \cite{Baune:2024ber}. The key results of this section are given by Theorems \ref{4.thm:4} and  \ref{4.thm:8} which show that flatness implies the interchange and Fay identities for Enriquez kernels. 
In the final subsection, we explain why the reverse statement, namely that the interchange and Fay identities for Enriquez kernels  imply  flatness, if it holds, has not been established.

\subsection{The Enriquez kernels and connection}
\label{secE.1}

Enriquez proved the existence and uniqueness, subject to certain monodromy and residue conditions, of a meromorphic flat connection $\cK_\text{E}$ on the configuration space $ {\text{Cf}}_n(\tilde\Sigma)$ of $n$ points $\xx= (x_1, \cdots, x_n) $ in the universal cover $\tilde \Sigma$ of $\Sigma$ \cite{Enriquez:2011}.  The resulting connection $\cK_\text{E} $ is a sum of $(1,0)$ forms $dx_i$ in each variable $x_i$, 
\begin{align}
\label{5.b.1}
\cK_\text{E} (\xx;a,b,t) &= \sum_{i=1}^n K_i(\xx;a,b,t) \, dx_i 
 \end{align}
where the triplet  $(a,b,t)$ obeys the structure relations of $\hat \mt_{h,n}$ in Definition \ref{2.def:1}. The flatness condition $d \cK_\text{E} -  \cK_\text{E} \wedge \cK_\text{E} =0$ on the configuration space $ {\text{Cf}}_n(\tilde\Sigma)$ is implemented by requiring the stronger conditions \cite{Enriquez:2011} where the two terms vanish separately  so that $d \cK_\text{E} =  \cK_\text{E} \wedge \cK_\text{E} =0$. The latter reduce to the following set of equations on the components $K_i$,
\bea
\label{5.b.2}
\p_i K_j - \p_j K_i = 0
\hskip 1in
{}  [ K_i, K_j] = 0
\eea
The decomposition of $K_i$ onto $\hat \mt_{h,n}$ may be carried out with the help of the same  Enriquez kernels $g^{I_1 \cdots I_r}{}_J(x,y)$ that were used to formulate the case with one variable and one puncture~\cite{Enriquez:2011}. A convenient way to provide these expressions is in terms of generating functions,
\bea
\label{5.b.3}
\bK_J(x,y;B) & = & \sum _{r=0}^\infty g^{I_1 \cdots I_r}{}_J(x,y) B_{I_1} \cdots B_{I_r}
= \sum _{\vI \in \cW_h} g^\vI{}_J(x,y) B_{\vI} 
\no \\
\XX (x,y;B) & =  & \sum _{r=0}^\infty \chi ^{I_1 \cdots I_r}  (x,y) B_{I_1} \cdots B_{I_r}
\, = \sum _{\vI \in \cW_h} \chi^\vI (x,y) B_{\vI}
\eea
where we set $B_i X = [b_i,X]$ for arbitrary $X \in \hat \mt_{h,n}$ and define $g ^\emptyset {}_J (x,y)= \om_J(x)$ when $r=0$ in the first line of (\ref{5.b.3}). The decomposition of $g^{I_1 \cdots I_r}{}_J(x,y)$ into the traceless part $\varpi$ and the trace part  $\chi$ with respect to the last pair of indices follows the structure given in (\ref{2.b.4}) for the DHS case, 
\bea
\label{5.b.4}
g^{I_1 \cdots I_r}{}_J(x,y) = \varpi ^{I_1 \cdots I_r}{}_J(x) - \chi ^{I_1 \cdots I_{r-1}} (x,y) \, \delta ^{I_r}_J
\eea
where $\varpi ^{I_1 \cdots I_r}{}_J(x)$ does not depend on $y$ and satisfies $\varpi ^{I_1 \cdots I_{r-1} J}{}_J(x)=0$.
The explicit form of Enriquez kernels has been studied in terms of Poincar\'e series \cite{Baune:2024biq}, DHS kernels \cite{DHoker:2024desz} and convolution integrals over $\mA$-cycles \cite{DHoker:2025dhv}, also see \cite{Ichikawa:2025kbi} for a discussion of degeneration limits.

\sm

 To express the components $K_i$ of (\ref{5.b.1}) in terms of generating functions, it is convenient to introduce an arbitrary auxiliary point $z \in \tilde \Sigma$ upon which $K_i$ does not depend,
\bea
\label{5.b.5}
K_i(\xx;a,b,t)  &= 
 {\bf K}_J(x_i,z; B_i) a_i^J + \sum_{j \neq i} \big \{  {\bf X}(x_i,x_j;B_i) - {\bf X}(x_i,z;B_i)  \big \}  t_{ij}  
\eea
Given the decomposition (\ref{5.b.5}), the left equation of (\ref{5.b.2}) is equivalent to the relation,
\bea
\label{5.b.6}
\p_y {\bf X}(x,y;B_i) t_{ij} - \p_x {\bf X}(y,x;B_j)  t_{ij} =0
\eea
which follows from the analogues of (\ref{2.b.7}) for DHS kernels and were proven in \cite{Enriquez:2011},
\bea
\label{5.b.7}
\p_y \chi^{I_1 \cdots I_r} (x,y) & = & (-)^r \p_x \chi ^{I_r \cdots I_1} (y,x)
\eea
The right equation of (\ref{5.b.2}) was proven in \cite{paper.1} as an essential ingredient in an alternative  proof of Enriquez's theorem, established originally in  \cite{Enriquez:2011} by different methods.

\subsection{Relating the DHS and Enriquez connections}
\label{secE.2}

The  multivariable DHS connection $\cJ_\text{DHS}$ and Enriquez connection $\cK_\text{E}$, both valued in the Lie algebra $\hat \mt_{h,n}$,  are related by the composition of a gauge transformation and an automorphism of $\hat \mt_{h,n}$. This equivalence is represented between the two top boxes in Figure 1. The precise statement is the subject of the following theorem, which was proven in~\cite{paper.1}, building on a corresponding theorem for the single-variable DHS and Enriquez connections~\cite{DHoker:2024desz}.

{\thm
\label{5.thm:1}
The DHS connection $\cJ_{\rm DHS}$ and the Enriquez connection $\cK_{\rm E}$, both in $n$ variables $\xx=(x_1, \cdots, x_n)$  and valued in the Lie algebra $\hat \mt_{h,n}$, are related as follows,
\bea
\label{5.e.1}
d- \cK_{\rm E}  (\xx;a,b,t) = \cU(\xx, \yy) ^{-1} \big ( d - \cJ_{\rm DHS} (\xx; \hat a, \hat b, \hat t) \big ) \cU(\xx,\yy)
\eea
where $(a,b,t)$ and $(\hat a, \hat b, \hat t)$ both satisfy the structure relations of Definition \ref{2.def:1} and the map $(a,b,t) \longrightarrow (\hat a, \hat b, \hat t)$ is an automorphism of $\hat \mt_{h,n}$.  The gauge transformation $\cU(\xx,\yy)$ and the automorphism $(a,b,t) \longrightarrow (\hat a, \hat b, \hat t)$
are given in section 5 of \cite{paper.1} and will not be needed here.}

\sm

The equivalence between the vanishing of the commutators of the components of the corresponding connections, exhibited in the top boxes of Figure 1, is captured in the following corollary of Theorem \ref{5.thm:1}.

{\cor
\label{5.cor:1}
The commutators of the $(1,0)$ components of the DHS and Enriquez connections are related to one another as follows for all $i,j \in \{ 1, \cdots, n\}$, 
\bea
\label{5.cor.1}
{} [ K_i(\xx; a,b,t) , K_j (\xx; a,b,t) ]  & = & 
\cU(\xx,\yy) ^{-1} \big [ J^{(1,0)}_i(\xx; \tilde a, \hat b, \hat t), J^{(1,0)}_j(\xx; \tilde a, \hat b, \hat t) \big ] \, \cU(\xx,\yy) 
\qquad
\eea
where  the map $(a,b,t) \to (\tilde a, \hat b , \hat t)$ is an automorphism of $\hat \mt_{h,n}$.  The vanishing of the commutators $[J^{(1,0)}_i , J^{(1,0)}_j]$  is equivalent to the vanishing of $[K_i, K_j]$ for all $i,j \in \{ 1, \cdots, n\}$. }

\begin{proof}
The relation between commutators in (\ref{5.cor.1}) immediately follows from the relation between the individual components which was proven in section 5.6 of \cite{paper.1},
\bea
\label{5.cor.1a}
K_i(\xx; a,b,t)  & = & 
\cU(\xx,\yy) ^{-1} J^{(1,0)}_i(\xx; \tilde a, \hat b, \hat t) \, \cU(\xx,\yy) 
\eea
It was proven in \cite{paper.1} that the map $(a,b,t) \to (\tilde a, \hat b, \hat t)$ is an automorphism of $\hat \mt_{h,n}$, 
 so that the relation between commutators holds for all $a,b,t$ satisfying the structure relations of (\ref{11.1}). This completes the proof of the corollary. 
\end{proof}
 
\subsection{The interchange identities for Enriquez kernels}
\label{secE.3}

The \textit{interchange identities} for Enriquez kernels provide an infinite family of relations that are bilinear in the Enriquez kernels and depend on two points $x_i,x_ j \in \tilde \Sigma$. The functions $\mQ^{ \vI}{}_K (i,j) = \mQ^{ \vI}{}_K (x_i,x_j)$  are $(1,0)$ forms in $x_i$ and $x_j$ defined by,
\bea
\label{5.c.2}
 \mQ^{ \vI}{}_K (i,j) = 
 \sum_{\vec{I} = \vP \vQ} \Big \{ 
 g^{\cS(\vQ) L }{}_K (i,j) \, g^{ \vP }{}_L(j,i) 
 +  g^{ \cS(\vQ) }{}_L(i,j) \, g^{ \vP L }{}_K (j,i)
\Big \}
\eea
and obtained by formally substituting $f^{\vI}{}_J(i,j) \rightarrow g^{\vI}{}_J(i,j)$ in the 
expression (\ref{2.c.2}) for $ \mP^{ \vI}{}_K (i,j) $. The functions $\mQ^{ \vI}{}_K (i,j) $  satisfy the symmetry property $\mQ^{ \vI}{}_K (j,i) =  \mQ^{ \cS(\vI)}{}_K (i,j) $ analogous to (\ref{2.c.3}) for DHS kernels. The interchange identities among Enriquez kernels were proven in section 9.2 and appendix C.7 of \cite{DHoker:2024ozn} and may be expressed as the following  system of equations,
\bea
\label{5.c.1}
 \mQ^{ \vI}{}_K (i,j) =0
 \eea
for arbitrary word $\vI  \in \cW_h$, arbitrary index $K=1,\cdots,h$ and distinct points $x_i, x_j \in \tilde \Sigma$.

\subsection{The Fay identities for Enriquez kernels}
\label{secE.3a}

The \textit{Fay identities} for Enriquez kernels are bilinear relations among Enriquez kernels that involve three distinct points $x_i,x_j, x_k \in \tilde \Sigma$.  In analogy with DHS kernels, they may be formulated in terms of two functions $\mG_0$ and $\mG_1$ which match the expressions (\ref{2.d.3}) and (\ref{2.d.2b}) for $\mF_0$ and $\mF_1$ after formally substituting $f^{\vI}{}_J(i,j) \rightarrow g^{\vI}{}_J(i,j)$ and $\partial_i \cG^{\vI}(i,j) \rightarrow  \chi^{\vI}(i,j)$. The function $\mG_0$ is  a $(1,0)$ form in $x_i$ and $x_j$ and a scalar in $x_k$,   defined by,
\bea
 \label{5.d.3}
\mG _0 ^{\vI |  \vJ | M}  {}_K   (i,j;k) 
& = &    \chi^\vI (i ; k, j ) \, \chi^\vJ (j ; k, i ) \, \delta^M_K
-  \sum_{\vI = \vP \vQ }   g^{\vP}{}_L(i,j) \, g^{(\vJ \, \shuffle \, L \vQ )M}{}_K(j,k) 
\no \\ &&
 -  \sum_{\vJ = \vR \vS } g^{\vR}{}_L(j,i) \, g^{(\vI \, \shuffle \, L  \vS)M}{}_K(i,k)
\eea
where we have used the following convenient shorthand in the first term on the right side,
\bea
 \label{5.d.1}
\chi^\vI (i ; j,k)  = \chi^\vI(i,j) -  \chi^\vI(i,k) 
\eea
The function $\mG_1$ is a $(1,0)$ form in $x_i$ and $x_j$ only, defined by,
\bea
 \label{5.d.2b}
\mG_1 ^{\vI |  \vJ | M}  {}_{\! K}   (i,j)
= - \sum_{\vJ = \vR \vS } 
\Big (  g^{\vR}{}_L(j,i) \, g^{\vI M \cS(\vS)  L}{}_{\! K}(i,j) 
+g^{\vI M \cS(\vS)}{}_L(i,j) \,  g^{\vR L}{}_{\! K}(j,i)  \Big )
\eea
In terms of these functions, the Fay identities for Enriquez kernels takes the form, 
\bea
\label{5.d.5}
\mG ^{\vI | \vJ | M} {}_K (i,j;k) = \mG_0 ^{\vI | \vJ | M} {}_K (i,j;k) + \mG_1 ^{\vI |  \vJ | M}  {}_{\! K}   (i,j) =0
\eea
for arbitrary $M,K=1,\cdots,h$, words $\vI,\vJ \in \cW_h$ and pairwise distinct points $x_i, x_j, x_k \in \tilde \Sigma$. 
The Fay identities for Enriquez kernels were conjectured in \cite{DHoker:2024ozn}. A full proof was given in \cite{Baune:2024ber} while a more direct proof of cases with $\vI = I$ a single letter and $\vJ \in \cW_h$ was presented in an appendix in \cite{DHoker:2025dhv}.

\subsection{Fay and interchange identities from flatness}
\label{fromflat}

In this subsection, we shall present the second main result of this work by deducing the interchange and Fay identities for  Enriquez kernels  in (\ref{5.d.5}) and (\ref{5.c.1}) from the flatness conditions $[K_i, K_j] = 0$ on the components of ${\cal K}_{\rm E}$.

{\thm 
\label{4.thm:4}
The commutator $[K_i ,K_j]$ of the $(1,0)_i$, $(1,0)_j$ components of the multivariable
Enriquez connection $\cK_{\rm E}$ in (\ref{5.b.5}) admits the following decomposition,
\bea
\left  [K_i ,K_j \right ] & = & 
 { 1 \over h} \sum _{\vI, \vJ}  \sum_{k \not = i,j} 
\mG ^{\vI |  \vJ | M}  {}_M   (i,j;k)  \Big [ B_{i \vI} \, t_{ik} \, , B_{j \vJ} \, t_{jk} \Big ]
 - \sum_{\vJ} \mQ^{\vec{J}}{}_K  (i,j) \, \big [ a_i^K , B_{i \vec{J}} \, t_{ij} \big ] 
\no \\ &&
+ \sum_{\vec{I} ,\vec{J}}  
\sum_{\vec{J} = \vR \vS}  \mI^{\vec{I} \shuffle \cS( \vS) | \vR | M }{}_K(i,j)
 \bigg [ B_{i \vI} \, \Big \{ B_{iM} \, a^K_i - { 1 \over h} \delta ^K_M \, B_{iN} a^N_i \Big \}   \, , B_{j \vJ} \,  t_{ij} \bigg ]
\no \\ &&
 + \sum _{\vI, \vJ}  \mg^{\vI|\vJ} (i,j) \, \Big [ B_{i \vI} \, t_{ij}  \, , B_{i \vJ} \, t_{ij} \Big ] 
  \label{4.thm.4} 
\eea
where the functions $\mG ^{\vI |  \vJ | M}  {}_K   (i,j;k) $ and $\mQ^{\vec{J}}{}_K  (i,j)$ are defined in (\ref{5.d.3}), (\ref{5.d.2b}), (\ref{5.d.5}) and  (\ref{5.c.2}), respectively. The traceless part  $\mI^{\vI | \vJ | M }{}_K(i,j)$ of $\mG ^{\vI |  \vJ | M}  {}_K   (i,j;k) $ is independent of $x_k$ and given by,
\bea
\mI^{\vI | \vJ | M }{}_K(i,j) = \mG^{\vI | \vJ | M }{}_K(i,j;k )  - { 1 \over h} \delta ^M_K \, \mG^{\vI | \vJ | N }{}_N(i,j;k)  
 \label{thm44.01}
\eea
while the combination $\mg^{\vI|\vJ} (i,j) $ in the last line of (\ref{4.thm.4}) is related to the coincident limit of $\mG ^{\vI |  \vJ | M}  {}_K   (i,j;k) $ and is given by, 
\bea
\label{4.f.f}
\mg^{\vI|\vJ} (i,j) =  \sum_{\vJ = \vR \vS}  \Bigg \{ \frac{1}{2} \sum_{\vI =  \vP \vQ}  
g^{\vP \shuffle \vR}{}_M(i,j) \, \mM^{ \cS(\vS)L \vQ }(j,i) 
 - \frac{1}{h}  
 \mG^{ \vI \shuffle \vR | \cS(\vS) | M }{}_L(i,j;j) \Bigg \} 
 \qquad
\eea
The function $\mM^\vI(j,i)$ is the counterpart for Enriquez kernels of the function  $\mL^\vI(j,i)$ introduced in Proposition~\ref{prop:dih} with the help of Proposition~\ref{2.prop:66} for DHS kernels. Its definition, 
\bea
\label{4.dih1}
\mM^\vI(j,i)  = \delta^{\vI + \cS(\vI)}(j) 
- \chi^{\vI + \cS(\vI)}(j,i)   - \sum_{\vI = \vX \vY \vZ}
g^{ (\vX \shuffle \cS(\vZ)) M (\vY + \cS( \vY)) }{}_M(j,i)
\qquad
\eea
involves the meromorphic counterpart $\delta ^\vI$ to the coincident limit $\gamma ^\vI$ in (\ref{2.h.2}) defined for~$\vI \not= \emptyset$, 
\bea
\delta ^\vI (x) = \lim _{y \to x} \left ( \chi^\vI(x,y) \right ) 
\label{4.f.k}
\eea
without any analogue of the angular singularity in the $y \to x$ behaviour
of $\partial_x \cG^{I_1}(x,y)$.} 

\begin{proof}
\vskip -0in
The proof of the theorem follows exactly the same steps as the proof of Theorem~\ref{2.thm:4} with the following substitutions for the kernels $f^\vI{}_J(x,y) $ and $ g^\vI{}_J(x,y)$ as well as for the components $J_i^{(1,0)}$ in (\ref{2.b.5}) and $K_i$ in (\ref{5.b.5}) and their decompositions, 
\begin{align}
f^\vI{}_J(x,y) & ~ \rightarrow ~  g^\vI{}_J(x,y) & \mP^{\vec{J}}{}_K  (i,j) 
	& ~ \rightarrow ~ \mQ^{\vec{J}}{}_K  (i,j)
\no \\
[J_i^{(1,0)}, J_j^{(1,0)}] & ~ \rightarrow ~ [K_i, K_j] & \mF ^{\vI |  \vJ | M}  {}_M   (i,j;k) 
 	& ~ \rightarrow  ~ \mG ^{\vI |  \vJ | M}  {}_M   (i,j;k)
\no \\
&& \mH^{\vec{I} \shuffle \cS( \vS) | \vR | M }{}_K(i,j) 
	& ~ \rightarrow ~ \mI^{\vec{I} \shuffle \cS( \vS) | \vR | M }{}_K(i,j)
\no \\
&& \mf^{\vI|\vJ} (i,j) & ~ \rightarrow ~ \mg^{\vI|\vJ} (i,j) 
\end{align} 
The statement (\ref{4.thm.4}) of the theorem then follows from the facts that  the proof of (\ref{2.thm.4}) in section \ref{sec:4} and appendices \ref{sec:C} and \ref{sec:D} does not rely on any functional relations of the kernels 
$f^\vI{}_J(x,y)$ other than the $y$-independence of its traceless part in (\ref{2.b.4}) (which holds in identical form for $g^\vI{}_J(x,y)$ and thus leads to the $k$ independence of (\ref{thm44.01})). Combinations such as 
$ \mP^{\vec{J}}{}_K  (i,j)$ and their counterparts $ \mQ^{\vec{J}}{}_K  (i,j)$,  which vanish on the support of the kernel identities,  are then generated by the identical combinatorial mechanisms, based on the structure relations of the Lie algebra $\mt_{h,n}$ for both  $J_i^{(1,0)}$ and $K_i$.
\end{proof}

The decomposition (\ref{4.thm.4}) of the commutator $[K_i ,K_j ]$ in the previous theorem
is the key for the following main theorem of this work.

{\thm 
\label{4.thm:8}
For $n \geq 3$, the vanishing of the  commutator $[K_i ,K_j ]$ of the  components of the multivariable
Enriquez connection implies the set of all interchange and Fay identities for Enriquez kernels,
\bea
\label{4.thm.9}
\left  [K_i ,K_j \right ] =0 
\qquad \Longrightarrow \qquad 
\left\{  \begin{array}{r} \mQ^{\vec{J}}{}_K  (i,j) =0 \\ \mG ^{\vI |  \vJ | M}  {}_K   (i,j;k) =0 
\end{array}\right.
\eea
where $\mG ^{\vI |  \vJ | M}  {}_K   (i,j;k) $ and $\mQ^{\vec{J}}{}_K  (i,j)$ were defined in (\ref{5.d.5}) and  (\ref{5.c.2}), respectively, for arbitrary $i \not= j \in \{ 1, \cdots, n \}$; $K,M \in \{ 1, \cdots, h \}$ and $\vI, \vJ \in \cW_h$. }

\begin{proof}
\vskip -0in
The proof proceeds by the same logic as that of the $\Longrightarrow$ implication in Theorem \ref{3.thm:3}.
By Lemma \ref{2.lem:5}, and assuming $n \geq 3$, the Lie algebra elements in $\mt_{h,n}$ on the right side of (\ref{4.thm.4}) are linearly independent such that each one of their coefficients has to vanish independently. In particular,
we arrive at
\bea
\mQ^{\vec{J}}{}_K  (i,j)  = 0 
\hskip 0.55in
\mG ^{\vI |  \vJ | L}  {}_L   (i,j;k)  = 0 
\hskip 0.55in
 \sum_{\vJ = \vR \vS} \mI^{\vec{I} \shuffle \cS( \vS) | \vR | M }{}_K(i,j) =0
 \label{thm44.02}
\eea
for $\vI, \vJ \in \cW_h$ and $K,M\in \{1,\cdots,h\}$
whose first equation matches the interchange identities among Enriquez kernels.
By the arguments in the proof of Lemma \ref{3.lem:6} and the decomposition (\ref{thm44.01})
of the quantity $\mG ^{\vI |  \vJ | M}  {}_K   (i,j;k)$ of (\ref{5.d.5}) into its trace with respect to $M,K$
and a traceless part $ \mI^{\vec{I} | \vJ | M }{}_K(i,j)$, the last two equations of (\ref{thm44.02})
amount to the Fay identities among Enriquez kernels.
\end{proof}

\subsection{Obstacles to proving equivalence}

We conclude this section by explaining the currently perceived  obstacles to establishing the inverse implication of (\ref{4.thm.4}), namely why the interchange and Fay identities of Enriquez kernels by themselves do not necessarily  imply the vanishing of $[K_i,K_j]$. The culprits are the coefficients $\mg^{\vI|\vJ} (i,j) $ of the $\hat \mt_{h,n}$ generators in the last line of (\ref{4.thm.4}) which, by equations (\ref{4.f.f}) to (\ref{4.f.k}), involve coincident limits of
Enriquez kernels.

\sm

The DHS counterparts of these coincident limits obey the decomposition (\ref{2.h.3})
in terms of modular tensors $\hat \mN^{I_1\cdots I_r}$ whose very existence rests on the
cyclic symmetry $\hat \mN^{I_1I_2\cdots I_r} = \hat \mN^{I_2\cdots I_rI_1}$ in (\ref{2.h.4}). 
This symmetry is a simple consequences of the integral representations
(\ref{defhatn2}) and (\ref{defhatn}) which, in turn,  follow from integrating lower-rank instances of (\ref{2.h.3})
over the Riemann surface $\Sigma$.

\sm

The meromorphic counterpart of the decomposition (\ref{2.h.3}) was conjectured in section 9.4 of \cite{DHoker:2024ozn} to take the following form (see (\ref{4.f.k}) for the definition of the coincident 
limit $\delta ^\vI (x)$),
\bea
\label{4.h.3}
\delta ^\vI (x) & = &  \sum_{\vI = \vX \vY \vZ} \Big ( \varpi^{\vX \shuffle \cS(\vZ)}{}_M(x) \, \mN^{M \vY} 
+ \varpi^{(\vX \shuffle \cS(\vZ)) M \vY }{}_M(x) \Big ) 
\eea 
in terms of quantities $\mN^{M \vY}$ that solely depend on the moduli of $\Sigma$. 
For $\vX \shuffle \cS(\vZ) = \emptyset$ we set $\varpi ^\emptyset {}_M(x)=\om_M(x) $ while for $\vI \not= \emptyset$ the function $\varpi ^\vI{}_M(x)$ is the traceless and $y$-independent part of $g^\vI{}_M(x,y)$ in (\ref{5.b.4}). However, consistency of (\ref{4.h.3}) with the $\mB_K$ monodromies in $x$ hinges on the cyclic symmetry $ \mN^{I_1I_2\cdots I_r} =  \mN^{I_2\cdots I_rI_1}$ which has not yet been derived beyond rank two. 

\sm

Given the conjectural status of (\ref{4.h.3}), the proof of Proposition \ref{prop:dih} which relies on (\ref{2.h.3}) cannot be adapted to the meromorphic case in an obvious way. Accordingly,
it is unclear how to prove the vanishing of the quantities $\mM^\vI(j,i) $ in (\ref{4.dih1}) without appealing to the flatness condition
$[K_i, K_j]=0$ and (\ref{4.thm.4}). As a result, the relation between the flatness conditions for the Enriquez connection
and the interchange and Fay identities of the Enriquez kernels 
is not an equivalence, but rather a one-sided implication, summarized by Theorem \ref{4.thm:8}.

\sm 

If the conjectural decomposition (\ref{4.h.3}) of the coincident limits $\delta ^\vI (x) $
holds, then the flatness condition $[K_i, K_j]=0$ implies the reflection parity
\bea
\mN^{M \vY} = \mN^{\theta(M \vY)}
\label{4.refl}
\eea
of the quantities $ \mN^{I_1\cdots I_r} $: the decomposition
(\ref{4.thm.4}) of the commutator, Lemma \ref{2.lem:5} on the
linear independence on the $\hat \mt_{h,n}$ generators and the 
Fay identities of Enriquez kernels imply the vanishing of the
quantities $\mM^\vI(j,i) $ in (\ref{4.dih1}). Together
with the assumed decomposition (\ref{4.h.3}) and the arguments in
the proof of Proposition \ref{prop:dih}, this
leads to the reflection properties (\ref{4.refl}). In conclusion, if the quantities
$ \mN^{I_1I_2\cdots I_r}$ in (\ref{4.h.3}) are cyclic (which is the consistency condition of
this decomposition), then flatness of ${\cal K}_{\rm E}$ guarantees that
they additionally obey (\ref{4.refl}) and thus exhibit full dihedral symmetry.

\newpage

\appendix

\section{Combinatorial definitions and identities}
\label{sec:A}

In section \ref{sec:A.1} of this appendix, we collect general definitions of combinatorial entities, including  the concatenation and shuffle products,  the antipode, and their representations on functions. In sections \ref{sec:A.2} and \ref{sec:A.3} we prove a number of specific combinatorial identities that will be of use in section \ref{sec:4} and in subsequent appendices. A comprehensive reference on the combinatorics of  free Lie algebras may be found in Reutenauer's book \cite{Reutenauer}.

\subsection{Basic definitions}
\label{sec:A.1}

Throughout, we consider an alphabet of $h$ letters, denoted by capitalized indices  $I=1,\cdots, h$, and denote a word of length $r$ in this alphabet by $\vI = I_1\cdots I_r$. The empty word has $r=0$ and is denoted by $\emptyset$.   The set of all words in an alphabet of $h$ letters is denoted by $\cW_h$. 

{\deff
\label{A.def:1}
The concatenation product is a binary operation $ \cW_h \times \cW_h \to \cW_h$ on words $\vI= I_1 \cdots I_r$ and $\vJ= J_1 \cdots J_s$ of lengths $r$ and $s$, respectively, that produces a word of length $r+s$ defined by concatenating the letters of both words as follows,
\bea
\label{A.def.1}
\vI \vJ =  I_1 \cdots I_r \,  J_1 \cdots J_s
\eea
The concatenation product is associative, has the empty set as its neutral element (since we have $\vI \emptyset = \emptyset \vI = \vI$), and makes $\cW_h$ into a non-commutative free monoid of rank $h$. }

\sm

For example, the sum over all deconcatenations $\vP \vQ$ of a word $\vI$ for an arbitrary function $\f$ of $\cW_h \times \cW_h$ may be written out explicitly as follows,
\bea
\sum _{\vI = \vP \vQ} \f (\vP, \vQ) = \sum _{\rho=0}^r \f( I_1 \cdots I_\rho, I_{\rho+1} \cdots I_r)
\eea

{\deff
\label{A.def:2}
The vector space of linear combinations of words in $\cW_h$ with coefficients in $\CC$ is denoted $\cL_h$. Extending the concatenation product to $\cL_h$ is achieved by imposing the distributive property on the concatenation product of words and their linear combinations,
\bea
(\a \vI + \b \vJ) \vK = \a \vI \vK + \b \vJ \vK \hskip 1in \vI, \vJ, \vK \in \cW_h, \qquad \a, \b \in \CC
\eea
and makes $\cL_h$ into a non-commutative free algebra with neutral elements $0$ and $\emptyset$ for the operations of the vector space sum and the concatenation product, respectively. }

\sm

For example, an arbitrary function $f$ on $\cW_h$ may be extended to a function on $\cL_h$ by imposing linearity, as illustrated below for  two arbitrary words $\vI, \vJ \in \cW_h$ and $ \a, \b \in \CC$,
\bea
\label{A.def.2}
f (\a \vI + \b \vJ) & = & \a f(\vI) + \b f(\vJ) 
\eea

{\deff
\label{A.def:3}
The shuffle product $\shuffle$ is a binary operation $\shuffle : \cL_h \times \cL_h \to \cL_h$ that is linear in each argument, associative and commutative. In view of the linearity property, the shuffle product may be defined on $\cL_h$ by its operation on pairs of words in $\cW_h$. The empty word $\emptyset$ plays the role of the neutral element for the shuffle product,
\bea
\label{A.def.3}
\vI \shuffle \emptyset = \emptyset \shuffle \vI = \vI
\eea
while for non-empty words $\vI = I_1\cdots I_r  \in \cW_h$ and $\vJ = J_1\cdots J_s \in \cW_h$, with $r,s\geq 1$, the shuffle product may be defined recursively in the total length $r+s$ of the words, 
\bea
\vI \shuffle \vJ &= & I_1 (I_2\cdots I_r  \shuffle \vJ) + J_1 (\vI \shuffle J_2\cdots J_s)  
\label{dfeq.01} 
\no \\
&= & (I_1\cdots I_{r-1} \shuffle \vJ) I_r + (\vI \shuffle J_1\cdots J_{s-1}) J_s
\eea}
For example, a function $f$ defined on $\cL_h$ by linearity via (\ref{A.def.2}),  evaluated on the shuffle product of two arbitrary words $\vI, \vJ \in \cW_h$, satisfies the recursive relation implied by (\ref{dfeq.01}),
\bea
f(\vI \shuffle \vJ) &= & f \big ( I_1 (I_2\cdots I_r  \shuffle \vJ) \big ) + f \big ( J_1 (\vI \shuffle J_2\cdots J_s) \big )
\no \\   
&= & f \big ( (I_1\cdots I_{r-1} \shuffle \vJ) I_r \big )  + f \big ( (\vI \shuffle J_1\cdots J_{s-1}) J_s \big )
\eea

{\DP 
\label{A.def:4}
The antipode $\cS$ is a linear operation  $\cS : \cL_h \to \cL_h$ whose action on $\cL_h$ may be defined by its action on words in $\cW_h$, 
\bea
\cS(I_1 I_2 \cdots I_r) = (-1)^r  I_r \cdots I_2 I_1
 \label{dfeq.03}
\eea
The antipode acts as an anti-automorphism on the concatenation product, 
\bea
\label{A.def.4a}
\cS( \vI \vJ) & = & \cS(\vJ) \, \cS(\vI)
\eea
and as an automorphism on the shuffle product, 
\bea
\label{A.def.4b}
\cS( \vI \shuffle \vJ) & = & \cS(\vI) \shuffle \cS(\vJ)
\eea
These property, together with the action of the antipode on a single letter $\cS(I) = - \cS(I)$,
recursively imply the explicit definition given earlier  in (\ref{dfeq.03}).}

\sm

For example, the action of the antipode for an arbitrary function $f$ of $\cW_h$, extended to $\cL_h$ by linearity via (\ref{A.def.2}), is given by,
\bea
\label{A.def.4c}
f \big (\cS(I_1 \cdots I_r) \big ) = (-)^r f(I_r \cdots I_1)
\eea 

{\DP
\label{A.def:5}
The deconcatenation coproduct $\Delta : \cL_h \to \cL_h \times \cL_h$ defined by, \bea
\label{A.def.5a}
\Delta (\vI) = \sum _{\vI = \vP \vQ} \vP \otimes \vQ
\eea
is a homomorphism for the shuffle product, and satisfies,
\bea
\label{A.def.5b}
\Delta (\vI \shuffle \vJ) = \Delta (\vI) \shuffle \Delta (\vJ)
\eea}
As an application, we have the following lemma which provides a useful formula  for the deconcatenation sum of the shuffle product.

{\lem
\label{lem.decsh}
For any function $\varphi: \cL_h \times \cL_h \rightarrow \cL_h$ and any pair of words $\vec{I},\vec{J} \in \cW_h$, the deconcatenation sum of the shuffle product $\vI \shuffle \vJ$ can be rearranged as follows,
\bea
\sum_{\vI \, \shuffle \, \vJ = \vX \vY} \varphi (\vX,\vY) 
= \sum_{\vI = \vP \vQ}  \sum_{\vJ = \vR \vS} \varphi (\vP\shuffle \vR, \vQ \shuffle \vS) 
\label{flt.81.b}
\eea
}

\begin{proof}
The lemma is a consequence of Proposition 1.9 of Reutenauer's book \cite{Reutenauer} using the homomorphism property of (\ref{A.def.5b}).  Indeed, since the shuffles are evaluated separately in each entry of a tensor product in (\ref{A.def.5a}), namely  $(\vP \otimes \vQ) \shuffle (\vR \otimes \vS) = (\vP \shuffle \vR) \otimes (\vQ \shuffle \vS)$,
the homomorphism property can be written in the form of (\ref{flt.81.b}).
\end{proof}

\subsection{Deconcatenation sums into two words}
\label{sec:A.2}

In this subsection, we prove three lemmas in which deconcatenation sums into two words are performed with summands that involve shuffle products and the antipode of words. These results will be of pervasive usage in the computations of section \ref{sec:4} and later appendices.

{\lem
\label{lem:sa}
The deconcatenation sum of the shuffle product with the antipode gives, 
\bea
\sum_{\vI = \vP \vQ } \cS(\vP) \shuffle \vQ = \delta _{\vI, \emptyset} 
\hskip 1in 
\delta_{\vI, \emptyset} = \left\{ \begin{array}{cl} \emptyset & \hbox{ for } \,  \vI= \emptyset \\
0 & \hbox{ for } \,  \vI \neq \emptyset \end{array} \right.
\label{std.sa}
\eea
}

\begin{proof}
\vskip -0.05in
The proof proceeds by induction on the length $r$ of the word $\vI=I_1\cdots I_r$. The relation (\ref{std.sa}) holds for $r=0$ since then $\vI= \vP=\vQ=\emptyset$ and for $r=1$ since the left side then evaluates to $\cS(I_1) \shuffle \emptyset + \emptyset \shuffle I_1 =0$. For the inductive step, we assume the validity of (\ref{std.sa}) for words of length $r \geq 1$ and consider the left side of (\ref{std.sa}) for  length $r{+}1$,
\bea
\sum_{I_1\cdots I_{r+1} = \vP \vQ } \cS(\vP) \shuffle \vQ
= \sum_{j=0}^{r+1}(-1)^j I_j\cdots I_2 I_1 \shuffle I_{j+1} I_{j+2}\cdots I_{r+1} 
\eea
Using the second recursion relation for the shuffle product in (\ref{dfeq.01}), we obtain two contributions for the summands with $j=1,\cdots,r$, the first factoring $I_1$ on the right, the other factoring $I_{r+1}$ on the right. 
Upon combining the terms with rightmost $I_1$ and $I_{r+1}$ with the $j=r{+}1$ and $j=0$ summands,
respectively, both classes of terms vanish by the induction assumption for length $r$, which completes the proof. \end{proof}

{\lem
\label{prop.jqmr}
The deconcatenation sum into two words of a summand consisting of a single letter $L$ interleaved between two shuffle products is given by, 
\begin{align}
\sum_{ \vI = \vec{P} \vec{Q} } (\vR \shuffle \vec{P}) L (\vS \shuffle \vec{Q}) = \vR L\vS \shuffle \vI  
\hskip 1in 
\vI,\vR,\vS \in \cW_h
\label{mainc27}
\end{align}}

\begin{proof}
\vskip -0.2in
We first establish the special cases of (\ref{mainc27}) for which either $\vS=\emptyset$ or $\vR=\emptyset$, 
\bea
\sum_{ \vI = \vP \vQ } (\vR \shuffle \vec{P}) L   \vec{Q} = \vR L \shuffle \vI
\hskip 1in
\sum_{ \vI = \vP \vQ }   \vec{P} L (\vS \shuffle \vec{Q}) = L\vS \shuffle \vI  
\label{spc27}
\eea
They are equivalent to one another upon applying the antipode. To prove the left equation of (\ref{spc27}) we proceed by induction on the length of the word $\vI$. The relation manifestly holds for $r=0$ since then $\vI=\vP=\vQ=\emptyset$. Assuming its validity for length $r$ we evaluate its right side for $\vI = I_1 \cdots I_{r+1}$ using the second relation of (\ref{dfeq.01}), 
\bea
\vR L \shuffle I_1 \cdots I_{r+1} & = & (\vR \shuffle I_1 \cdots I_{r+1}) L + (\vR L \shuffle I_1 \cdots I_r) I_{r+1}
\no \\ & = &
 (\vR \shuffle I_1 \cdots I_{r+1}) L 
+  \sum_{j=0}^r (\vR \shuffle I_1\cdots I_j) L I_{j+1} \cdots  I_{r+1}  
\eea
Upon including the first term on the second line as the $j=r+1$ contribution to the sum, we recover the first equation in (\ref{spc27}) for length $r+1$, thereby completing its proof. 

\sm

We  shall now prove (\ref{mainc27}) by induction on the total number $N = r{+}t$ of letters in $\vI = I_1\cdots I_r$ and $\vS = S_1\cdots S_t$. The case $r=0$, which includes the case $N=0$,  holds trivially since then $\vI=\vP=\vQ=\emptyset$. For  $t=0$ we recover the left equation in (\ref{spc27}) which was proven already. Henceforth we consider the case $r, t \geq 1$, assume that the relation holds for total length $N$, and evaluate the right side of (\ref{mainc27}) for total length $r{+}t=N{+}1$, 
\begin{align}
\vR L S_1\cdots S_t \shuffle I_1 \cdots I_r  &= 
 (\vR L S_1\cdots S_{t-1} {\shuffle} I_1 \cdots I_r ) S_t
{+}(\vR L S_1\cdots S_t {\shuffle} I_1 \cdots I_{r-1})I _r 
\notag  \\
&= \sum_{j=0}^r (\vR \shuffle I_1\cdots I_j) L (S_1\cdots S_{t-1} \shuffle I_{j+1} \cdots I_r ) S_t \notag \\
&\quad
+ \sum_{j=0}^{r-1} (\vR \shuffle I_1\cdots I_j) L (S_1\cdots S_t \shuffle I_{j+1} \cdots I_{r-1} ) I_r
\label{flt.113}
\end{align}
We have used the second case of (\ref{dfeq.01}) in the first equality.  In passing from the first line on the right side to the next lines, we have used the length-$N$ instance of (\ref{mainc27}) as an inductive assumption.
The factors multiplying $L$ to the right on the last two lines combine into the shuffle product $(S_1 \cdots S_t \shuffle I_{j+1} \cdots I_{r})$ using (\ref{dfeq.01}) which gives the left side of (\ref{mainc27}) for $r{+}t=N{+}1$. We note that the term $(S_1\cdots S_{t} \shuffle I_{j+1} \cdots I_r ) \rightarrow (S_1\cdots S_{t} \shuffle I_{j+1} \cdots I_{r-1} )I_r$ is absent for $j=r$ since the  factor $I_{j+1}\cdots I_r$ is then empty. This completes the proof of Lemma \ref{prop.jqmr}.
\end{proof}

 {\lem
\label{C.words}
The deconcatenation sum into two words of a summand consisting of a letter $L$ concatenated with a double shuffle product involving an antipode is given by, 
\bea
 \label{invid.01}
 \sum_{\vI = \vP \vQ}  \vQ \shuffle \big(\vR \shuffle  \cS(\vP) \big)L =  \vR L \vI 
 \hskip 1in 
 \vI, \vR \in \cW_h
\eea
}

\begin{proof}
\vskip -0.2in
The proof proceeds by induction on the length $r$ of $\vI = I_1 \cdots I_r$. The case $r=0$ holds since then $\vI = \vP = \vQ=\emptyset $. Assuming that the relation  holds for words $\vI$ of length $r$, we evaluate the right side of (\ref{invid.01}) for length $r+1$ by using the induction assumption to recast the result as follows,
\bea
\label{flt.75}
 \vR L I_1\cdots I_r I_{r+1} &=&  \sum_{j=0}^{r}
(-)^j  \Big( I_{j+1}\cdots  I_{r}  \shuffle ( \vR  \shuffle I_j\cdots I_1) L \Big) I_{r+1} 
  \no \\
 &= &   \sum_{j=0}^{r}
(-)^j  \Big( I_{j+1}\cdots  I_r I_{r+1}  \shuffle ( \vR  \shuffle I_j\cdots I_1) L
\no \\ && \hskip 0.6in
 -  \big \{  I_{j+1}\cdots I_r I_{r+1}  \shuffle ( \vR  \shuffle I_j\cdots I_1 ) \big \}   L \Big)  
\eea
We may extend to summation range to include $j=r+1$ as the summand cancels for this assignment. Having done so, the contribution from the second line sums to zero, as may be seen by using the associativity and commutativity  of the shuffle product to remove the inner parentheses and the applying Lemma \ref{lem:sa}. The contribution from the first line gives (\ref{invid.01}) for words $\vI$ of length $r+1$.  This completes the inductive proof of the lemma
\end{proof}

\subsection{Deconcatenation sums into more than two words}
\label{sec:A.3}

Here, we shall state and prove the lemma that enters the proof of Proposition \ref{D.prop:5}
needed for the simplification of $\cC^{(2 \mS)}_{ij}$ in (\ref{D.99}).

{\lem
\label{matchT}
The deconcatenation sum for  words $\cS(\vJ), \vI $ interleaved by a letter $M$,\footnote{We note that the contribution of $\vY=\emptyset$ to the sum in (\ref{dcapp.01}) vanishes in view of  Lemma \ref{lem:sa} as the deconcatenation $\theta(\vJ) M \vI = \vX \vZ$ cannot be empty.} 
\begin{align}
&\sum_{\cS(\vJ)M \vI = \vX \vY \vZ} \big( \vX \shuffle \cS(\vZ) \big) N \big( \vY + \cS(\vY) \big)
= \Lambda^{M | \vI | \vJ | N}_{(x)}+ \Lambda^{M | \vI | \vJ | N}_{(y)}+ \Lambda^{M | \vI | \vJ | N}_{(z)}
\label{dcapp.01}
\end{align}
can be decomposed into the following three contributions,  
\bea
\label{dcapp.02} 
 \Lambda^{M | \vI | \vJ  | N}_{(x)} & = &
  \sum_{\vI = \vW \vX \vY \vZ}
  \big( \cS(\vJ) \shuffle \cS(\vZ) \big) M  \big(\vW \shuffle \cS(\vY)  \big) N  \big(\vX + \cS(\vX)  \big) 
\no \\
\Lambda^{M | \vI | \vJ  | N}_{(y)} & = &
  \sum_{\vI = \vP \vQ} \sum_{\vJ = \vX \vY}   
   \big(\cS(\vQ) \shuffle \cS(\vY) \big) N  \big(   \cS(\vX) M \vP -  \cS(\vP) M \vX  \big)
\no \\
   \Lambda^{M | \vI | \vJ  | N}_{(z)} & = &
   -   \sum_{\vJ = \vW \vX \vY \vZ} \big( \cS(\vI) \shuffle \cS(\vZ) \big) M \big(\vW \shuffle \cS(\vY) \big) N  \big(\vX + \cS(\vX) \big) 
\eea
Under swapping $\vI$ and $\vJ$ the contributions transform as follows,
\bea
\label{dcapp.02a} 
\Lambda^{M | \vI | \vJ  | N}_{(z)} = - \Lambda^{M | \vJ | \vI  | N}_{(x)}
\hskip 1in 
\Lambda^{M | \vJ | \vI  | N}_{(y)} = - \Lambda^{M | \vI | \vJ  | N}_{(y)}
\eea
as a result of the anti-symmetry of the  left side of (\ref{dcapp.01}) under swapping $\vI \leftrightarrow \vJ$.}

\begin{proof}
\vskip -0in
We shall distinguish three cases corresponding to whether the letter $M$ of the deconcatenation sum $\cS(\vJ)M \vI = \vX \vY \vZ$  belongs to the subword $\vX$, $\vY$ or $\vZ$, resulting in the respective  contributions  $ \Lambda^{M | \vI | \vJ | N}_{(x)}$, $\Lambda^{M | \vI | \vJ | N}_{(y)}$  and $ \Lambda^{M | \vI | \vJ | N}_{(z)}$. The $z$-contribution 
is related to the $x$-contribution by (\ref{dcapp.02a}), so it will suffice to prove the formulas for the $x$- and $y$-contributions. 

\sm

$\bullet$ When  $M$ belongs to $\vX$ we use the parametrization $\vX = \vT M \vU$ and deconcatenate 
$\cS(\vJ)M \vI =  \vT M \vU \vY \vZ$ by setting $ \vI =\vU \vY \vZ$ and $\cS(\vJ) = \vT$ so that sum becomes, 
\begin{align}
\sum_{ \cS(\vJ)M \vI = \vX \vY \vZ\atop{M \in \vX, \vY \not= \emptyset}} \! \! \! \! \! \big( \vX \shuffle \cS(\vZ) \big) N \big( \vY {+} \cS(\vY) \big) &=  \sum_{\vI = \vU \vY \vZ \atop {\vY \neq \emptyset} }  \big(  \cS(\vJ) M \vU \shuffle \cS(\vZ) \big) N \big( \vY {+}\cS(\vY) \big)
\notag \\
&=  \sum_{\vI = \vW \vX \vR \atop {\vX \neq \emptyset} }  \big(  \cS(\vJ) M \vW \shuffle \cS(\vR) \big) N \big( \vX {+}\cS(\vX) \big)
\label{dcapp.04}
\end{align}
In going from the first line to the second, we have relabeled $\vY \to \vX$, $\vU \to \vW$ and $\vZ \to \vR$. The words to the left of the product $N( \vX {+}\cS(\vX))$ may be rearranged using the relation, 
\bea
\cS(\vJ) M \vW  \shuffle\cS(\vR)  = \sum_{\vR = \vY \vZ}
( \cS( \vJ )  \shuffle  \cS( \vZ)) M (\vW  \shuffle \cS(\vY))
\eea 
which follows from Lemma \ref{prop.jqmr}. Substituting this relation into the second line of (\ref{dcapp.04}) gives 
the first line of (\ref{dcapp.02}) which completes the proof of case $x$.

$\bullet$ When $M$ belongs to $\vY$ we use the parametrization $\vY = \vT M \vU$ and deconcatenate 
$\cS(\vJ)M \vI = \vX \vT M \vU \vZ$ by setting $\cS(\vJ) = \vX \vT$ and $ \vI = \vU \vZ$,
\begin{align}
\sum_{\cS(\vJ)M \vI = \vX \vY \vZ\atop{M \in \vY}} \! \! \! \! \! \big( \vX \shuffle \cS(\vZ) \big) N \big( \vY {+} \cS(\vY) \big) 
&=  \sum_{ \vI = \vU \vZ} \, \sum_{\cS(\vJ) = \vX \vT}  \! \!\big( \vX \shuffle \cS(\vZ) \big) 
N \big( \vT M \vU + \cS(\vT M \vU)  \big) 
\label{dcapp.03}
\end{align}
We recover the expression (\ref{dcapp.02}) for $  \Lambda^{M | \vI | \vJ  | N}_{(y)}$  by renaming the summation variables $\vU \to \vP $, $\vZ \to \vQ$, $\vX \to \cS(\vY)$ and $\vT \to \cS(\vX)$, which completes the proof of case $y$. 

\sm

The three cases of $\cS(\vJ)M \vI = \vX \vY \vZ$ for $M \in \vX$, $M \in \vY$ and $M \in \vZ$ on the left sides of (\ref{dcapp.04}), (\ref{dcapp.03}) and its $z$-contribution counterpart cover the full summation range on the left side of (\ref{dcapp.01}). Hence, the above identification of the  $x$-, $y$- and $z$-contributions  reproduces the right
side of (\ref{dcapp.01}) which completes the proof of the lemma.
\end{proof}

\newpage

\section{The generalized Leibniz rule}
\label{sec:B}

The elements $B_{i I}$ act as derivations on the Lie algebra $\hat \mt _{h,n}$. The index labeling the point~$i$ will play no role in this appendix and will be suppressed. As a derivation, the element $B_I$ obeys Leibniz's rule,  
\bea
\label{B.a.1}
B_{I} [X, Y] =  [  B_{I} X,  Y  ] +  [  X,  B_{I}  Y  ] \hskip 1in X, Y \in \hat \mt _{h,n}
\eea
When expressed in terms of the elements $b_I$ with the help of $B_I X= [ b_I, X]$, (\ref{B.a.1}) is equivalent to the Jacobi identity. The product $B_\vI = B_{I_1} \cdots B_{I_r}$, which is associated with a word $\vI$ of length $r$, does not act as a derivation for $r \geq 2$ but satisfies a generalized Leibniz rule. 

{\prop
\label{B.prop:1}
The elements $B_\vI= B_{I_1} \cdots B_{I_r}$ obey the generalized Leibniz rule,\footnote{The symbol $\delta _{\vI, \vP \shuffle \vQ}$ is defined to  equal one when the word $\vI$ is an element of the set of all words produced by the shuffle product $\vP \shuffle \vQ$, and zero otherwise. }
\bea
B_\vI \, \big [ X,Y \big ] = \sum_{\vP, \vQ} \delta_{\vI , \vP\shuffle \vQ} \, \big [  B_\vP \, X,  B_\vQ  \, Y \big  ]
\label{B.a.2}
\eea
for arbitrary $X,Y \in \hat \mt_{h,n}$. This identity is equivalent to the following relation for an arbitrary sequence of coefficients $C^{\vI}$,
\bea
\sum_{\vI} C^{\vI} B_\vI \, \big [ X,Y \big ] = \sum_{\vP, \vQ} C^{\vP \shuffle \vQ} \, \big [  B_\vP \, X,  B_\vQ  \, Y \big ]
\label{B.a.3}
\eea
}

\begin{proof}
\vskip -0.2in
We shall prove (\ref{B.a.2}) by induction on the length $r$ of the word $\vI$. For $r=0$ the relations $\vI=\vP=\vQ= \emptyset$ reduce  both sides of (\ref{B.a.2}) to $[X,Y]$, while for $r=1$ the identity (\ref{B.a.2}) reduces to (\ref{B.a.1}), both of which hold true. The induction process is initialized by assuming (\ref{B.a.2}) to hold for all words $\vI$ of length less than or equal to $r$. We then need to prove the validity of (\ref{B.a.2}) for arbitrary words of length $r{+}1$, which may  be parametrized by $J \vI$ for an arbitrary word $\vI$ of length $r$ and an extra letter $J$, for which (\ref{B.a.2}) becomes,
\bea
B_{J \vI}\, \big [ X, Y \big ] = \sum_{\vP, \vQ} \delta_{J\vI , \vP\shuffle \vQ} \, \big [  B_{ \vP} X,  B_{ \vQ}  Y \big ]
\label{B.a.4}
\eea
In order to demonstrate this relation we evaluate its right hand side. There is no contribution from $\vP=\vQ=\emptyset$. Isolating the contributions from $\vP= \emptyset$ and $\vQ= \emptyset$, respectively, we have, 
\bea
\text{RHS} =  \big [ X, B_{J \vI}\, Y \big ] + \big [ B_{J \vI}\,  X, Y \big ] +
\sum_{ \vP \not = \emptyset, \vQ\not= \emptyset} \delta_{J\vI , \vP\shuffle \vQ} \, \big [  B_{\vP} X,  B_{ \vQ}  Y \big ]
\label{B.a.5}
\eea
Since $\vP, \vQ \not= \emptyset$ in (\ref{B.a.5}), we may parametrize them by factoring out a single letter to the left, $\vP = P_1 \vP'$ and $\vQ = Q_1 \vQ'$ which results in the shuffle product $\vP \shuffle \vQ= P_1(\vP' \shuffle \vQ) + Q_1(\vP \shuffle \vQ') $ and the following relation for the Kronecker symbol of (\ref{B.a.5}),
\bea
\label{B.a.6}
\delta_{J\vI , \vP\shuffle \vQ} = \delta_{J,P_1} \, \delta_{\vI , \vP' \shuffle \vQ} 
+ \delta_{J, Q_1} \delta_{\vI , \vP\shuffle \vQ'}
\eea
Inserting this expression into the right side of (\ref{B.a.5}) and carrying out the sums over $P_1, Q_1$,
\bea
\text{RHS} & = & \big [ X, B_{J \vI}\, Y \big ]  + \big [ B_{J \vI}\,  X, Y \big ] + 
\sum_{ \vP', \vQ \not= \emptyset} \delta_{\vI , \vP' \shuffle \vQ} \, \big [  B_{J} B_{\vP' } \, X,  B_{\vQ}  \, Y \big ]
\no \\ &&
+\sum_{ \vP \not= \emptyset , \vQ'} \delta_{\vI , \vP \shuffle \vQ'} \, \big [  B_{\vP } \, X,  B_{J} B_{\vQ' } \, Y \big ]
\label{B.a.7}
\eea
Including the first term on the right side as the $\vP = \emptyset$ contribution to the fourth term (denoting the summation variable with this new range by $\vP'$), and the second term as the $\vQ=\emptyset$ contribution to the third term (denoting the summation variable  by $\vQ'$), we obtain, 
\bea
\text{RHS} = 
\sum_{ \vP', \vQ'} \delta_{\vI , \vP' \shuffle \vQ'} \, \big [  B_{J} B_{\vP' } \, X,  B_{\vQ'}  \, Y \big ]
+\sum_{ \vP', \vQ'} \delta_{\vI , \vP' \shuffle \vQ'} \, \big [  B_{\vP' } \, X,  B_{J} B_{\vQ' } \, Y \big ]
\label{B.a.8}
\eea
Using the inductive assumption for length $r$, the right side equals $B_J B_{\vI} [X,Y] = B_{J \vI} [X,Y]$, which reproduces the left side of (\ref{B.a.4}). This completes the proof of Proposition \ref{B.prop:1}. \end{proof}

{\prop
\label{B.prop:2} 
A useful alternative to the generalized Leibniz rule is as follows,
\bea
\label{B.a.9}
\sum _{\vI} \cC^\vI \, [ B_{\vI} X, Y ] = \sum _{\vP, \vQ} \cC^{\vP \shuffle \vQ} \, B_\vP [ X, B_{\cS (\vQ)} Y] 
= \sum _{\vP, \vQ} \cC^{\vP \shuffle \cS(\vQ)} \, B_\vP [ X, B_{ \vQ} Y] 
\eea
for arbitrary coefficients $\cC^\vI$.}

\begin{proof}
To prove (\ref{B.a.9}) we apply the Leibniz rule of Proposition \ref{B.prop:1} to $B_\vP$ on the right side, 
\bea
\sum _{\vP, \vQ}  \cC^{\vP \shuffle \cS(\vQ)}  B_\vP [ X, B_{\vQ} Y] 
 = \sum_{\vR, \vS, \vQ}  \cC^{(\vR \shuffle \vS) \shuffle \cS(\vQ)} [ B_\vR X, B_\vS B_{\vQ} Y]
 \eea
Using the formula $B_\vS B_{\vQ}= B_{\vS \vQ}$  and introducing a redundant summation over $\vT= \vS \vQ$, 
\bea
\sum _{\vP, \vQ} \cC^{\vP \shuffle \cS(\vQ)} \, B_\vP [ X, B_{\vQ} Y] 
= \sum_{\vR, \vT}  \sum_{\vT=\vS \vQ} \cC^{\vR \shuffle \vS \shuffle \cS (\vQ)} \,  [ B_{\vR} X, B_\vT Y] 
\eea
In view of Lemma \ref{lem:sa}, the sum of $\cC^{\vR \shuffle \vS \shuffle \cS (\vQ)}$ over $\vS$ and $\vQ$ gives zero when $\vT \not= \emptyset$ and $\cC^\vR$ when $\vT=\emptyset$, which readily proves the alternative Leibniz formula. \end{proof}

\newpage

\section{Proof of Proposition \ref{prop:dih}}
\label{sec:F}

In this appendix, we shall prove Proposition \ref{prop:dih} which states that the combination, 
\bea
\mL^\vI(x,y) = \hf^{\vec{I} + \cS(\vec{I})}(x) - \p_x \cG ^{\vec{I} + \cS(\vec{I})}(x,y)
- \sum_{\vec{I}= \vec{X} \vec{Y} \vec{Z}}
f^{ (\vec{X} \shuffle \cS(\vec{Z})  ) M ( \vec{Y} + \cS( \vec{Y})  )  }{}_M(x,y)
 \label{flt.97}
\eea
vanishes for all $\vI\neq \emptyset$ provided that $\hf^\vI(x)$ admits the decomposition (\ref{2.h.3}) whose coefficients satisfy the dihedral symmetry relations of (\ref{2.h.4}).  Our proof of Proposition \ref{prop:dih} will proceed by decomposing the DHS kernels on the right side of (\ref{flt.97}) into their traces and traceless parts and separately showing agreement with the left side.

\subsection{Establishing  an auxiliary result}

Starting with equation (\ref{2.h.3}) for $\hf^\vI(x)$, the antipode $\hf^{\cS(\vI)}(x)$ evaluates as follows for $\vI \neq \emptyset$, 
\bea
\label{F.1}
\gamma ^{\cS(\vI)}  (x) & = & 
 \sum_{\vI = \vX \vY \vZ} \Big ( \p_x \Phi ^{\vX \shuffle \cS(\vZ)}{}_M(x) \, \hat \mN^{M \cS(\vY)} 
+ \p_x \Phi ^{(\vX \shuffle \cS(\vZ)) M \cS(\vY) }{}_M(x) \Big ) 
\qquad
\eea
where we set $\p_x \Phi ^\emptyset {}_M(x) = \om_M(x)$. We note that contributions 
from $\vY= \emptyset$ to both terms in the summand
vanish in view of $\sum_{\vI = \vX \vZ} \p_x \Phi ^{(\vX \shuffle \cS(\vZ) )\vJ}{}_M(x) = 0$
for arbitrary $\vI,\vJ \in {\cal W}_h$ with $\vI \neq \emptyset$, see (\ref{std.sa}).
To obtain (\ref{F.1}),  we have used the freedom to change summation variables $\vX \to \cS(\vZ), \vY \to \cS(\vY)$ and $\vZ \to \cS(\vX)$ in the deconcatenation of $\cS(\vI) = \vX \vY \vZ$ for $\gamma ^{\cS(\vI)}  (x) $. As a result, we have the following formula for the symmetrized combination,
\bea
\label{F.2}
\gamma ^{\vI + \cS(\vI)}  (x) & = &  
\sum_{\vI = \vX \vY \vZ}  \p_x \Phi ^{\vX \shuffle \cS(\vZ)}{}_M(x) \, ( \hat \mN^{M \vY} + \hat \mN^{M \cS(\vY)} )
\no \\ &&
+ \sum_{\vI = \vX \vY \vZ;  \vY \not= \emptyset}  \p_x \Phi ^{(\vX \shuffle \cS(\vZ)) M  ( \vY + \cS(\vY)  ) }{}_M(x) 
\eea
where we have freely  inserted the superfluous restriction $\vY \not= \emptyset$ for later convenience. 
Combining the relations for dihedral symmetry of $\hat \mN$, we have, 
\bea
\label{F.3}
\hat \mN^{M \vY} +  \hat \mN^{M \cS(\vY)}  =0
\eea
so that all the dependence on $\hat \mN$ in (\ref{F.2}) cancels and the formula  reduces to,
\bea
\label{F.4}
\gamma ^{\vI + \cS(\vI)}  (x)
=  \sum_{\vI = \vX \vY \vZ;  \vY \not= \emptyset}  \p_x \Phi ^{(\vX \shuffle \cS(\vZ)) M  ( \vY + \cS(\vY)  ) }{}_M(x) 
\eea

\subsection{Completing the proof}

Eliminating $\hf^{\vI + \cS(\vI)}(x)$ from $\mL^\vI(x,y)$ in (\ref{flt.97})  using (\ref{F.4}) gives,
\bea
\label{F.5}
\mL^\vI(x,y) & = &  
 \sum_{\vec{I}= \vec{X} \vec{Y} \vec{Z};   \vec{Y}\neq \emptyset }
 \big( \partial_x \Phi^{ (\vec{X} \shuffle \cS(\vec{Z})  ) M ( \vec{Y} + \cS( \vec{Y})  )  }{}_M(x) 
 - f^{ (\vec{X} \shuffle \cS(\vec{Z})  ) M ( \vec{Y} + \cS( \vec{Y})  )  }{}_M(x,y)   \big)  
 \no \\ && - \p_x \cG ^{\vec{I} + \cS(\vec{I})}(x,y)
 \no \\
& = & - \p_x \cG ^{\vec{I} + \cS(\vec{I})}(x,y)
- \sum_{\vec{I}= \vec{X} \vec{Y} \vec{Z}; \vec{Y}\neq \emptyset }
 f^{ (\vec{X} \shuffle \cS(\vec{Z})  ) M ( \vec{Y} + \cS( \vec{Y})  )  }{}_M(x,y) \Big | _{\p \cG}
\eea
where $|_{\p \cG}$ instructs us to retain only the trace part of $f$, since the traceless part cancelled against the $\p \Phi$ contributions in passing to the last line of (\ref{F.5}). Since $\vY \not= \emptyset$ and the contribution from $\vY$ being a single letter manifestly cancels, we see that $\vY$ must be a word of at least two letters, say $L$ and $N$,  and may be parametrized by $\vY = L \vW N$ for arbitrary word $\vW$, so that,  
\bea
\label{F.6a}
\mL^\vI(x,y) & = & - \p_x \cG ^{\vec{I} + \cS(\vec{I})}(x,y)
- \sum_{\vec{I}= \vec{X} L \vec{W} N \vec{Z}}
 f^{ (\vec{X} \shuffle \cS(\vec{Z})  ) M ( L \vW N  + N \cS( \vW) L  )  }{}_M(x,y) \Big | _{\p \cG}
\eea
Retaining only the trace part, and suppressing the arguments $(x,y)$ for brevity, gives,
\bea
\label{F.6b}
\mL^\vI & = & - \p_x \cG ^{\vec{I} + \cS(\vec{I})}
+ \sum_{\vec{I}= \vec{X} L \vec{W} N \vec{Z}}
 \Big ( \p_x \cG ^{ (\vec{X} \shuffle \cS(\vec{Z})  ) N  L \vW }
 + \p_x \cG^{ (\vec{X} \shuffle \cS(\vec{Z})  ) L N \cS( \vW)  }
 \Big ) 
\eea
The first term in the summand effectively gives a sum over an arbitrary word $\vY = L \vW \not= \emptyset$, while the second term gives a sum over $\vY = \vW N \not= \emptyset$, 
\bea
\label{F.6c}
\mL^\vI & = & - \p_x \cG ^{\vec{I} + \cS(\vec{I})}
+ \sum_{\vec{I}= \vec{X} \vY N \vec{Z}} \p_x \cG ^{ (\vec{X} \shuffle \cS(\vec{Z})  ) N  \vY }
- \sum_{\vec{I}= \vec{X} L \vY  \vec{Z}} \p_x \cG^{ (\vec{X} \shuffle \cS(\vec{Z})  ) L  \cS( \vY)  }
\eea
Here, we have lifted the restriction $\vY \not = \emptyset$ in both sums since the $\vY = \emptyset$ contribution manifestly cancels between the sums. 

\sm

Let us focus on the first sum in (\ref{F.6c}) and separate the deconcatenation for  $\vX = \emptyset$ from those with $\vX \not= \emptyset$ which we may parametrize by  $\vX  \rightarrow \vX M$,
\begin{align}
 \sum_{\vec{I}= \vec{X} \vec{Y} N \vec{Z}  }
\p_x \cG^{ (\vec{X} \shuffle \cS(\vec{Z})  ) N  \vec{Y}   }  
&=  \sum_{\vec{I}= \vec{X} M \vec{Y} N \vec{Z}  }
\p_x \cG ^{ (\vec{X} M\shuffle \cS(\vec{Z})  ) N  \vec{Y}   }   + 
 \sum_{\vec{I}=  \vec{Y} N \vec{Z}  }
\p_x \cG ^{  \cS(\vec{Z})  N  \vec{Y}   }     \label{flt.102} \\
&=  \sum_{\vec{I}= \vec{X} M \vec{Y} N \vec{Z}  }
\p_x \cG ^{ (\vec{X} M\shuffle \cS(\vec{Z}) N )   \vec{Y}  -  (\vec{X} \shuffle \cS(\vec{Z}) N )  M \vec{Y}   }   + 
 \sum_{\vec{I}=  \vec{Y} N \vec{Z} }
\p_x \cG ^{  \cS(\vec{Z})  N  \vec{Y}   }  \notag \\
&= - \sum_{\vec{I}= \vec{X}  \vec{Y} \vec{Z} \atop{ \vec{X}, \vZ\neq \emptyset }}
\p_x \cG ^{ (\vec{X} \shuffle \cS(\vec{Z})  )   \vec{Y}   }
+ \sum_{\vec{I}= \vec{X}  \vec{Y}  \vec{Z} \atop{ \vec{Y} ,\vZ\neq \emptyset }}
\p_x \cG ^{   (\vec{X} \shuffle \cS(\vec{Z})  )  \vec{Y}   }
-  \sum_{\vec{I}=  \vec{Y}  \vec{Z} \atop{ \vec{Z}\neq \emptyset }}
\p_x \cG^{  \cS(\vec{Z})    \vec{Y}   }  
\notag
 \end{align}
 We have used $ (\vec{X} M\shuffle \cS(\vec{Z})  ) N
 = \vec{X} M\shuffle \cS(\vec{Z}) N    -  (\vec{X} \shuffle \cS(\vec{Z}) N )  M$ in 
 passing to the second line and in the last step absorbed letters of the combinations
 $\vX M$, $M \vY$ and $\cS(\vZ)N$ into redefinitions of the accompanying words.
 
 \sm
 
 The deconcatenations into three words $\vX,\vY,\vZ$ in the last line of (\ref{flt.102})
 only differ on the restrictions on the non-empty words and otherwise cancel. 
The non-cancelling part can be isolated by restricting the first and second term
of the following difference to empty $\vY$ and empty $\vX$, respectively
 \begin{align}
 {-} \sum_{\vec{I}= \vec{X}  \vec{Y} \vec{Z} \atop{ \vec{X}, \vZ\neq \emptyset }}
\p_x \cG ^{ (\vec{X} \shuffle \cS(\vec{Z})  )   \vec{Y}   }
+ \sum_{\vec{I}= \vec{X}  \vec{Y}  \vec{Z} \atop{ \vec{Y} ,\vZ\neq \emptyset }}
\p_x \cG ^{   (\vec{X} \shuffle \cS(\vec{Z})  )  \vec{Y}   }
&=  - \sum_{\vec{I}= \vec{X}   \vec{Z} \atop{ \vec{X}, \vZ\neq \emptyset }}
\p _x \cG ^{ \vec{X} \shuffle \cS(\vec{Z})    }
+ \sum_{\vec{I}=   \vec{Y}  \vec{Z} \atop{ \vec{Y} ,\vZ\neq \emptyset }}
\p_x \cG^{   \cS(\vec{Z})   \vec{Y}   } \notag\\
&= \p_x \cG^ {\vI} + \p_x \cG ^{\cS(\vI)} + \sum_{\vec{I}=   \vec{Y}  \vec{Z} \atop{ \vec{Y} ,\vZ\neq \emptyset }}
\p_x \cG ^{   \cS(\vec{Z})   \vec{Y}   }
 \label{flt.103} 
 \end{align}
 We have used that the full deconcatenation sum $ \sum_{\vec{I}= \vec{X}   \vec{Z} }
\vec{X} \shuffle \cS(\vec{Z})  $ vanishes for $\vI \not= \emptyset$, and therefore
reinstated the contributions $ \p_x \cG ^{\vI}$ and $\p_x \cG ^{\cS(\vI)}$ from empty
 $\vZ$ and $\vX$. Inserting (\ref{flt.103}) into (\ref{flt.102}) results in
 \begin{align}
 \sum_{\vec{I}= \vec{X} \vec{Y} N \vec{Z}  }
\p_x \cG^{ (\vec{X} \shuffle \cS(\vec{Z})  ) N  \vec{Y}   }  
&=  \p_x \cG^{\vI} + \p_x \cG^{\cS(\vI)}
+ \sum_{\vec{I}=   \vec{Y}  \vec{Z} \atop{ \vec{Y} ,\vZ\neq \emptyset }}
\p_x \cG^{   \cS(\vec{Z})   \vec{Y}   }
-  \sum_{\vec{I}=  \vec{Y}  \vec{Z} \atop{ \vec{Z}\neq \emptyset }}
\p_x \cG^{  \cS(\vec{Z})    \vec{Y}   }   = \p_x \cG^{\vI} 
\label{flt.104}
\end{align}
By the logic of the previous steps, the difference of the
 two deconcatenation sums $\sum_{\vec{I}=  \vec{Y}  \vec{Z}}$
 comes from the empty word $\vY = \emptyset$ which is only
 allowed in the last term and therefore cancels $ \p_x \cG^{\cS(\vI)}$.
 
 \sm
 
In summary, the manipulations in (\ref{flt.102}) to (\ref{flt.104}) identify
the first sum $  \sum_{\vec{I}= \vec{X} \vec{Y} N \vec{Z}  }$
$ \p_x \cG^{ (\vec{X} \shuffle \cS(\vec{Z})  ) N \vec{Y}   }  $ 
in (\ref{F.6c}) as $  \p_x \cG^{\vI} $.
Repeating the analysis for the last term in (\ref{F.6c}),
\begin{align}
\sum_{\vec{I}= \vec{X}  L\vec{Y} \vec{Z} } 
\p_x \cG^{ (\vec{X} \shuffle \cS(\vec{Z})  ) L   \cS( \vec{Y})  } &=
 \sum_{\vec{I}= \vec{X}  \vec{Y} \vec{Z} \atop{ \vec{X}, \vZ\neq \emptyset }}
\p_x \cG^{ (\vec{X} \shuffle \cS(\vec{Z})  )   \cS(\vec{Y} )  }
- \sum_{\vec{I}= \vec{X}  \vec{Y}  \vec{Z} \atop{ \vX, \vec{Y}  \neq \emptyset }}
\p_x \cG^{   (\vec{X} \shuffle \cS(\vec{Z})  )  \cS(\vec{Y})   }
+  \sum_{\vec{I}=  \vec{X}  \vec{Y} \atop{ \vec{X}\neq \emptyset }}
\p_x \cG^{ \vX  \cS(\vec{Y})    }  
\notag\\
&= {-} \p_x \cG^{\vI} - \p_x \cG^{\cS(\vI)} + \p_x \cG^{\vI} = - \p_x \cG^{\cS(\vI)}
\label{flt.105}
\end{align}
and combining with (\ref{flt.104}) and (\ref{F.6c}) yields the desired identity 
$\mathfrak{L}^{\vI}(x,y)=0$ which concludes our proof of
Proposition \ref{prop:dih}.

\newpage

\section{Proof of Proposition \ref{3.lem:2} for $\cC^{(1)}_{ij}$}
\label{sec:C}

The starting point is the expansion of $\cC^{(1)}_{ij}$ given in (\ref{3.c.1}). To simplify the commutators, we shall make use of the identity given in the following lemma.
{\lem
\label{distijk}
For an arbitrary triplet $i,j,k$ of pairwise distinct points, arbitrary word $\vJ \in \cW_h$ and index $M=1,\cdots,h$, the following identity holds, 
\bea
\big[ a_i^M , B_{j \vJ} \, t_{jk} \big] 
= \sum_{\vec{J} = \vec{P} L \vec{Q}} \delta^M_L \, B_{k \cS(\vQ)} \, [ t_{ik} , B_{j \vP} \, t_{jk} ]
\label{capp.01}
\eea
where the sum is over all $s$ de-concatenations of the word $\vJ = J_1 \cdots J_s$ into three subwords words given by 
$\vP = J_1\cdots J_{i-1}$, a single letter $L$,  and $\vQ = J_{i+1} \cdots J_s$ for $i=1, \cdots, s$. } 

\subsection{Proof of Lemma \ref{distijk}}

The relation (\ref{capp.01}) is proven by induction on the length $s$ of the word $\vJ=J_1\cdots J_s$. For  $s=0$ we have $\vJ = \emptyset$ and the left side of (\ref{capp.01}) vanishes in view of the relations on the last line of (\ref{11.1}) while its right side vanishes since no such de-concatenations of $\vJ=\emptyset$ exist. For the inductive step, we shall assume that (\ref{capp.01}) holds for all words $\vJ$ of length less or equal to $s\geq 0$. It then remains to prove the validity of (\ref{capp.01}) for all  words of length $s{+}1$. 

\sm

To do so, consider the left side of (\ref{capp.01}) for an arbitrary word $R \vJ$ of length $s{+}1$, written as a concatenation of an arbitrary single letter word $R$ and an arbitrary word $\vJ$ of length $s$. Using the relation $B_{j R \vJ} = B_{jR} B_{j \vJ}$ and the Leibniz rule for the action of $B_{jR}$,
\bea
\big[ a_i^M , B_{j R \vJ} \, t_{jk} \big] 
= B_{jR} \big[ a_i^M , B_{j \vJ} \, t_{jk} \big] - \big[  B_{jR} \, a_i^M , B_{j \vJ} \, t_{jk} \big] 
\label{capp.62} 
\eea
The first term may be expressed using the induction hypothesis, while the second may be simplified by using the relation $B_{jR} \, a_i^M = \delta ^M_R \, t_{ij}$ on the second line of (\ref{11.1}), so that we find, 
\bea
\label{capp.625} 
\big[ a_i^M , B_{j R \vJ} \, t_{jk} \big] 
= B_{jR}  \sum_{\vec{J} = \vec{P} L \vec{Q}} \delta^M_L \, B_{k \cS(\vQ)} [  t_{ik} , B_{j \vP} \, t_{jk} ]
- \delta^M_R \,  [t_{ij} , B_{j \vJ} \, t_{jk}]
\eea
The last term can be rewritten via the following rearrangement,
\begin{align}
 [t_{ij} , B_{j \vJ} t_{jk}] &=  [t_{ij} , B_{k \cS( \vJ)} t_{j k}] =  B_{k \cS( \vJ)}\, [t_{ij} ,  t_{jk}]
 = -  B_{k \cS( \vJ)}\, [t_{ik} ,  t_{jk}]
 \label{capp.63}
\end{align}
so that, using $B_{jR} t_{ik}=0$, equation (\ref{capp.625}) becomes, 
\bea
\label{capp.626} 
\big[ a_i^M , B_{j R \vJ} \, t_{jk} \big] 
=  \sum_{\vec{J} = \vec{P} L \vec{Q}} \delta^M_L \, B_{k \cS(\vQ)} [  t_{ik} , B_{j R \vP} \, t_{jk} ]
+ \delta^M_R \, B_{k \cS( \vJ)}\, [t_{ik} ,  t_{jk}]
\eea
Recasting the result in terms of de-concatenations of the word $R \vJ$ of length $s{+}1$,
\bea
\label{capp.627} 
\big[ a_i^M , B_{j R \vJ} \, t_{jk} \big] 
=  \sum_{R \vec{J} = \vec{P} L \vec{Q}} \delta^M_L \, B_{k \cS(\vQ)} [  t_{ik} , B_{j R \vP} \, t_{jk} ]
\eea
the last term in (\ref{capp.626}) corresponds to the de-concatenation in (\ref{capp.627}) for which $\vP=\emptyset$.
This completes the inductive step and establishes the identity (\ref{capp.01}).

\subsection{Proving  Proposition \ref{3.lem:2} using Lemma \ref{distijk}}
\label{sec:B.1}

To prove  Proposition \ref{3.lem:2} with the help of Lemma \ref{distijk}, we use the identity $\big[ B_{i \vI} \, a_i^M , B_{j \vJ} \, t_{jk} \big]  = B_{i \vI} \,  \big[ a_i^M , B_{j \vJ} \, t_{jk} \big] $ in (\ref{3.c.1}) and then substitute the expression of (\ref{capp.01}) for $\big[ a_i^M , B_{j \vJ} \, t_{jk} \big] $,  
\begin{align}
\cC_{ij} ^{(1)} = \sum_{k \neq i,j} \sum_{\vI,\vP,\vQ}  
f^\vI{}_M(i,j) \, \p_j \cG ^{\vP M \cS(\vQ)}(j ; k, i)  B_{k \vQ} \, \big[ B_{i \vI} \, t_{ik} , B_{j \vP} \, t_{jk} \big] 
- ( i \leftrightarrow j)
\end{align}
Applying the generalized Leibniz rule of (\ref{B.a.9}) to move $B_{k \vQ}$ inside the commutator we get, 
\begin{align}
\cC_{ij} ^{(1)} = \sum_{k \neq i,j} \sum_{\vI,\vP,\vX,\vY}  
f^\vI{}_M(i,j) \, \p_j \cG ^{\vP M \cS(\vX \shuffle \vY)}(j ; k, i)  \big[ B_{k \vX} B_{i \vI} \, t_{ik} , B_{k \vY} B_{j \vP} \, t_{jk} \big] 
- ( i \leftrightarrow j)
\end{align}
Commuting $B_k$ past $B_i$ and $B_j$, using the antipode identities $B_{k \vX} \, t_{ik} = B_{i \, \cS (\vX)} \, t_{ik}$ and $B_{k \vY} \, t_{jk} = B_{j \, \cS (\vY)} \, t_{jk}$ and changing summation variables $\vX,\vY$ to $\cS(\vX),\cS(\vY)$, we obtain,
\bea
\cC_{ij} ^{(1)} = 
\sum_{k \neq i,j} \sum_{\vI,\vP, \vX,\vY}  
f^\vI{}_M(i,j) \, \p_j \cG^{\vP M  ( \vX \shuffle \vY)}(j; k, i)    \, \big[ B_{i \vI \vX} \, t_{ik} , B_{j \vP \vY} \, t_{jk} \big] 
- ( i \leftrightarrow j)
\eea
Introducing auxiliary sums over $\vR= \vI \vX$ and $\vJ= \vP \vY$, we rearrange the sums as follows,
\bea
\cC_{ij} ^{(1)} =  \sum_{k \neq i,j} \, \sum_{\vR , \vJ}  \, 
\sum_{\vR = \vI \vX} \,  \sum_{\vJ = \vP \vY} f^{\vI}{}_M(i,j) \, 
\p_j \cG ^{\vP M  ( \vX \shuffle \vY)}(j; k, i) 
 \big[ B_{i \vR} \, t_{ik} , B_{j \vJ} \, t_{jk} \big] 
- ( i \leftrightarrow j)
\qquad
\eea
The next step is to use the following shuffle  identity,
\bea
\sum _{\vJ = \vP \vY} \vP M ( \vX \shuffle \vY) = \vJ \shuffle M \vX
\eea
which is the second identity in (\ref{spc27}) with $L$, $\vS$, $\vI$ renamed
to $M$, $\vX$, $\vJ$ to obtain the following  expression, 
\begin{align}
\cC_{ij} ^{(1)} =   \sum_{k \neq i,j} \sum_{\vR,\vJ}  
\sum_{\vR = \vI \vX} f^\vI{}_M(i,j) \, \p_j \cG ^{\vJ \shuffle M  \vX }(j; k, i)  
 \big[ B_{i \vR} \, t_{ik} , B_{j \vJ} \, t_{jk} \big] 
 - ( i \leftrightarrow j)
\label{simc.03}
\end{align}
A final change of variables $\vI \to \vP$ followed by $\vR \to \vI$ and $\vX \to \vQ$, as well as the corresponding change of variables for the term in which $i$ and $j$ have been swapped, gives the expression for $\cC^{(1)}_{ij}$ in (\ref{3.c.1}) and thereby completes the proof of Proposition \ref{3.lem:2}.

\newpage

\section{Proof of Proposition \ref{3.lem:3} for $\cC^{(2)}_{ij}$}
\label{sec:D}
 
The starting point for deriving the expression (\ref{3.d.2}) of Proposition \ref{3.lem:3} is the series expansion given for $\cC^{(2)}_{ij}$ in  (\ref{3.d.1}) and repeated here for convenience, 
 \bea
 \label{D.a.1}
\cC_{ij} ^{(2)}  & = & 
\sum_{\vI, \vJ} f^\vI{}_M (i,j) f^\vJ {}_N (j,i)  \big [ B_{i\vI} \, a^M_i, B_{j\vJ} \, a_j^N \big ]
\eea
Our strategy will be to reduce the expression to a simpler form by converting all its contributions to having only one or zero exposed $a$ factors, and to arrange these simplified contributions in a \textit{standard form} of linearly independent Lie algebra elements.

\subsection{Preparatory material}

We begin by establishing Proposition \ref{bigprop} below which will be instrumental in expressing $\cC^{(2)}_{ij}$ in a convenient basis of Lie algebra elements. Its proof will require the use of Lemmas \ref{lem.ijij1}  and \ref{funstep} which will be provided later in this subsection.

{\prop
\label{bigprop}
For arbitrary $i \not = j \in \{ 1, \cdots, n\}$,  $M,N \in \{ 1,\cdots,h\} $ and $\vI,\vJ \in \cW_h$, the 
commutator in (\ref{D.a.1}) may be evaluated as follows, 
\begin{align}
\big[ B_{i \vI} \, a_i^M , B_{j \vJ} \, a_j^N \big] = \mR^{MN}_{i\vI | j \vJ} + B_{i \vI} \, \mS^{MN} _{i|j \vJ}
\label{capp.04}
\end{align}
in terms of the following contributions,
\begin{align}
 \mR^{MN}_{i\vI | j \vJ} &= - \sum_{\vJ = \vP L \vQ} \delta^M_L B_{ i \vI} \, 
 \big[ B_{i \vQ}  a_i^N , B_{i \cS(\vP)} t_{ij}\big] 
  - \sum_{\vI = \vP L \vQ} \delta^N_L B_{ i \vP} \, 
 \big[ B_{i \vQ}  a_i^M , B_{i \cS(\vJ)} t_{ij}\big]
 \label{capp.05} \\
\mS^{MN} _{i|j \vJ}&= \sum_{\vJ = \vP K \vQ L \vR} B_{j \vP} \, \bigg\{
 {-} \delta^N_K \delta^M_L \,  \big[ B_{i \vR}   t_{ij} , B_{i \cS(\vQ)} t_{ij}\big]
\label{capp.06} \\
&\hskip 1.2in 
+ \delta^M_K \delta^N_L \delta_{\vR \not= \emptyset}   \sum_{\vX,\vY} \delta_{\vQ, \vX \shuffle \vY}
 \big[ t_{ij} ,  B_{i \vX}  \big( B_{i \vR}+B_{i \cS(\vR)} \big) B_{ i \cS(\vY) } t_{ij}\big]
\bigg\} \notag
\end{align}  
where $\delta_{\vR \not= \emptyset} = 1- \delta_{\vR, \emptyset}$ and $\delta_{\vR, \emptyset}$ was defined in (\ref{std.sa}). Furthermore,  $\sum_{\vJ = \vP L \vQ}$ instructs us to sum over all possible deconcatenations of the word $\vJ$ into three words:  $\vP$, the one-letter word $L$,  and $\vQ$, and similarly for the deconcatenation sum  $\sum_{\vJ = \vP K \vQ L \vR}$ into five words. Finally, we note that the Lie algebra element $\mS^{MN} _{i|j \vJ}$ is independent of the word $\vI$.} 

\begin{proof}
\vskip -0.1in
We prove the proposition by induction on the length $r$ of the word $\vI$. The base case $r=0$ itself is a non-trivial identity that will be proven in Lemma~\ref{funstep} with the help of Lemma~\ref{lem.ijij1} which, in turn,  will both be proven later in this subsection. Assuming the validity of the base case $r=0$, we now proceed with the inductive step and assume that the decomposition (\ref{capp.04}) in terms of (\ref{capp.05}) and (\ref{capp.06}) holds for all words $\vI$ of length $r$.  To prove that these relations then hold for length $r+1$, we apply $B_{iT}$ to both sides of (\ref{capp.04}), where $T$ is a single letter, and use Leibniz's rule to evaluate the left side, 
\bea
\label{D.a.2}
\big[ B_{i T \vI} \, a_i^M , B_{j \vJ} \,  a_j^N \big] 
+ \big[ B_{i  \vI} \, a_i^M , B_{j \vJ} \,  t_{ij} \big] \delta ^N_T 
= B_{i T} \, \mR^{MN}_{i\vI | j \vJ} + B_{i T \vI} \, \mS^{MN} _{i|j \vJ}
\eea
Evaluating $B_{i T} \, \mR^{MN}_{i\vI | j \vJ} $ gives,
\bea
\label{D.a.3}
B_{i T} \, \mR^{MN}_{i\vI | j \vJ} = 
- \sum_{\vJ = \vP L \vQ} \delta^M_L B_{ i T \vI} \, 
 \big[ B_{i \vQ}  a_i^N , B_{i \cS(\vP)} t_{ij}\big] 
  - \sum_{\vI = \vP L \vQ} \delta^N_L B_{ i T \vP} \, 
 \big[ B_{i \vQ}  a_i^M , B_{i \cS(\vJ)} t_{ij}\big]
 \eea
Moving the term $\big[ B_{i  \vI} \, a_i^M , B_{j \vJ} \,  t_{ij} \big] \delta ^N_T$ on the left side of (\ref{D.a.2}) to the right side, and combining it with the second sum in (\ref{D.a.3}) completes the sum into the deconcatenation of an arbitrary word of length $r+1$ parametrized by $ T \vI$. This completes the proof of the inductive step as well as of the proposition (recall, however, that we borrowed the validity of the base case $r=0$ from Lemma~\ref{funstep} which remains to be proven below). 
\end{proof}

\subsubsection{Auxiliary lemmas}

The auxiliary Lemmas~\ref{lem.ijij1} and \ref{funstep} in this subsection will provide the proof of the base case $r=0$ for the proof by induction of Proposition \ref{bigprop}. 

{\lem
\label{lem.ijij1}
For arbitrary $i \not= j \in \{ 1, \cdots, n\}$,  $N \in \{ 1,\cdots,h\}$ and $\vJ \in \cW_h$  we have, 
\bea
\big[ t_{ij} , B_{j\vec{J}} \, a_j^N \big] = \big[ B_{i\vec{J}} \, a_i^N , t_{ij} \big] 
+ \sum_{\vJ = \vP L \vQ \atop{\vQ \neq \emptyset}} \delta^N_L \sum_{\vX,\vY} \delta_{\vP,\vX\shuffle \vY}
\big[ B_{i \vX} \big( B_{i\vQ} {+} B_{i \cS(\vQ)} \big) B_{i \cS(\vY)}  t_{ij}, t_{ij} \big]\ \
 \label{flt.83} 
 \eea
where $\delta _{\vI, \vP \shuffle \vQ}$ was defined in Proposition \ref{B.prop:1}.} \\
 
\begin{proof} 
\vskip -0.2in
This lemma is proven by induction on the length $s$ of the word $\vJ = J_1\cdots J_s$.
For the base case of $s=0$, the empty word $\vJ$ does not admit any deconcatenations into
$ \vP L \vQ$ including the letter $L$. Accordingly, (\ref{flt.83}) reduces to $[t_{ij},a_j^N] = [a_i^N, t_{ij}]$
at $s=0$ which holds by the structure relation of $\mt_{h,n}$ in the first line of (\ref{2.prop.1}).

\sm

For the inductive step, we assume (\ref{flt.83}) to hold for all words $\vJ$ of length $s$ and deduce its validity for an arbitrary  word $\vJ \rightarrow T \vJ$ of length $s{+}1$. Applying the combination $(B_{iT}{+}B_{jT})$ to both side of  (\ref{flt.83}), using the Leibniz rule and the structure relations $(B_{iT}{+}B_{jT}) t_{ij} = 0$, $[ B_{iT} , B_{j\vec{J}} ] = 0$ and $B_{iT} \, a_j^N= \delta^N_T t_{ij}$ of (\ref{11.1}), we obtain, 
\begin{align}
\big[ t_{ij} , B_{j T \vec{J}} \,  a_j^N \big] &
= - \delta^N_T \, \big[   t_{ij}, B_{j\vec{J}} \, t_{ij} \big]  
+ (B_{iT}{+} B_{jT})  \big[ B_{i\vec{J}}  \, a_i^N , t_{ij} \big] 
  \label{capp.67}  \\
&\quad + (B_{iT}{+} B_{jT}) 
 \sum_{\vJ = \vP L \vQ \atop{\vQ \neq \emptyset}} \delta^N_L \sum_{\vX,\vY} \delta_{\vP,\vX\shuffle \vY}
\big[ B_{i \vX} \big( B_{i\vQ} {+} B_{i \cS(\vQ)} \big) B_{i \cS(\vY)}  t_{ij}, t_{ij} \big]
\notag
\end{align}
Together with the relation $B_{j\vec{J}} \, t_{ij}  = B_{i\cS(\vec{J} )} \, t_{ij} $, and the structure relations mentioned earlier, we may  cast the right side of the  first line of (\ref{capp.67}) into the following form, 
\bea
\delta^N_T \, \big[  B_{j\vec{J}} \, t_{ij} , t_{ij}\big]  
{+}  \big[  (B_{iT}{+} B_{jT})  B_{i\vec{J}} \, a_i^N , t_{ij} \big] 
= \delta^N_T \, \big[  \big( B_{i\vec{J}} {+} B_{i \cS(\vec{J} )}  \big)  t_{ij} ,     t_{ij}\big] 
{+} \big[ B_{i T \vec{J}} \, a_i^N , t_{ij} \big] \ \
  \label{capp.68} 
\eea
Similarly, moving $(B_{iT}{+} B_{jT})$ inside the commutator using $(B_{iT}{+}B_{jT}) t_{ij} = 0$, the second line of (\ref{capp.67}) in turn becomes,
\bea
 \label{capp.69} 
&&
\sum_{\vJ = \vP L \vQ \atop{\vQ \neq \emptyset}} \delta^N_L \sum_{\vX,\vY} \delta_{\vP,\vX\shuffle \vY}
\Big( \big[ B_{i T \vX} \big( B_{i\vQ} {+} B_{i \cS(\vQ)} \big) B_{i \cS(\vY)}  t_{ij}, t_{ij} \big] 
+ \big[ B_{i \vX} \big( B_{i\vQ} {+} B_{i \cS(\vQ)} \big) B_{i \cS(T \vY)}  t_{ij}, t_{ij} \big]  \Big) 
\no \\ && \qquad =
  \sum_{\vJ = \vP L \vQ \atop{\vQ \neq \emptyset}} \delta^N_L \sum_{\vX,\vY} \delta_{T \vP,\vX\shuffle \vY}
 \big[ B_{i \vX} \big( B_{i\vQ} {+} B_{i \cS(\vQ)} \big) B_{i \cS(\vY)}  t_{ij}, t_{ij} \big] 
\no \\ && \qquad =
 \sum_{T \vJ = \vP L \vQ \atop{\vQ \neq \emptyset}} \delta^N_L \sum_{\vX,\vY} \delta_{\vP,\vX\shuffle \vY}
 \big[ B_{i \vX} \big( B_{i\vQ} {+} B_{i \cS(\vQ)} \big) B_{i \cS(\vY)}  t_{ij}, t_{ij} \big]  
\no \\ && \qquad \qquad
 - \delta^N_T  \sum_{\vX,\vY} \delta_{\emptyset,\vX\shuffle \vY}
  \big[ B_{i \vX} \big( B_{i\vJ} {+} B_{i \cS(\vJ)} \big) B_{i \cS(\vY)}  t_{ij}, t_{ij} \big] 
\eea
The last step makes use of  the following deconcatenation sum for the word $T \vJ$ and an arbitrary function $\f$ on $\cW_h \times \cW_h$,\footnote{Note that the only dependence on the letter $L$ in the summand on the left side is through $\delta^M_L$, the function $\f$ being independent of $L$.}
\bea
  \sum_{T \vec{J} = \vec{P} L \vec{Q}} \delta^M_L \, \varphi(\vP,\vQ) =  \delta^M_T \, \varphi(\emptyset,\vJ)
  + \sum_{\vec{J} = \vec{P} L \vec{Q}} \delta^M_L \, \varphi(T \vP,\vQ)
\label{decRJ}
\eea
In assembling the right side of (\ref{capp.67}) from (\ref{capp.68}) and (\ref{capp.69}), the terms
$ \delta^N_T[  ( B_{i\vec{J}} {+} B_{i \cS(\vec{J} )})  t_{ij} ,     t_{ij}] $ cancel one another, and we are left with,
\begin{align}
\big[ t_{ij} , B_{j T \vec{J}} \, a_j^N \big] &=  \big[ B_{i T \vec{J}} \,  a_i^N , t_{ij} \big]
+ \! \! \!   \sum_{T \vJ = \vP L \vQ \atop{\vQ \neq \emptyset}} \! \! 
\delta^N_L \sum_{\vX,\vY} \delta_{\vP,\vX\shuffle \vY}
 \big[ B_{i \vX} \big( B_{i\vQ} {+} B_{i \cS(\vQ)} \big) B_{i \cS(\vY)}   t_{ij}, t_{ij} \big]   \
  \label{capp.70} 
\end{align}
which is the statement (\ref{flt.83}) of the lemma for an arbitrary word $T \vJ$ of length $s+1$, thereby 
completing the inductive step and the proof of the lemma.
\end{proof}

The lemma below establishes the $r=0$ base case for Proposition \ref{bigprop}. 

{\lem
\label{funstep}
For arbitrary $i \not = j \in \{1, \cdots, n \}$,   $\vJ \in \cW_h$ and $M,N \in \{1,\cdots,h\}$, the structure relations of $\mt_{h,n}$ imply the $r=0$ case of Proposition \ref{bigprop}, 
\begin{align}
\label{funstep.1}
\big[ a_i^M , B_{j \vJ} \, a_j^N \big] 
=   - \sum_{\vJ = \vP L \vQ} \delta^M_L \,  \big[ B_{i \vQ}  a_i^N , B_{j \vP} t_{ij}\big]  + \mS^{MN}_{i | j \vJ}
\end{align}
where $\mS^{MN}_{i | j \vJ}$ was defined in (\ref{capp.06}) of Proposition \ref{bigprop}.} \\

\begin{proof}
\vskip -0.2in
The lemma is proven by induction on the length $s$ of the word $\vJ = J_1\cdots J_s$. The base case at $s=0$ follows from the structure relation $[a_i^M, a_j^N] = 0$ of (\ref{11.1})  and the fact that the summation ranges on the right side of (\ref{capp.06}) and (\ref{funstep.1}) are empty for $\vJ = \emptyset$. For the inductive step, we assume (\ref{funstep.1}) to hold for all words $\vJ$ of length $s$ and use this assumption  to evaluate both side of (\ref{funstep.1}) for all words of length $s{+}1$ parametrized by $\vJ \rightarrow T\vJ$. 

\sm

The left side of (\ref{funstep.1}) may be evaluated for a word $T \vJ$ using Leibniz's rule for $B_{jT}$,
\bea
\big[ a_i^M , B_{jT \vJ} \, a_j^N \big]  
=  B_{jT} \,  \big[ a_i^M , B_{j\vJ} \,  a_j^N \big] - \delta^M_T \,  \big[t_{ij}, B_{j \vJ} \, a_j^N \big]
\label{D.a.4}
\eea
and then applying the inductive assumption to represent the first term on the right side.
\bea
\big[ a_i^M , B_{jT \vJ} \, a_j^N \big]  
=  - \sum_{\vJ = \vP L \vQ} \delta^M_L \,  B_{jT} \big[ B_{i \vQ}  a_i^N , B_{j \vP} t_{ij}\big]  
+ B_{jT} \mS^{MN}_{i | j \vJ} - \delta^M_T \,  \big[t_{ij}, B_{j \vJ} \, a_j^N \big]
\label{D.a.5}
\eea 
Using Leibniz's rule for $B_{jT}$ under the sum, we obtain, 
\bea
\big[ a_i^M , B_{jT \vJ} \, a_j^N \big]  
& =  & - \sum_{\vJ = \vP L \vQ} \delta^M_L \,  \delta^N_T  \big[ B_{i \vQ}  t_{ij}  , B_{j \vP} t_{ij}\big]  
- \sum_{\vJ = \vP L \vQ} \delta^M_L \,  \big[ B_{i \vQ}  a_i^N , B_{j T \vP} t_{ij}\big] 
\no \\ &&
+ B_{jT} \mS^{MN}_{i | j \vJ} - \delta^M_T \,  \big[t_{ij}, B_{j \vJ} \, a_j^N \big]
\label{D.a.6}
\eea 
To relate $B_{jT} \mS^{MN}_{i | j \vJ}$ to $\mS^{MN}_{i | j T \vJ}$ we evaluate (\ref{capp.06}) for the word $\vJ \to T \vJ$ by decomposing  the deconcatenation sum over $T \vJ = \vP K \vQ L \vR$ in the expression for $\mS^{MN} _{i | j \vJ}$ of (\ref{capp.06}) into deconcatenation sums over $ \vJ = \vP K \vQ L \vR$ and $\vJ = \vQ L \vR$ using the following formula for an arbitrary function $\f$ of three words and one letter,\footnote{Note that the only dependence on the letter $K$ in the summand on the left side is through $\delta^M_K$, the function $\f$ being independent of $K$.}
\bea
\label{decTJ}
\sum_{T\vec{J} = \vec{P} K \vec{Q} L\vR}  \! \!  \delta^M_K \, \varphi(\vP,\vQ,L,\vR) 
=\! \! \! \sum_{\vec{J} = \vec{P} K \vec{Q} L\vR}  \! \!  \delta^M_K \, \varphi(T\vP,\vQ,L,\vR)
 +\delta^M_T \! \sum_{\vec{J} =  \vec{Q} L\vR} \! \varphi(\emptyset,\vQ,L,\vR)
 \qquad
\eea
The result is as follows,
\bea
\label{D.a.7}
\mS^{MN} _{i | j T \vJ} & = &
B_{jT} \, \mS^{MN} _{i | j  \vJ} 
- \sum _{\vJ = \vQ L \vR} \delta ^M_L \, \delta ^N_T \, \big [ B_{i \vR} \, t_{ij}, B_{i \cS(\vQ)} \, t_{ij} \big ]
\no \\ &&
+ \delta ^M_T \sum _{\vJ = \vQ L \vR} \delta ^N_L  \, \delta _{\vR \not= \emptyset} \sum_{\vX, \vY} \delta _{\vQ, \vX \shuffle \vY} \big [ t_{ij}, B_{i \vX} \big ( B_{i \vR} + B_{i \cS(\vR)} \big ) B_{i \cS(\vY)} t_{ij} \big ]
\eea
The sum on the second line may be expressed in terms of two commutators with the help of Lemma \ref{lem.ijij1}, and we obtain the alternative expression, 
\bea
\label{D.a.8}
\mS^{MN} _{i | j T \vJ} & = &
B_{jT} \, \mS^{MN} _{i | j  \vJ} 
- \sum _{\vJ = \vQ L \vR} \delta ^M_L \delta ^N_T \big [ B_{i \vR} \, t_{ij}, B_{i \cS)\vQ)} \, t_{ij} \big ]
+ \delta ^M_T \big [ B_{i \vJ} a^N_i + B_{j \vJ} a_j ^N , t_{ij}]
\eea
Using the above relation to eliminate $B_{jT} \, \mS^{MN} _{i | j  \vJ}$ from (\ref{D.a.6}), we obtain, 
\bea
\label{D.a.9}
\big[ a_i^M , B_{jT \vJ} \, a_j^N \big]  
& =  &
- \sum_{\vJ = \vP L \vQ} \delta^M_L \,  \big[ B_{i \vQ}  a_i^N , B_{j T \vP} t_{ij}\big] 
- \delta ^M_T \big [ B_{i \vJ} a^N_i  , t_{ij}] + \mS^{MN} _{i | j T \vJ}
\eea 
where the contribution from the first term on the right side of (\ref{D.a.6}) is cancelled by the second term on the right of (\ref{D.a.8}) and the last term in (\ref{D.a.6}) is cancelled by the last term in the commutator of (\ref{D.a.8}). Finally, using the relation (\ref{decRJ}), we recognize the sum of the first two terms on the right side of (\ref{D.a.9}) as the first term on the right side of (\ref{funstep.1}), evaluated for $\vJ \to T \vJ$. This concludes the proof of the inductive step and of Lemma \ref{funstep}. 
\end{proof}

\subsection{Decomposing $\cC^{(2)}_{ij}$}

Substituting the expression for the commutator in (\ref{capp.04}) of Proposition \ref{bigprop} into the expression for $\cC^{(2)}_{ij}$ in (\ref{D.a.1}), we obtain the following decomposition, 
\bea
\label{D.b.1}
\cC^{(2)}_{ij} = \cC^{(2\mR )}_{ij} + \cC^{(2 \mS)}_{ij}
\eea
where the components are given by,
\bea
\label{D.b.2}
\cC^{(2\mR )}_{ij} & = & \sum _{\vI, \vJ} f^\vI {}_M (i,j) \, f^\vJ {}_N (j,i) \, \mR^{MN} _{i \vI | j \vJ}
\no \\
\cC^{(2\mS )}_{ij} & = & \sum _{\vI, \vJ} f^\vI {}_M (i,j) \, f^\vJ {}_N (j,i) \, B_{i \vI} \, \mS^{MN} _{i | j \vJ}
\eea
with the expressions for $\mR^{MN} _{i \vI | j \vJ}$ and $\mS^{MN} _{i | j \vJ}$ given in (\ref{capp.05}) and (\ref{capp.06}), respectively.

\subsection{Decomposing $\cC^{(2\mR )}_{ij} $}
\label{apd3}

In this subsection, we shall prove the following decomposition formula for the function $\cC^{(2\mR)}_{ij}$ onto the linearly independent Lie algebra elements  identified in Lemma \ref{2.lem:5}. 

{\prop
\label{D.prop:4}
We have the following expression for $\cC_{ij}^{(2\mR)}$, 
\bea
 \label{D.c.4}
\cC^{(2\mR )}_{ij} & = & 
\sum _{\vJ} \cR^{\emptyset | \vJ} _{ij} {}_K \big [ a^K _i , B_{j \vJ} \, t_{ij} \big ]
+ \sum _{ \vI, \vJ}  \cR^{\vI M | \vJ} _{ij} {}_K  \Big [ B_{i \vI} \, \Big ( B_{iM} a^K _i 
- { 1 \over h} B_{iN} a_i^N \delta ^K_M \Big ) , B_{j \vJ} \, t_{ij} \Big ]
\no \\ &&
- { 1 \over h} \sum _{k \not= i,j} \sum _{ \vI, \vJ} \cR^{\vI M | \vJ} _{ij} {}_M   \big [ B_{i \vI} \, t_{ik}, B_{j \vJ} \, t_{ij} \big ]
- { 1 \over h} \sum _{ \vI, \vJ} \cR^{\vI M | \vJ} _{ij} {}_M  \big [ B_{i \vI} \, t_{ij} , B_{j \vJ} \, t_{ij} \big ] 
\eea
where $\cR^{\vP | \vQ} _{ij} {}_S$ is given by (\ref{D.c.3}).}

\begin{proof}
\vskip 0in
To prove the proposition, we substitute the expression of (\ref{capp.06}) for $\mR^{MN} _{i \vI | j \vJ}$ into the first formula of (\ref{D.b.2}), 
\bea
 \label{D.c.1}
\cC^{(2\mR )}_{ij} & = &
- \sum _{\vI, \vP. \vQ} f^\vI {}_L (i,j) \, f^{\vP L \vQ}  {}_M (j,i) B_{i \vI} \big [ B_{i \vQ} \,  a_i^M , B_{j \vP} \, t_{ij} \big ]
\no \\ &&
- \sum _{\vJ, \vP, \vQ} \, f^{ \vP L \vQ} {}_M (i,j) f^\vJ{}_L (j,i) B_{i \vP} \big [ B_{i \vQ} \, a_i ^M , B_{j \vJ} \, t_{ij} \big ]
\eea
Using the Leibniz rule to move $B_{i \vI}$ and $B_{i \vP}$ inside their respective commutators, we obtain, 
\bea
 \label{D.c.2}
\cC^{(2\mR )}_{ij} & = &
- \sum _{\vP. \vQ. \vX, \vY} f^{ \vX \shuffle \cS(\vY)} {}_L (i,j) \, f^{\vP L \vQ}  {}_M (j,i) 
B_{i \vI} \big [ B_{i \vX \vQ} \,  a_i^M , B_{j \vP \vY} \, t_{ij} \big ]
\no \\ &&
- \sum _{\vP,  \vQ, \vX, \vY} f^{ (\vX \shuffle \cS(\vY))  L \vQ} {}_M (i,j) \, f^\vP{}_L (j,i) 
\big [ B_{i \vX \vQ} \, a_i ^M , B_{j \vP \vY} \, t_{ij} \big ]
\eea 
Recasting the sums in terms of the deconcatenations $\vI = \vX \vQ$ and $\vJ = \vP \vY$, we obtain, 
\bea
 \label{D.c.3a}
\cC^{(2\mR )}_{ij} & = & \sum _{ \vI, \vJ} \cR^{\vI | \vJ} _{ij} {}_M \big [ B_{i \vI} \, a^M _i , B_{j \vJ} \, t_{ij} \big ]
\eea
After letting $\vX \to \vP$, $\vP \to \vR$ and $\vY \to \vS$, the coefficient may be identified with the function $\cR^{\vI | \vJ} _{ij} {}_M$ given in (\ref{D.c.3}).  Shortly, we shall evaluate $\cR ^{ \vI | \vJ}_{ij} {}_K$ in terms of functions that bring its expression closer to the interchange and Fay identities. To do so, we shall distinguish between the cases $\vI = \emptyset$ and $\vI \not= \emptyset$ which may be parametrized by the index $\vI \to \vI M$ for an arbitrary single letter $M$ and an arbitrary word $\vI$. As a result, the expression (\ref{D.c.3a}) may be decomposed as follows, 
\bea
 \label{D.c.3b}
\cC^{(2\mR )}_{ij} & = & 
\sum _{\vJ} \cR^{\emptyset | \vJ} _{ij} {}_K \big [ a^K _i , B_{j \vJ} \, t_{ij} \big ]
+ \sum _{ \vI, \vJ}  \cR^{\vI M | \vJ} _{ij} {}_K  \Big [ B_{i \vI} \, \Big ( B_{iM} a^K _i - { 1 \over h} B_{iN} a_i^N \delta ^K_M \Big ) , B_{j \vJ} \, t_{ij} \Big ]
\no \\ &&
+ { 1 \over h} \sum _{ \vI, \vJ} \cR^{\vI M | \vJ} _{ij} {}_M  \big [ B_{i \vI} \, B_{iN} a^N _i , B_{j \vJ} \, t_{ij} \big ]
\eea
where we have further decomposed the contribution for the index $\vI M$ into its traceless part in the indices $M$ and $K$ in the second term and its trace part in the third term. Using the structure relation (\ref{2.lem.4}) of Lemma \ref{2.lem:4}, we recast the commutator in the last term as follows, 
\bea
\big [ B_{i \vI} \, B_{iN} a^N _i , B_{j \vJ} \, t_{ij} \big ]
& = &
- \big [ B_{i \vI} \, t_{ij} , B_{j \vJ} \, t_{ij} \big ] - \sum _{k \not= i,j} \big [ B_{i \vI} \, t_{ik}, B_{j \vJ} \, t_{ij} \big ]
\eea
which gives  the final result in the form of Proposition \ref{D.prop:4}.
\end{proof}

\subsection{Decomposing $\cC^{(2\mS )}_{ij} $}

In this subsection, we shall prove the following decomposition formula for the function $\cC^{(2\mS)}_{ij}$ defined by (\ref{D.b.2}) onto the linearly independent Lie algebra elements  identified in Lemma~\ref{2.lem:5}. 

{\prop
\label{D.prop:5}
The function $\cC^{(2\mS )}_{ij} $ may be expressed as follows,
\bea
\label{D.prop.5}
\cC^{(2\mS )}_{ij} & = & 
\half \sum _{\vI, \vJ, \vK} f^\vK {}_M (i,j) \, \cT_{ij} ^{M | \vI | \vJ} B_{i \vK} \, \big [ B_{i \vI} \, t_{ij}, B_{i \vJ} \, t_{ij} \big ]
\eea
where the combination $\cT_{ij} ^{M |\vI |\vJ} $ is given by,
\bea
\label{D.prop.5a}
\cT_{ij} ^{M |\vI |\vJ} & = & - \sum _{\cS(\vJ) M \vI = \vX \vY \vZ; \vY \not= \emptyset} 
f^{(\vX \shuffle \cS(\vZ)) N ( \vY + \cS(\vY))} {}_N(j,i)
\eea}

\begin{proof}
\vskip -0.05in
To start off, we recast the second equation in (\ref{D.b.2}) as follows,
\bea
\label{D.q.1}
\cC^{(2\mS )}_{ij} & = & \sum _{\vK} f^\vK {}_M (i,j) \, B_{i \vK} \, \Psi ^M_{ij} 
\hskip 1in
\Psi ^M_{ij} = \sum _{\vJ} f^\vJ {}_N (j,i) \, \mS^{MN} _{i | j \vJ}
\eea
where the combination $\mS^{MN} _{i | j \vJ}$ was defined by (\ref{capp.06}) of Proposition \ref{bigprop}. Next, we 
rearrange the sum over $\vJ$ in $\Psi ^M_{ij}$ as follows,
\bea
\label{D.q.2}
\Psi ^M_{ij}
& = &
- \sum_{\vP , \vQ , \vR} f^{\vP L \vQ M \vR} {}_L (j,i) \  B_{j \vP}   \big[ B_{i \vR}   t_{ij} , B_{i \cS(\vQ)} t_{ij}\big]
 \\ &&
+  \sum_{\vP, \vX, \vY, \vR\not= \emptyset} f^{\vP M (\vX \shuffle \vY) L \vR} {}_L(j,i) 
B_{j \vP}  \big[ t_{ij} ,  B_{i \vX}  \big( B_{i \vR}+B_{i \cS(\vR)} \big) B_{ i \cS(\vY) } t_{ij}\big] 
 \no 
\eea
Using Leibniz's rule for $B_{j \vP}$ in both terms, 
\bea
\label{D.q.3}
\Psi ^M_{ij}
& = &
- \sum_{ \vQ , \vR, \vX, \vY} f^{(\vX \shuffle \vY)  L \vQ M \vR} {}_L(j,i) \,   
\big[ B_{i  \vR \cS(\vX) }  \, t_{ij} , B_{i \cS(\vY \vQ)} t_{ij}\big]
 \\ &&
+  \sum_{\vT, \vU, \vX, \vY,  \vR\not= \emptyset} f^{(\vT \shuffle \vU)  M (\vX \shuffle \vY) L \vR} {}_L(j,i) 
 \big[ B_{i \cS(\vT)} t_{ij} ,  B_{i \vX}  \big( B_{i \vR}+B_{i \cS(\vR)} \big) B_{ i \cS(\vU \vY) } t_{ij}\big] 
 \no 
\eea
Changing summation variables $\vX \to \cS(\vX)$, $\vY \to \cS(\vY)$, $\vQ \to \cS(\vQ)$ in the first line and $\vT \to \cS(\vI)$, $\vU \to \cS(\vU)$, $\vY \to \cS(\vY)$ in the second line, we obtain, 
\bea
\label{D.q.4}
\Psi ^M_{ij}
& = &
- \sum_{ \vQ , \vR, \vX, \vY} f^{(\cS(\vX) \shuffle \cS(\vY))  L \cS(\vQ) M \vR} {}_L(j,i) \,   
\big[ B_{i  \vR \vX }  \, t_{ij} , B_{i \vQ \vY} t_{ij}\big]
 \\ &&
+  \sum_{\vI, \vU, \vX, \vY,  \vR\not= \emptyset} f^{(\cS(\vI) \shuffle \cS(\vU))  M (\vX \shuffle \cS(\vY)) L \vR} {}_L(j,i) 
 \big[ B_{i \vI} t_{ij} ,  B_{i \vX}  \big( B_{i \vR}+B_{i \cS(\vR)} \big) B_{ i \vY \vU } t_{ij}\big] 
 \no 
\eea
Recasting the sum on the first line in terms of $\vI = \vR \vX$ and $\vJ = \vQ \vY$, and in the second line in terms of $\vJ = \vX(\vR + \cS(\vR)) \vY \vU$ we obtain, after relabeling of the summation variables, 
\bea
\label{D.q.5}
\Psi ^M_{ij}
& = &
 \sum _{\vI, \vJ} \big[ B_{i  \vI }  \, t_{ij} , B_{i \vJ} t_{ij}\big] \bigg \{ 
 - \sum_{ \vI = \vP \vQ} \sum_{\vJ = \vR \vS} f^{(\cS(\vQ) \shuffle \cS(\vS))  L \cS(\vR) M \vP} {}_L(j,i) \,   
 \\ && \hskip 1.2in
+ \sum_{\vJ = \vX \vR \vY \vU; \vR \not= \emptyset} f^{(\cS(\vI) \shuffle \cS(\vU))  M (\vX \shuffle \cS(\vY)) L (\vR + \cS(\vR)) } {}_L(j,i) \bigg \}
 \no 
\eea
The anti-symmetry under swapping $\vI$ and $\vJ$ of the commutator $\big[ B_{i  \vI }  \, t_{ij} , B_{i \vJ} \, t_{ij}\big]$ guarantees that only the anti-symmetric part of the large parentheses enters into the above formula. Defining the anti-symmetric combination,
\bea
\label{D.q.6}
\cT_{ij} ^{M |\vI |\vJ} & = &
 \sum_{ \vI = \vP \vQ} \sum_{\vJ = \vR \vS} f^{(\cS(\vQ) \shuffle \cS(\vS))  L \big ( \cS(\vP) M \vR - \cS(\vR) M \vP \big)} {}_L(j,i) \,   
 \\ && \quad
+  \bigg\{ \sum_{\vJ = \vX \vR \vY \vZ; \vR \not= \emptyset} f^{(\cS(\vI) \shuffle \cS(\vU))  M (\vX \shuffle \cS(\vY)) L (\vR + \cS(\vR)) } {}_L(j,i) - ( \vI \leftrightarrow \vJ)  \bigg\} 
\no
\eea
we obtain the expression (\ref{D.prop.5}) for $\cC^{(2\mS )}_{ij}$. In particular, using Lemma \ref{matchT} to combine the three terms on the right of (\ref{D.q.6}), we obtain the considerably simplified expression (\ref{D.prop.5a}) for $\cT_{ij} ^{M |\vI |\vJ} $ and thereby conclude the proof of Proposition \ref{D.prop:5}.
\end{proof}

\subsection{Simplifying $\cR ^{ \vI | \vJ}_{ij} {}_K$}

In this section, we shall simplify the quantity  $\cR ^{ \vI | \vJ}_{ij} {}_K$ which is defined by (\ref{D.c.3}),
governs the expression (\ref{D.c.4}) for $\cC^{(2\mR )}_{ij}$ and will eventually produce the combinations
$\Xi ^{\vI |\vJ}(i,j)$ in (\ref{3.f.2}) that cancel between ${\cal C}^{(2)}_{ij}$ and ${\cal C}^{(0)}_{ij}{+}{\cal C}^{(1)}_{ij}$.
The first double sum in the expression for $\cR$ in (\ref{D.c.3})  may be simplified to a single sum using the first identity of (\ref{spc27}) evaluated for $\vR=  \cS(\vS)$,  so that $\cR ^{ \vI | \vJ}_{ij} {}_K$ in (\ref{D.c.3}) becomes, 
\bea
\label{D.d.2}
\cR ^{ \vI | \vJ}_{ij} {}_K = 
- \sum _{\vJ = \vR \vS}  f^{ \vI \shuffle \cS(\vS)  L } {}_K (i,j) \, f^\vR {}_L (j,i) 
 - \sum_{\vI= \vP \vQ} \sum _{\vJ = \vR \vS}  f^{\vP \shuffle \cS(\vS)} {}_L (i,j) \, f^{\vR L \vQ} {}_K (j,i) 
\eea
We shall now evaluate the different components of $\cR ^{ \vI | \vJ}_{ij} {}_K$ that enter into the decomposition of $\cC^{(2\mR)}_{ij}$ given in (\ref{D.c.4}) of Proposition \ref{D.prop:4}. When $\vI= \emptyset$ in (\ref{D.d.2}), comparison of the resulting expression with the one for $\mP^\vJ {}_K(i,j)$   in (\ref{2.c.2})  establishes the proposition below.  

{\prop
\label{D.prop:2}
The component $\cR ^{ \emptyset | \vJ}_{ij} {}_K$ is given as follows, 
\bea
\label{D.prop.2}
\cR ^{ \emptyset | \vJ}_{ij} {}_K & = & - \mP^\vJ {}_K (i,j)
\eea
where $\mP^\vJ {}_K (i,j)$ was defined in (\ref{2.c.2}). }

\sm

When $\vI \not= \emptyset$ in (\ref{D.d.2}), we  use the parametrization $\vI \to \vI M$ for a single letter $M$ and an arbitrary word $\vI$. We may project onto the traceless and trace components in the indices $K$ and $M$, both of which enter into the decomposition of $\cC^{(2\mR)}_{ij}$ in (\ref{D.c.4}) of Proposition \ref{D.prop:4}.  To evaluate the traceless and trace parts  we shall make use of two lemmas.  

{\lem
\label{D.lem:10}
The expression for $\cR ^{ \vI M | \vJ}_{ij} {}_K$ may be recast as follows with the help of $\mF_1$, 
\bea
\label{D.d.6}
\cR ^{ \vI M | \vJ}_{ij} {}_K & = &  \sum _{\vJ = \vR \vS} \mF_1^{\vI \shuffle \cS(\vS) | \vR | M } {}_K (i,j) 
- \sum _{\vJ = \vR \vS}   f^{ (\vI  \shuffle \cS(\vS)  L)M  } {}_K (i,j) \, f^\vR {}_L (j,i)  
 \\ && 
- \sum_{\vI = \vP \vQ} \sum _{\vJ = \vR \vS}  f^{\vP \shuffle \cS(\vS)} {}_L (i,j) \, f^{\vR L \vQ M} {}_K (j,i) 
\no \qquad
\eea}

\begin{proof}
\vskip -0.2in
To prove the lemma, we begin by decomposing the deconcatenation sum of the composite index $\vI M = \vP' \vQ'$ in (\ref{D.d.2}) into the contributions $(\vP' , \vQ')= (\vP, \vQ M)$ and $(\vP' , \vQ')= (\vI M, \emptyset)$ with $\vI = \vP \vQ$, and we obtain,
\bea
\label{D.d.3}
\cR ^{ \vI M | \vJ}_{ij} {}_K & = & 
- \sum _{\vJ = \vR \vS}  \Big ( f^{  (\vI M \shuffle \cS(\vS)  )  L } {}_K (i,j) 
+ f^{ (\vI  \shuffle \cS(\vS)  L)M  } {}_K (i,j) \Big ) \, f^\vR {}_L (j,i)  
 \\ && 
 - \sum _{\vJ = \vR \vS}  f^{\vI M \shuffle \cS(\vS)} {}_L (i,j) \, f^{\vR L} {}_K (j,i) 
  - \sum_{\vI = \vP \vQ} \sum _{\vJ = \vR \vS}  f^{\vP \shuffle \cS(\vS)} {}_L (i,j) \, f^{\vR L \vQ M} {}_K (j,i) 
\no \qquad
\eea
where we have used the shuffle relation $\vI M \shuffle \cS(\vS) L = (\vI M \shuffle \cS(\vS)) L + (\vI  \shuffle \cS(\vS) L)M $ to further decompose the first term of (\ref{D.d.2}) for $\vI \to \vI M$.  To prove (\ref{D.d.6}), we perform the deconcatenation sum of the expression for $\mF_1$ given by (\ref{2.d.2b}),  
\bea
\label{D.d.4}
\sum _{\vJ = \vR \vS} \mF_1^{\vI \shuffle \cS(\vS) | \vR | M } {}_K (i,j) & = &
 - \sum _{\vJ = \vR \vS}  \sum_{\vR = \vX \vY } 
\Big (  f^{\vX}{}_L(j,i) \, f^{(\vI \shuffle \cS(\vS) ) M \cS(\vY)  L}{}_{\! K}(i,j) 
\no \\ && \hskip 1in
+f^{(\vI \shuffle \cS(\vS) ) M \cS(\vY)}{}_L(i,j)   f^{\vX L}{}_{\! K}(j,i)  \Big )
\eea
Recasting the sum over $\vR, \vS, \vX, \vY$ as a deconcatenation of $\vJ$ into three words  $\vJ = \vX \vY \vS$ we may rearrange this sum in terms of a deconcatenation of $\vJ$ into $\vJ = \vX \vT$ and $\vT= \vY \vS$. The deconcatenation sum for $\vT=\vY\vS$ may be carried out using the formula,
\bea
\label{D.d.5}
\sum_{\vT = \vY \vS} \big(\vI \shuffle \cS(\vS) \big) M \cS(\vY)= \vI M \shuffle \cS(\vT)
\eea
which is obtained from the left equation in (\ref{spc27}) by letting $\vI \to \cS(\vT)$, $\vP \to \cS(\vS)$, $\vQ \to \cS(\vY)$,  $\vR \to \vI$and $L \to M$. As a result, we obtain the following simplified formula, 
\bea
\label{D.d.10}
\sum _{\vJ = \vR \vS} \mF_1^{\vI \shuffle \cS(\vS) | \vR | M } {}_K (i,j) & = &
- \sum_{\vJ = \vR \vS} \Big ( f^{( \vI M \shuffle \cS(\vS) ) L} {}_K(i,j) \, f^\vR{}_L(j,i) 
\no \\ && \hskip 0.6in
+ f^{ \vI M \shuffle \cS(\vS) } {}_L(i,j) \, f^{\vR L} {}_K(j,i)  \Big )
\eea
Eliminating the first terms on the right of the first and second lines of (\ref{D.d.3}) in favor of $\mF_1$ using (\ref{D.d.10})  gives (\ref{D.d.6}), which completes the proof of the lemma.
\end{proof}

{\lem
\label{D.prop:1}
The function $\cR $ may be expressed in terms of a deconcatenation sum of the coincident limit of the combination 
$\mF$ in (\ref{2.d.3}) and (\ref{2.d.5}), 
\bea
\label{D.cor.1}
\cR ^{ \vI M | \vJ}_{ij} {}_K & = & 
\sum_{\vJ = \vR \vS}  \mF^{\vI \shuffle \cS(\vS) | \vR | M}{}_K(i,j;j)
- \p_i \p_j \cG ^{\vI \shuffle \cS(\vJ)} (i,j)  \, \delta ^M_K 
\no \\ &&
- \delta ^M _K \sum _{\vI = \vP \vQ} \sum _{\vJ = \vR \vS} 
f^{\vP \shuffle \cS(\vS)} {}_L(i,j) \, \Big ( \hf^{\vR L \vQ} (j)  - \p_j \cG ^{\vR L \vQ} (j,i) \Big ) 
\eea
where the quantities $\hf^{\vI} (j) $ are defined by (\ref{2.h.2}).
}

\begin{proof}
\vskip -0.05in
To prove the lemma, we use the decomposition  $\mF= \mF_0+\mF_1$ given in (\ref{2.d.5}), the formula (\ref{D.d.6}) of Lemma \ref{D.lem:10}, and the coincident limit of $\mF_0$ given in (\ref{2.h.5}). It is now straightforward to carry out the deconcatenation sum, and we obtain, 
\bea
\label{D.lem.6a}
\sum _{\vJ = \vR \vS} \mF_0^{\vI \shuffle \cS(\vS) | \vR | M } {}_K(i,j;j) 
& = & 
- \sum _{\vJ = \vR \vS} f^{(\vI \shuffle \cS(\vS) L) M} {}_K(i,j) \, f^\vR{}_L(j,i) 
+  \p_i \p_j \cG^{\vI \shuffle \cS(\vJ)} (i,j) \, \delta ^M_K 
\no \\ &&
- \sum_{\vI = \vP \vQ} \sum _{\vJ = \vR \vS} f^{\vP \shuffle \cS(\vS)} {}_L(i,j) \, \phi^{\vR L \vQ M}{}_K(j)
\eea
where $\phi^{\vI M}{}_K(x)$ is defined after (\ref{2.h.5}). Combining (\ref{D.d.6}),  (\ref{D.lem.6a}) and (\ref{2.b.4}) gives (\ref{D.cor.1}), thereby proving Lemma \ref{D.prop:1}. 
\end{proof}

With the help of Lemmas \ref{D.lem:10} and \ref{D.prop:1} we are ready to establish the following expressions for $\cR ^{ \vI M | \vJ}_{ij} {}_K$, which we decompose into its traceless and trace parts in the indices $M$ and $K$.

{\prop
\label{D.prop:3}
The traceless  and trace parts of $\cR ^{ \vI M | \vJ}_{ij} {}_K$ are given as follows,
\bea
\label{D.prop.3}
\cR ^{ \vI M | \vJ}_{ij} {}_K - { 1 \over h} \cR ^{ \vI N | \vJ}_{ij} {}_N \, \delta^M_K  & = & 
 \sum_{\vJ = \vR \vS}  \mH^{\vI \shuffle \cS(\vS) | \vR | M}{}_K(i,j)
\no \\ 
\cR_{ij} ^{ \vI M | \vJ}{}_M & = & - \sum _{\vJ = \vR \vS} \Xi^{\vI \shuffle \theta (\vS) | \vR} (i,j)
\eea
where $\mH^{\vI | \vJ | M}{}_K (i,j)$ and  $\Xi ^{\vI | \vJ} (i,j)$ were defined in (\ref{2.d.11}) and (\ref{3.f.2}), respectively.}

\begin{proof}
\vskip -0in
The first line is established  by taking the traceless projection of (\ref{D.cor.1}) in Lemma \ref{D.prop:1}, which kills the second term on the first line and the entire second line, and then using the fact that the traceless part of $\mF^{\vI |\vJ|M}{}_K(i,j;k)$ is independent of $x_k$. 

\sm

To establish the second line of (\ref{D.prop.3}), we carry out the deconcatenation sum of $\Xi ^{\vI | \vJ}(i,j)$, 
\bea
\label{D.70}
\sum_{\vJ = \vR \vS} \Xi ^{\vI \shuffle \cS(\vS) | \vR} (i,j) 
& =  &
- \sum _{\vJ = \vR \vS} \mF_1^{\vI \shuffle \cS(\vS) | \vR | L}{}_L(i,j) 
\no \\ &&
 - h \sum_{\vJ = \vR \vS} \,  \sum_{\vI \shuffle \cS(\vS) = \vX \vY }  f^{\vX}{}_L(i,j) \, \p_j \cG^{\vR \shuffle L \vY} (j,i) 
 \no \\ &&
-  h \sum_{\vJ = \vR \vS}  \sum_{\vR = \vX \vY } f^{\vX}{}_L(j,i) \, \p_i \cG^{\vI \shuffle \cS(\vS) \shuffle L  \vY}(i,j)
\qquad
\eea
and match with the trace of the expression (\ref{D.d.6}) for $-\cR ^{ \vI M | \vJ}_{ij} {}_K$ rather than (\ref{D.cor.1}).
The first term in (\ref{D.70}) on the right precisely matches the opposite of the  trace of the first term in (\ref{D.d.6}) and requires no further evaluation. To simplify the second term, we use Lemma \ref{lem.decsh} to evaluate the deconcatenation of the shuffle product $\vI \shuffle \cS(\vS)$, which becomes, 
\bea
 - h \sum_{\vJ = \vR \vS} \,  \sum_{\vI = \vP \vQ } \,  \sum _{\cS(\vS) = \vU \vV} 
 f^{\vP \shuffle \vU}{}_L(i,j) \, \p_j \cG^{\vR \shuffle L (\vQ \shuffle \vV)} (j,i) 
\eea
The deconcatenations of $\vJ$ and $\cS(\vS)$ amount to a deconcatenation of $\vJ$ into a triple product, $\vJ = \vR \cS(\vV) \cS(\vU)$ and may be rearranged as a sum over $\vJ = \vT \cS(\vU)$ and $\vT= \vR \cS(\vV)$, 
\bea
 - h  \sum_{\vI = \vP \vQ } \sum_{\vJ = \vT \cS(\vU)} \, \sum_{\vT = \vR \cS(\vV)} 
 f^{\vP \shuffle \vU}{}_L(i,j) \, \p_j \cG^{\vR \shuffle L (\vQ \shuffle \vV)} (j,i) 
\eea
The deconcatenation sum over $\vR$ and $\vV$ may be performed with the help of the identity,
\bea
\label{D.71}
\sum_{\vT = \vR \cS(\vV)} \vR \shuffle L (\vQ \shuffle \vV)= \vT L \vQ
\eea
which is obtained by taking the antipode of  (\ref{invid.01}) in Lemma \ref{C.words}. The result is given by,
\bea
 - h  \sum_{\vI = \vP \vQ } \sum_{\vJ = \vT \cS(\vU)} 
 f^{\vP \shuffle \vU}{}_L(i,j) \, \p_j \cG^{\vT L \vQ} (j,i) 
\eea
which, upon changing summation variables $\vT \to \vR$ and $\cS(\vU) \to \vS$ gives the opposite of the trace of the  third term in (\ref{D.d.6}). To simplify the third term in (\ref{D.70}), we recognize the double sum as a deconcatenation into three words, $\vJ = \vX \vY \vS$ and rearrange it as a sum over $\vJ = \vX \vT$ and $\vT=\vY \vS$. The sum over $\vY$ and $\vS$ may be performed with the help the special case $\vQ= \emptyset$ of (\ref{D.71}). The third term in (\ref{D.70}) then becomes, 
\bea
-  h \sum_{\vJ = \vX \vT}  f^{\vX}{}_L(j,i) \, \p_i \cG^{\vI \shuffle \cS(\vT) L }(i,j)
\eea
and gives the opposite of the trace of the second term (\ref{D.d.6}), which completes the proof of the second line in (\ref{D.prop.3}) and Proposition \ref{D.prop:3}. 
\end{proof}

Using the results of  Propositions \ref{D.prop:2} and \ref{D.prop:3}, we recast the expressions for $\cC^{(2 \mR)} _{ij} $, defined in (\ref{D.b.2}) and  given in (\ref{D.c.3a}) and (\ref{D.c.3}), in terms of the functions $\mP$, $\mH$ and $\Xi$, which gives equation (\ref{3.d.3}) of Proposition \ref{3.lem:3}.

\subsection{Combining $\cC^{(01)}_{ij}$ and $\cC^{(2 \mS)}_{ij}$}

We begin by recalling the definition of $\cC^{(01)}_{ij}$ and the simplified form of $\cC^{(2\mS )}_{ij}$
obtained in Proposition \ref{D.prop:5}, 
\bea
\label{D3.f.4}
\cC^{(01)}_{ij} & = &  
  - {1 \over h} \sum_{\vI, \vJ}   \cR^{\vI M |\vJ} _{ij}{}_M  \,   \big [ B_{i \vI} \, t_{ij} , B_{j \vJ} \,  t_{ij} \big ] 
 \no \\
\cC^{(2\mS )}_{ij} & = & 
\half \sum _{\vK} f^\vK {}_M (i,j) \,  \sum _{\vI, \vJ} \cT_{ij} ^{M |\vI |\vJ} B_{i \vK} \, \big[ B_{i  \vI }  \, t_{ij} , B_{i \vJ} t_{ij}\big] 
\eea
Substituting the expression  for $\cR^{\vI M |\vJ} _{ij}{}_M$ obtained by taking the trace of (\ref{D.cor.1}) of Lemma~\ref{D.prop:1}  into the first line of (\ref{D3.f.4}), we obtain, 
\bea
\label{D3.f.5}
\cC^{(01)}_{ij} & = & 
- { 1 \over h} \sum_{\vI, \vJ} \big [ B_{i \vI} \, t_{ij} , B_{j \vJ} \,  t_{ij} \big ]  
    \sum_{\vJ = \vR \vS}  \mF^{\vI \shuffle \cS(\vS) | \vR | M}{}_M(i,j;j) + \cC^{(01a)}_{ij} 
\eea
where $\cC^{(01a)}_{ij} $ is defined by, 
\bea 
\cC^{(01a)}_{ij}  & = & 
\sum_{\vI, \vJ} \big [ B_{i \vI} \, t_{ij} , B_{j \vJ} \,  t_{ij} \big ]   \bigg \{ \p_i \p_j \cG^{\vI \shuffle \cS(\vJ)} (i,j)  
\no \\ &&
+ \sum _{\vI = \vP \vQ} \sum _{\vJ = \vR \vS} 
f^{\vP \shuffle \cS(\vS)} {}_M(i,j) \, \Big ( \hf ^{\vR M \vQ} (j)  - \p_j \cG^{\vR L \vQ} (j,i) \Big ) 
\bigg \}
\eea
Converting $B_{j \vJ} \,  t_{ij} = B_{i \cS(\vJ)} t_{ij}$ and changing variables $\vJ \to \cS(\vJ)$ and $\cS(\vR) \leftrightarrow \vS$ we obtain, 
\bea 
\cC^{(01a)}_{ij}  & = & 
\sum_{\vI, \vJ} \big [ B_{i \vI} \, t_{ij} , B_{i \vJ} \,  t_{ij} \big ]   \bigg \{ \p_i \p_j \cG^{\vI \shuffle \vJ} (i,j)  
\no \\ &&
+ \sum _{\vI = \vP \vQ} \sum _{\vJ = \vR \vS} 
f^{\vP \shuffle \vR} {}_M(i,j) \, \Big ( \hf^{\cS(\vS) M \vQ} (j) - \p_j \cG ^{\cS(\vS) M \vQ} (j,i) \Big ) 
\bigg \}
\eea
Anti-symmetry of the commutator $\big [ B_{i \vI} \, t_{ij} , B_{i \vJ} \,  t_{ij} \big ] $ under swapping $\vI $ and $\vJ$ and commutativity of the shuffle product conspire to cancel the term in $\p_i \p_j {\cal G}^{\vI \shuffle \vJ} (i,j)  $  and anti-symmetrize the second term in $\vQ$ and $\vS$ as follows, 
\bea 
\cC^{(01a)}_{ij}  & = & 
\half \sum_{\vP, \vQ, \vR, \vS} \big [ B_{i \vP \vQ} \, t_{ij} , B_{i \vR \vS} \,  t_{ij} \big ]    
f^{\vP \shuffle \vR} {}_M(i,j) \sum_{\vW} \delta _{\vW, \cS(\vS) M \vQ} 
\no \\ && \hskip 1.5in \times 
 \Big ( \hf ^{\vW + \cS(\vW) } (j)  - \p_j \cG ^{\vW + \cS(\vW) } (j,i) \Big ) 
\eea
where we have eliminated $\vI, \vJ$ in favor of $\vP, \vQ, \vR,\vS$. 
Proposition \ref{prop:dih} defines the~combination,
\bea
\mL^\vW (j,i) = \hf^{\vW + \cS(\vW) } (j)  - \p_j \cG^{\vW + \cS(\vW) } (j,i) + \cT^{M | \vQ | \vS}_{ij}
\eea
for $\vW = \cS(\vS) M \vQ$, so that $\cC^{(01a)}_{ij} $ becomes, 
\bea 
\label{c01a}
\cC^{(01a)}_{ij}  & = & 
\half \sum_{\vP, \vQ, \vR, \vS} \big [ B_{i \vP \vQ} \, t_{ij} , B_{i \vR \vS} \,  t_{ij} \big ]    
f^{\vP \shuffle \vR} {}_M(i,j)  \Big( \mL^{\cS(\vS) M \vQ} (j,i)   -  \cT^{M | \vQ | \vS}_{ij}   \Big) 
\eea
The contributions from $\cT^{M | \vQ | \vS}_{ij} $ may be recast as $- \cC^{(2\mS)}_{ij}$ 
upon recognizing that the sum over $\vP$ and $\vR$ amounts to a Leibniz rule for $B_{i\vK}$ in (\ref{D.prop.5}).
By combining the resulting form of (\ref{c01a}) with (\ref{D3.f.5}), we are left with,
\bea
\label{D.99alt}
\cC^{(01)}_{ij} + \cC^{(2 \mS)}_{ij} & = & 
\sum_{\vI, \vR, \vS} \big [ B_{i \vI} \, t_{ij} , B_{i \vR \vS} \,  t_{ij} \big ]  \bigg \{ 
 - { 1 \over h}     \mF^{\vI \shuffle \vR | \cS(\vS) | M}{}_M(i,j;j) 
 \no \\ && \hskip 1.6in 
+ \half \sum_{\vI = \vP\vQ}  g^{\vP \shuffle \vR} {}_M(i,j) \,  \mL^{\, \cS(\vS) M \vQ} (j,i)  \bigg \}
\eea
or, equivalently, (\ref{D.99}) with $\mf$ given by (\ref{3.f.f}), completing our proof of Proposition~\ref{3.lem:3}.

\newpage

\section{Proof of Lemma \ref{2.lem:5}}
\label{sec:E}

In this appendix, we provide a proof of Lemma \ref{2.lem:5} which, we recall, states the linear independence of the Lie algebra elements in the following four sequences, for fixed values $i \not= j \in  \{ 1, \cdots, n \}$ for $n \geq 3$, and indexed by arbitrary words $\vI, \vJ \in \cW_h$, 
\begin{enumerate}
\itemsep=0in
\item  $\big [ B_{i \vI} \, t_{ij}  \, , B_{i \vJ} \, t_{ij} \big ]$
for any antisymmetrized pair of $\vI,\vJ$;
\item $\big [ B_{i \vI} \, t_{ik} \, , B_{j \vJ} \, t_{jk} \big ]$ further indexed by $k \not = i,j \in \{ 1, \cdots, n\}$;
\item $\big [ a_i^K , B_{i \vec{J}} \, t_{ij} \big ] $ further index by $K \in \{ 1,\cdots,h \}$;
\item $ \big [ B_{i \vI} \,  \big\{ B_{iK} \, a^M_i - h^{-1} \delta ^M_K \, B_{iN} a^N_i  \big\}   \, , B_{i \vJ} \, t_{ij} \big ]$ 
further indexed by  $M,K \in \{ 1,\cdots, h\} $.
\end{enumerate}

\begin{proof}
The proof of linear independence is simplified by exploiting the assumption that the indices $i \not= j$ are fixed, so that the  elements in the four sequences belong to a subalgebra of $\mt_{h,n}$ generated by $a_i^I, a_j^J, b_{iI}, b_{jJ}, t_{ij}, t_{ik}$ and $ t_{jk}$ for all values of $k \not= i,j$ and $I,J =1, \cdots, h$.

\sm

$\bullet$  The elements of sequence 1 involve only the generators $b_{iI}$ and $t_{ij}$ for all $I = 1, \cdots, h$.  As may be verified by inspecting the structure relations (\ref{11.1}), these generators are not subject to any relations in $\mt_{h,n}$, and therefore freely generate a  subalgebra of $\mt_{h,n}$. As a result, the elements in sequence 1 cannot be subject to any linear interrelations. 

\sm

$\bullet$ For $n \geq 3$, the elements in sequence 2 obey the following identity, 
\bea
\label{4.lem.3}
\big [ B_{i \vI} \, t_{ik} \, , B_{j \vJ} \, t_{jk} \big ] = \big [ B_{k \cS (\vI)} \, t_{ik} \, , B_{k \cS (\vJ)} \, t_{jk} \big ] 
\eea
For fixed $k \not= i,j$ the elements in sequence 2 involve only the generators $B_{k I}, t_{ik}$ and $t_{jk}$. As may be verified by inspecting the structure relations (\ref{11.1}), these elements are not subject to any relations and therefore freely generate a  subalgebra of $\mt_{h,n}$. As a result, for fixed $k \not= i,j$, the elements in sequence 2 must be linearly independent. 

\sm

Let us now assume the existence of a linear relation of the following form, involving elements of both sequences 1 and 2 for at least one value of $k$,
\bea
\label{4.lem.4}
\sum _{\vI, \vJ} \Big ( \cC^{\vI |\vJ} \big [ B_{i \vI} \, t_{ij}  \, , B_{i \vJ} \, t_{ij} \big ]  
+ \sum_{k \not = i,j} \cD ^{\vI |\vJ} _k \big [ B_{k  \vI} \, t_{ik} \, , B_{k \vJ} \, t_{jk} \big ]  \Big )=0
\eea
where $\cC^{\vI |\vJ}$ and $\cD ^{\vI |\vJ} _k$ are arbitrary coefficients. Here, we have re-expressed the elements of sequence 2 using (\ref{4.lem.3}) and dropped the action of the antipode without loss of generality. Applying $B_{\ell L}$ to (\ref{4.lem.4}) for arbitrary $\ell \not= i,j$ and arbitrary $L$, we find that the first term under the sum contributes zero, while the only non-vanishing contribution from the second term is for $k = \ell$. Therefore, the sum over $k$ collapses to the term $k =\ell$ and we obtain, 
\bea
\label{4.lem.5}
\sum _{\vI, \vJ}   \cD ^{\vI |\vJ} _\ell B_{\ell L}  \big [ B_{\ell  \vI} \, t_{i\ell} \, , B_{\ell  \vJ} \, t_{j\ell} \big ]  =0
\eea
This combination is composed only of elements in sequence 2 for fixed value of $\ell$ which, by our arguments of the preceding paragraph, must all be linearly independent, so that we have $\cD ^{\vI |\vJ} _\ell=0$ for all $\vI, \vJ,\ell$. The remaining combination of (\ref{4.lem.4}) involves solely elements in sequence 1 and therefore $\cC^{\vI |\vJ}=0$. This concludes our proof that the union of the elements in sequences 1 and 2 are linearly independent. 

\sm
$\bullet$ Next, for $n \geq 3$, let us assume the existence of a linear relation that includes all four sequences,
\bea
\label{4.lem.6}
&& \sum _{\vI, \vJ} \cC^{\vI |\vJ} \big [ B_{i \vI} \, t_{ij}  \, , B_{i \vJ} \, t_{ij} \big ]
+ \sum _{\vI, \vJ} \sum_{k \not= i,j} \cD^{\vI |\vJ}_k \big [ B_{i \vI} \, t_{ik} \, , B_{j \vJ} \, t_{jk} \big ]
+ \sum _{\vI} \cE^\vI _K \, [ a^K _i, B_{i \vI} \, t_{ij} ]
\no \\ && \hskip 0.6in
+ \sum_{\vI, \vJ} \cF^{\vI | \vJ | K} {}_M \Big [ B_{i \vI} \, B_{iK} \, a^M_i  \, , B_{i \vJ} \, t_{ij} \Big ]
=0
\eea
We shall leave the coefficients $\cF^{\vI | \vJ | K} {}_M$ arbitrary and prove that the condition for linear independence of all four sequences is equivalent to $\cF^{\vI | \vJ | K} {}_M$ being traceless in $K$, $M$.

\sm
 
Applying $B_{\ell L}$ with arbitrary $\ell \not= i,j$ to (\ref{4.lem.6}), we see that the contributions from the first term and from  the second term vanish for $k \not = \ell$, so that we obtain,
\bea
\label{4.lem.7}
 \sum _{\vI, \vJ}  \cD^{\vI |\vJ}_\ell B_{\ell L} \big [ B_{i \vI} \, t_{i\ell} \, , B_{j \vJ} \, t_{j\ell} \big ]
 + \sum _{\vI} \cE^\vI _L \, [ t_{i\ell} , B_{i \vI} \, t_{ij} ]
 + \sum_{\vI, \vJ} \cF^{\vI | \vJ | K} {}_L \Big [ B_{i \vI} \,  B_{iK}  t_{i\ell}  , B_{i \vJ} \, t_{ij} \Big ]
=0
\qquad
\eea
Since $i,j,\ell$ are mutually distinct, the first term may be rearranged by bringing $B_{i \vI} $ and $B_{j \vJ}$ outside of the commutator, using the second relation $[t_{i\ell}, t_{j\ell}]=- [ t_{i\ell}, t_{ij}]$ of (\ref{2.prop.1}),  moving $B_{\ell L}$ and $B_{j \vJ}$ back into the commutator, and converting $B_\ell$ and $B_j$ to $B_i$, 
\bea
\label{4.lem.9} 
\sum _{\vI, \vJ}  \cD^{\vI |\vJ}_\ell B_{i \vI} \,  \big [ B_{i L}  t_{i\ell} \, , B_{i \cS(\vJ)} t_{ij} \big ]
+ \sum _{\vI} \cE^\vI _L \, [ t_{i\ell} , B_{i \vI} \, t_{ij} ] 
 + \sum_{\vI, \vJ} \cF^{\vI | \vJ | K} {}_L \Big [ B_{i \vI K} \, t_{i\ell} , B_{i \vJ} \, t_{ij} \Big ]=0
 \qquad 
\eea
There are no relations between the generators $B_{i I}, t_{ij}$ and $t_{i\ell}$ so that the subalgebra of $\mt_{h,n}$ in which (\ref{4.lem.9}) takes values  is generated freely. The elements in the first and third terms all have at least one $B_i$ applied to $t_{i\ell}$ while no $B_i$ is applied to $t_{i\ell}$ in the second term. Therefore, the two sets of terms produce different elements of sequence 2 and must vanish separately. Thus we must have $\cE^\vI _L=0$ as well as the relation, 
\bea
\label{4.lem.10} 
\sum _{\vI, \vJ}  \cD^{\vI |\vJ}_\ell B_{i \vI} \,  \big [ B_{i L}  t_{i\ell} \, , B_{i \cS(\vJ)} t_{ij} \big ]
 + \sum_{\vI, \vJ} \cF^{\vI | \vJ | K} {}_L \Big [ B_{i \vI } \, B_{iK}  t_{i\ell}  , B_{i \vJ} \, t_{ij} \Big ]=0
\eea
Using Leibniz's formula for $B_{i \vI}$ and suitably relabeling the indices gives, 
\bea
\label{4.lem.11} 
\sum _{\vI, \vJ} \sum _{\vJ= \vR \vS}  \cD^{\vI \shuffle \vR |\cS(\vS)}_\ell  \big [ B_{i \vI} B_{i L}  t_{i\ell} \, , B_{i \vJ} t_{ij} \big ]
 + \sum_{\vI, \vJ} \cF^{\vI | \vJ | K} {}_L \Big [ B_{i \vI } \, B_{iK}  t_{i\ell}  , B_{i \vJ} \, t_{ij} \Big ]=0
\eea 
Since the algebra of $B_{iI}, t_{ij}$ and $t_{i \ell}$ is freely generated, this relation implies,
\bea
\sum _{\vJ= \vR \vS}  \cD^{\vI \shuffle \vR | \theta(\vS)}_\ell  \delta ^K_L  + \cF^{\vI | \vJ | K} {}_L =0
\eea 
so that the only non-trivial solutions feature $\cF^{\vI | \vJ | K}{}_M$ proportional to $\delta ^K_M$ with a coefficient that is independent of $K$ and $M$. 
This relation corresponds to the trace relation on the third line of  (\ref{11.1}). Therefore we have established that linear independence of the four sequences of Lemma \ref{2.lem:5} is equivalent to $\cF^{\vI | \vJ | K}{}_M$ being traceless in $K,M$, and the concatenation sum in the first term being zero. The vanishing of the deconcatenation sum for all $\vI$ and $\vJ$ is equivalent to $\cD^{\vI |\vJ}_\ell=0$ for all $\vI, \vJ$ by the arguments in the proof of Lemma \ref{3.lem:6}, thereby completing the proof of the lemma.  
\end{proof}

\newpage

\end{document}